\documentstyle[12pt,epsf,rotating]{article}
\begin{document}
\newcommand {\epem}  {e$^+$e$^-$}
\newcommand {\qq} {q$\overline{\mathrm{q}}$}
\newcommand {\pp} {p$\overline{\mathrm{p}}$}
\newcommand {\gluglu} {gg}
\newcommand {\durham} {$k_\perp$}
\newcommand {\bbbar} {b$\overline{\mathrm{b}}$}
\newcommand {\qbar} {$\overline{\mathrm{q}}$}
\newcommand {\ecm} {\mbox{$E_{\mathrm{c.m.}}$}}
\newcommand {\ejet} {\mbox{$E_{\mathrm{jet}}$}}
\newcommand {\mnch} {\langle n_{\,\mathrm{ch.}} \rangle}
\newcommand {\gincl} {g$_{\,\mathrm{incl.}}$}
\newcommand {\nff} {n_f}
\newcommand {\ratio} {r}
\newcommand {\ng} {n_{\mathrm{G}}}
\newcommand {\nf} {n_{\mathrm{F}}}
\newcommand {\alpmz} {\alpha_S(\mathrm{M_Z})}
\newcommand {\lms} {\Lambda_{\overline{\mathrm{MS}}}}
\newcommand {\rsoft} {$r_{soft}$}
\newcommand {\pperp} {$p_{\perp}$}

\begin{flushright}
FIAN TD31/00\\
UCRHEP-E273\\
April 7, 2000 \\
(revised September 18, 2000)
\end{flushright}
\vspace*{0.5cm}
\begin{center}
  {\large \bf  HADRON MULTIPLICITIES}\\
\bigskip
  {\large  I.M. Dremin$^{1}$, \,\, J.W. Gary$^{2}$}\\
\bigskip
{\it $^{1}$Lebedev Physical Institute, Moscow 117924, Russia\\
$^{2}$Department of Physics, University of California, 
  Riverside CA 92521, USA}

\vspace*{.4cm}
\begin{abstract}
\noindent
We review results on hadron multiplicities
in high energy particle collisions. 
Both theory and experiment are discussed.
The general procedures used to describe particle multiplicity 
in Quantum Chromodynamics (QCD) are summarized.
The QCD equations for the generating functions of 
the multiplicity distributions are presented
both for fixed and running coupling strengths.
The mean multiplicities of gluon and quark jets,
their ratio, higher moments,
and the slopes of multiplicities
as a function of energy scale,
are among the main global features of
multiplicity for which QCD results exist.
Recent data from high energy {\epem} experiments,
including results for separated quark and gluon jets,
allow rather direct tests of these results.
The theoretical predictions are generally quite 
successful when confronted with data. 
Jet and subjet multiplicities are described. 
Multiplicity in limited regions of
phase space is discussed in the context
of intermittency and fractality. 
The problem of
singularities in the generating functions is formulated. 
Some special features of average multiplicities 
in heavy quark jets are described.
\end{abstract}

\end{center}

\newpage
\tableofcontents

\newpage
\section{Introduction}

The number of hadrons created in high energy inelastic
collisions varies from one event to another. 
The distribution of the number of produced hadrons
(the multiplicity distribution, for short) 
provides a basic means to characterize the events.
The multiplicity distribution contains information 
about multiparticle correlations in an integrated form,
providing a general and sensitive means to probe
the dynamics of the interaction.

Here, we would like to review the main theoretical approaches
and experimental results concerning hadron multiplicities.
Much precise experimental information on multiplicities
has become available in recent years from {\epem},
hadron, and nucleus collisions.
The experimental progress has stimulated 
additional theoretical developments.
Therefore, the time seems appropriate for a review.
The theoretical results summarized here represent
an expanded and updated 
version of an earlier work~\cite{ufn}.

After introducing basic definitions and 
notation in Section~\ref{sec-definitions},
we turn in Section~\ref{sec-phenomenology} to describe 
some phenomenological approaches to multiplicity.
The phenomenological methods are based on
simplified ideas about particle emission,
exploiting widely used distributions from 
probability theory (see e.g. Ref.~\cite{6}). 
Among them,
the negative binomial distribution is especially
popular because it describes experimental data 
reasonably well for various reactions
over a wide energy interval.


The discovery of asymptotic freedom,
in conjunction with the parton model 
used to characterize deep inelastic and {\epem} data,
led to the development of the theory of strong interactions,
Quantum Chromodynamics (QCD).
QCD provides a means to apply perturbative techniques
to hadronic processes with large transferred momenta,
with quarks and gluons playing the role of partons~\cite{1}-\cite{5}.
The current state of affairs concerning multiplicity 
in quark and gluon jets as predicted by QCD and 
measured by experiment forms the bulk of our review,
presented 
in Sections~\ref{sec-equations}-\ref{sec-heavy}.

Lattice calculations imply that hadronization
--~the transition of quarks and gluons to hadrons at
the final stage of evolution~--
is an inherent property of QCD. 
Nonetheless,
we are still unable to treat hadronization 
in an unambiguous manner since the problem of confinement 
has yet to be solved mathematically.
Simplified estimates suggest
either that hadronization does not drastically 
alter the parton level results
or else that its effects can be estimated
from the energy dependence of experimental observables.
Phenomenologically, 
the distributions of partons and hadrons 
are often found to be remarkably similar.
This implies that the study of the partonic stage of
an event is of utmost importance since the 
properties of high energy multihadronic events are
primarily determined at that level.

The probabilistic equations of QCD
for the parton cascade can be solved within
the perturbative approach. 
The parton shower framework allows
the use of sub-series of the perturbative expansion,
with terms ordered according to their energy behavior. 
The zero order approximation already involves
an arbitrary number of produced particles.
Higher order approximations lead to detailed predictions 
for multiplicity distributions which
can be tested through comparison with experiment.
Before proceeding to details in the following sections,
however,
we wish to note two problematic areas and then briefly 
comment on the main topics we address.

First,
QCD yields results on partons,
and not hadrons,
as already mentioned. 
Therefore an assumption about hadronization must be
made in any comparison of experiment with theory.
A common assumption is
local parton-hadron duality (LPHD)~\cite{8},
which states that parton distributions
are simply renormalized in the process of hadronization,
without changing their shape.
LPHD originated from the
idea of soft preconfinement~\cite{9}, 
whereby partons group in colorless
clusters without disturbing the initial spectra. 
Phenomenological models of
hadronization have been incorporated into Monte Carlo 
simulations of inelastic processes and in most cases
support the approximate property of LPHD.

Second,
perturbative analysis has limitations 
in the context of any specific approximation.
Formally speaking, perturbation theory should be used
only when the coupling constant is very small
compared to unity.
This condition is fulfilled by QCD only for
extremely large transferred momenta, however,
often for energies much larger than present experimental conditions.
In any QCD process,
the energies of cascading partons degrade during their evolution, 
and a proper accounting for soft partons, 
their recoil due to interaction, 
and energy-momentum conservation laws,
should be included.
All these considerations are neglected in 
the lowest order approximation,
for which only processes with a rather large gradient of energies and
emission angles at each stage of the evolution are considered 
(the so-called double-logarithmic approximation, 
DLA\footnote{Also referred to as leading order (LO).}). 
Account is taken of soft partons and
strict transverse momentum ordering in
subsequent terms of the perturbative series, 
such as the modified leading-logarithmic approximation 
(MLLA\footnote{Also referred to as next-to-leading-order (NLO).}). 
Recoil effects and conservation laws can be incorporated
at the next-to-next-to-leading order (NNLO),
the next-to-next-to-next-to-leading order (3NLO),
and higher order stages.
The conservation laws are more accurately included,
the higher the order of the perturbative approximation.

In most cases these corrections are under control,
being about 10-20\% percent of the main terms
at present energies. 
In spite of their rather small total contribution, 
they are often very important and change the qualitative theoretical
description in regions where the corresponding functions are small. 
For example, they are crucial for the proper description 
of the multiparticle production process.
This manifests itself mathematically as a new
expansion parameter equal to the product of the QCD 
coupling constant (or, more precisely, of its square root) 
and the rank of the moment of the distribution. 
Thus it is for large ranks, i.e.~high multiplicities,
that the corrections are most important.
These problems are discussed in detail 
in Sections~\ref{sec-equations}-\ref{sec-exactsolutions}.

The first results on the multiplicity distributions of partons
were obtained using the double-logarithmic
approximation (for reviews, see Refs.~\cite{5,10}).
They yielded asymptotic Koba-Nielsen-Olesen (KNO) scaling~\cite{11}. 
According to the KNO hypothesis~\cite{7}, 
the multiplicity distribution depends only on the ratio of
the number of particles to the average multiplicity 
(see Section~\ref{sec-phenomenology}). 
KNO scaling failed to be valid in the higher order approximations,  
however,
i.e.~the asymptotic realm is too far from present energies. 
At the same time, 
the width of the asymptotic KNO function was found to
be much larger than experimental distributions. 
This problem was resolved~\cite{12,13}
through consideration of the higher order effects mentioned above.
The increase of average multiplicity with energy
was found to depend on the coupling constant in a manner
which is faster than any logarithmic function and 
slower than any power-like one 
(if the running coupling constant is used),
in agreement with experimental findings.
One can now state that an overall agreement between 
theory and data has been achieved,
at least qualitatively.

Moreover, some qualitative predictions of 
perturbative QCD (pQCD) were found to be unexpectedly 
well suited for ``soft'' hadronic processes as well. 
Given the large value of the expansion parameter at small scales,
this may seem puzzling.
However, higher order corrections should account for
ever softer partons in a consistent manner,
implying a more general correspondence 
between soft and hard processes than is usually
considered in theoretical schemes.
For example,
the newly discovered feature of the oscillation
of cumulant moments (see Section~\ref{sec-hoapprox}),
prompted by solutions of the QCD equations,
were experimentally observed in {\epem},
hadron-, and nucleus-induced reactions.
These oscillations were found to be extremely sensitive
to small details of the multiplicity distributions.

Many discussions are devoted to the value of the 
ratio of average multiplicities between gluon and quark jets,~$r$. 
The value obtained in the
double-logarithmic approximation is~$r$=9/4. 
The simplest corrections reduce this value by about~10\%.
An even larger decrease of $r$ is predicted by 
the exact solution of the equations for the generating functions 
in the case of a fixed coupling constant, 
by the analytic higher order approximations 
with a running coupling constant,
and by computer solutions of these equations. 
These results are discussed in 
Sections~\ref{sec-psolutions}-\ref{sec-exactsolutions}.
A proper accounting for the boundary between the perturbative 
and non-perturbative regions appears to be crucial
for this quantity. 
The energy dependence of the average multiplicities,
i.e.~their slopes,
is considered in Section~\ref{sec-psolutions}. 
The ratios of the slopes and of the corresponding curvatures
are found to be less sensitive to higher order corrections 
than the ratio of multiplicities itself.
Some attempts to account for non-perturbative effects in 
evolving jets are briefly described in Section~\ref{sec-nonperturbative}.

The behavior of the moments of multiplicity is 
strongly influenced by the nature of the singularities 
of the generating function, which are not yet known. 
Approaches to this problem are discussed
in Section~\ref{sec-singularities}.

Experimental measurements of multiplicity as they relate
to tests of the QCD predictions
are described in Section~\ref{sec-experiment}.
Especially noteworthy are recent data from {\epem}
annihilation experiments operating on the Z$^0$ peak,
namely the four experiments ALEPH, DELPHI, L3 and OPAL
at the LEP storage ring at CERN
and the SLD experiment at the SLC collider at SLAC.
The simplicity of the hadronic production process in {\epem} events,
along with the large data samples collected by these experiments,
has resulted in an unprecedented level of experimental precision 
for multiplicity related quantities,
leading to the possibility, for example,
of examining multiplicity in small phase space windows.
Recent progress in distinguishing quark and gluon jets 
by these experiments has made it possible 
to study their properties separately. 
In particular, the multiplicity distributions for each set have 
recently been analyzed for the first time~\cite{bib-opal98}.

The evolution of jets can be studied by resolving subjets,
and by determining the subjet multiplicity rate as a
function of the resolution scale.
Theoretical and experiment results on subjet multiplicities
are described in Section~\ref{sec-subjets}.
This is followed in Section~\ref{sec-threejets}
by a brief discussion of particle multiplicity 
in {\epem} three-jet events.

It is of interest to study multiplicity not only in total 
phase space but also in small subregions. 
These studies are generally focused on the intermittency 
phenomenon and on the fractality of particle distributions 
within a selected phase space volume
(for a recent review, see~\cite{14}), 
related to a relative widening of the multiplicity distribution 
for smaller phase space volumes. 
Intermittency gives rise to an increase of multiplicity moments 
in a power-like manner as the phase space window decreases.
Such tendencies have been experimentally observed. 
Quantum Chromodynamics provides a qualitative description
of the increase of the moments,
relates the intermittency exponents (or fractal dimensions) 
directly to the QCD anomalous dimension
(i.e.~the coupling constant), 
and clearly delineates the region of applicability 
of the regularities,
indicating the scales at which one should consider the
coupling constant to be running or at which it can be
treated as approximately fixed. 
Theoretical and experimental aspects of
intermittency and fractality
are described in Section~\ref{sec-intermittency}.

A deeper understanding of specific features of multiplicity
can be gained if inclusive distributions of 
particles and their mutual correlations are considered. 
The quantum mechanical origin of the 
interacting partons reveals itself in various interference effects. 
They lead to the hump-backed plateau of rapidity distributions, 
to correlations of partons in energies and azimuthal angles, 
to the string (or drag) effect in three-jet events,
and to interference phenomena in the production of 
heavy bosons and lepton pairs at large transverse momenta.
We do not describe these results here,
instead referring the reader to monographs~\cite{1,5} 
and to a recent review~\cite{khoc}.
Nonetheless, we discuss one interference effect, namely
the suppression of the forward production 
of accompanying particles in processes with heavy quarks. 
Inclusion of this topic is justified here because it directly affects
the relation between the mean multiplicities 
in heavy- and light-quark jets. 
A summary of theoretical and experimental results 
on multiplicity in heavy quark jets is
given in Section~\ref{sec-heavy}. 

To keep this review to a reasonable length,
we do not describe the interactions of hadrons, nuclei 
or polarized quarks except in passing.
These topics merit their own review. 
Our main purpose is to present a coherent,
updated overview of the theoretical and experimental status
of multiplicity in high energy hadron jets.
We apologize to the authors of papers
whose contributions have not been mentioned.
These omissions are unintentional.

\section{Definitions and notation}
\label{sec-definitions}

\noindent 
The multiplicity distribution is defined by the formula
\begin{equation}
  P_{n} = \frac {\sigma_{n}}{\sum_{n=0}^{\infty}\sigma_{n}}=\frac {\sigma _n}
{\sigma _{inel}} ,    \label{1}
\end{equation}
where $\sigma_{n}$ is the cross section of an
$n$-particle production process 
(the so-called topological cross section), 
$\sigma _{inel}$ is the inelastic cross section,
and the sum is over all possible values of $n$ so that
\begin{equation}
  \sum_{n=0}^{\infty}P_{n} = 1 .          \label{2}
\end{equation}
It is often more convenient to represent the 
multiplicity distribution by its moments, 
i.e.~by another set of numbers obtained from it by a definite
algorithm. 
All such sets can be obtained from 
the so-called generating function
defined by the formula
\begin{equation}
  G(y,z) = \sum_{n=0}^{\infty }P_{n}(y)(1+z)^{n}  ,    \label{3}
\end{equation}
which substitutes an analytic function of $z$ in place of the set of numbers 
$P_{n}(y)$ at a fixed energy $y$.

In what follows, use will often be made of the (normalized) factorial moments
$F_{q}$ and cumulants $K_{q}$ determined from the generating function $G(z)$
by the relations
\begin{equation}
  F_{q} = \frac {\sum_{n} P_{n}n(n-1)...(n-q+1)}{(\sum_{n} P_{n}n)^{q}} =
\frac {1}{\langle n \rangle ^{q}}\cdot \frac {d^{q}G(z)}{dz^{q}}\vline _{z=0}, 
\label{4}
\end{equation}
\begin{equation}
  K_{q} = \frac {1}{\langle n \rangle ^{q}}\cdot \frac {d^{q}\ln G(z)}{dz^{q}}
\vline _{z=0}, \label{5}
\end{equation}
where
\begin{equation}
\langle n \rangle = \sum_{n=0}^{\infty }P_{n}n              \label{6}
\end{equation}
is the average multiplicity. 
The expression for $G(z)$ can be written as
\begin{equation}
  G(z) = \sum _{q=0}^{\infty } \frac {z^q}{q!} \langle n \rangle ^{q} F_{q} 
  \;\;\;\; ( F_0 = F_1 = 1 ),    \label{7}
\end{equation}
\begin{equation}
  \ln G(z) = \sum _{q=1}^{\infty } \frac {z^q}{q!} \langle n \rangle ^{q} K_{q} 
  \;\;\;\; ( K_1 = 1 ).    \label{8}
\end{equation}
The distribution $P_{n}$ and its ordinary moments $C_{q}$
can be derived from the generating function $G(z)$ 
using the formulas
\begin{equation}
  P_{n} = \frac {1}{n!}\cdot \frac {d^{n}G(z)}{dz^{n}}\vline _{z=-1} ,  
  \label{9}
\end{equation}
\begin{equation}
  C_{q} = \frac {\sum _{n=0}^{\infty }P_{n}n^{q}}{\langle n \rangle ^{q}} =
\frac {1}{\langle n \rangle ^{q}}\cdot \frac {d^{q}G(e ^{z}-1)}{dz^{q}}
\vline _{z=0} . \label{10}
\end{equation}
All the moments are connected by definite relations that 
can be derived from their definitions given above.
For example, the factorial moments and cumulants are related 
to each other through the identities
\begin{equation}
  F_{q} = \sum _{m=0}^{q-1} C_{q-1}^{m} K_{q-m} F_{m} ,       \label{11}
\end{equation}
which are nothing other than the relations between the derivatives 
of a function and its logarithm at the point where 
the function itself equals unity. 
Here
\begin{equation}
  C_{q-1}^{m} = \frac {(q-1)!}{m!(q-m-1)!} = \frac {\Gamma (q)}{\Gamma (m+1)
\Gamma (q-m)} = \frac {1}{mB(q,m)}     \label{12}
\end{equation}
are the binomial coefficients, and $\Gamma$ and $B$ denote 
the gamma and beta functions, respectively. 
There are only numerical coefficients in the 
recurrence relations~(\ref{11}).
Therefore iterative solution yields all cumulants if
the factorial moments are given, and vice versa. 
In this sense, cumulants
and factorial moments are equally suitable. 
The relations between them for the lowest ranks are
\begin{eqnarray}
  F_{1}&=&K_{1}=1, \nonumber  \\
  F_{2}&=&K_{2}+1, \nonumber  \\
  F_{3}&=&K_{3}+3K_{2}+1, \nonumber \\
  F_{4}&=&K_{4}+4K_{3}+3K_{2}^{2}+6K_{2}+1, \nonumber \\
  F_{5}&=&K_{5}+5K_{4}+10K_{3}K_{2}+10K_{3}+15K_{2}^{2}+10K_{2}+1. 
  \label{12a}
\end{eqnarray}

The physical meaning of these moments can be seen 
from their definitions if they are presented in the form of 
integrals of correlation functions (for a review, see Ref.~\cite{14}). 
Let the single symbol $y$ represent all kinematic variables needed 
to specify the position of each particle in
the phase space volume~$\Omega $. 
A sequence of inclusive 
$q$-particle differential cross sections $d^{q}\sigma /dy_{1}\ldots dy_{q}$
defines the factorial moments in the following manner:
\begin{equation}
  F_{q}=\frac {1}{ \sigma _{inel}\langle n\rangle ^q}\int _{\Omega }dy_{1} 
  \ldots \int _{\Omega } dy_{q}\frac {d^{q}\sigma }{dy_{1}\ldots dy_{q}}. 
\label{12b}
\end{equation}
Therefore, the factorial
moments include in integrated form all correlations within the system of
particles under consideration. 
The corresponding expressions for cumulants are
more complicated as can be seen from relations~(\ref{11}) and~(\ref{12a}).
Therefore we do not present them here but instead remark that 
analogous relations exist in quantum field theory where formulas similar 
to eqs.~(\ref{4}) and~(\ref{5}) define the complete set of Feynman graphs 
(both connected and disconnected) and the subset of connected diagrams,
respectively (see e.g. Ref.~\cite{1}). 
Thus, factorial moments represent integral characteristics of any correlation
between the particles whereas cumulants of rank $q$ represent genuine
$q$-particle correlations not reducible to the product of lower order
correlations.\footnote{This interpretation is valid, however, only for
moments with a rank smaller than the average multiplicity at a given energy
(for more details, see Ref.~\cite{14}).}
To be more precise, 
all $q$ particles are connected to each other in the $q$th cumulant 
and cannot be split into disconnected groups. 
One can say they form a $q$-particle
cluster which is not divisible into smaller clusters, 
in analogy with Mayer cluster decomposition in statistical mechanics.

It is a common feature of distributions in particle physics that their
factorial moments and cumulants increase rapidly 
as the rank $q$ increases.
Therefore it is often convenient to consider their ratio 
introduced in Ref.~\cite{13}:
\begin{equation}
  H_{q} = \frac {K_q}{F_q}       ,      \label{13}
\end{equation}
which does not share this feature.
At the same time the $H_q$ ratios exhibit
all the qualitative features of the cumulants. 
As will be discussed below, 
the ratios $H_{q}$ appear in a natural
manner as the solution of the QCD equations 
for the generating functions~\cite{13}.

From eq.~(\ref{4}), 
one can derive the relation 
between the factorial moments $F_q$
and the ordinary moments $C_q$ 
with the same or lower ranks.
The two types of moments are found to differ by terms
which depend only on the mean multiplicity so that, for example,
\begin{equation}
  F_{2} = \frac {\langle n(n-1)\rangle }{\langle n \rangle ^2} 
  = C_2 - \langle n \rangle ^{-1} .     \label{14}
\end{equation}
Thus ordinary and factorial moments have a different 
energy dependence,
since $\langle n\rangle ^{-1}$ is energy dependent.
In general,
factorial moments differ from ordinary moments by
lower order correlation terms suppressed by the inverse power
of the mean multiplicity to the corresponding power.
Thus ordinary and factorial moments coincide asymptotically.

The generating function contains the same
physical information as the multiplicity distribution. 
This information is also contained 
in the unnormalized moments and their ratio. 
For normalized moments, 
the average multiplicity at a given energy also needs to be specified.
Note that higher rank moments emphasize higher multiplicity events,
i.e.~multiplicities $n\geq q$ contribute to factorial moments
of (integer) rank $q$, as seen from eq.~(\ref{4}). 
If the distribution is truncated at some $n\,$=$\,n_{max}$, 
all factorial moments with rank $q>n_{max}$ are zero,
whereas they are positive for smaller~$q$. 
The cumulants may be either positive or negative
if the distribution is truncated.

An emphasis on low multiplicity events can be made
by considering
derivatives of the generating functions at $z$=-1.
For example, the so-called combinants~\cite{kagy,he},
defined by
\begin{equation}
  Q_n=\frac {1}{n!}\frac {d^n\ln G(z)}{dz^{n}} \vline _{z=-1} ,
     \label{comb}
\end{equation}
measure correlations within the low multiplicity 
part of the distribution.

Of potential interest would be moments with an exponentially
damped high-multiplicity tail,
defined by
\begin{equation}
  D_q=\frac {1}{\langle n\rangle^q}\frac {d^qG(e^z-1)}{dz^q}
  \vline _{z=-1}
  \label{dq}
\end{equation}
(see the expression for the ordinary moments $C_q$ given by 
eq.~(\ref{10})). 
Events with multiplicity $n\,$=$q$ are enhanced in these moments. 
This type of moment has not yet been studied, however.

To this point we have assumed that the rank of a
moment is a positive integer. 
However, the definitions (\ref{4}), (\ref{5}) and (\ref{10}) 
can be easily generalized to non-integer moments~\cite{15},
by re-expressing the factorial moments as
\begin{equation}
  F_{q} = \frac {1}{\langle n\rangle ^{q}}\sum _{n=0}^{\infty }P_n \frac
  {\Gamma (n+1)}{\Gamma (n-q+1)} ,       \label{15}
\end{equation}
which is valid for any real value of~$q$. 
This result can also be obtained
using a more general approach,
by forming the derivative of any real order $q$ 
of the generating function,
i.e.~by using fractional calculus.
Fractional moments have not yet found much application
in particle physics
(see section~\ref{sec-nbd}, however).
Therefore,
we refer the reader to Refs.~\cite{ufn}, \cite{16}-\cite{imdr}
for further discussion.

The method of generating functions is a particular 
realization of the more general method of generating functionals. 
The latter approach considers the probability to find a certain
$n$-particle distribution inside the phase space volume,
in place of $P_n$ in eq.~(\ref{3}), 
and therefore utilizes functions $z(y_i )$ 
instead of the variable~$z$,
where the $z(y_i)$ depend on the phase space variables~$y_i$. 
The corresponding distributions and 
correlation functions are obtained by taking the variational derivatives of 
the generating functional over the probing functions~$z(y_i )$. 
The moments of the multiplicity distribution are related 
to the integrals of the correlation functions,
as seen from eq.~(\ref{12b}). 
For more details, see Refs.~\cite{14, khoc}.

\section{Phenomenology}
\label{sec-phenomenology}

\subsection{KNO scaling and $F$ scaling}

A principal phenomenological issue is the energy dependence 
of multiplicity for different colliding particles and nuclei. 
One of the most successful assumptions about the shape of the
multiplicity distribution at high energies is the hypothesis 
that its energy dependence is determined by the 
average multiplicity in such a manner that 
$P_n$ may be represented as:
\begin{equation}
  P_{n} = \frac {1}{\langle n \rangle}
    f(\frac {n}{\langle n \rangle}) . \label{20}
\end{equation}
This property is called KNO scaling after the names of its
authors~\cite{7},
who proposed it on the basis of the Feynman 
plateau of rapidity distributions. 
Earlier, a similar relation was obtained in the framework of
conformal field theories~\cite{pol}. 
KNO scaling is usually considered to be an asymptotic property,
i.e.~valid in the limit $\langle n\rangle\rightarrow\infty$. 
The multiplicity distribution is restored from the asymptotic
function $f(x)$ through the Poisson transform
\begin{equation}
  P(n)=\int _{0}^{\infty }f(x) \frac {(\langle n\rangle x)^{n}}{n!}
  e^{-\langle n\rangle x}dx .    \label{inv}
\end{equation}
The function $f(x)$ is unspecified.

The normalization conditions are
\begin{equation}
  \int _{0}^{\infty }f(x)dx = 1, \;\;\;\;\; 
      \int _{0}^{\infty }xf(x)dx=1.\label{21}
\end{equation}
It is clear that the ordinary moments of the KNO distribution,
eq.~(\ref{20}),
do not depend on energy and are just functions of their rank~$q$:
\begin{equation}
  C_{q} = \int _{0}^{\infty }x^{q}f(x)dx = independent\; of\; E .   \label{22}
\end{equation}
In contrast, the factorial moments of the 
KNO distribution do depend on energy, 
as follows from eq.~(\ref{14}) for $F_2$, 
tending to constant values only at asymptotically high energies. 
Constancy of the factorial moments is called $F$ scaling.
Therefore $F$~scaling coincides with KNO scaling asymptotically. 
The generating function depends on the average multiplicity 
in both cases as seen 
from the definitions (\ref{3}), (\ref{7}) and~(\ref{8}).

If Feynman scaling is approximate,  KNO scaling is violated, 
suggesting the replacement of eq.~(\ref{20}) by other functions.
In particular, it has been proposed~\cite{hegy}
to utilize the set of relations with arbitrary functions~$f_k$
\begin{equation}
  P(n)=\frac {\langle n^{k}\rangle ^{2}}{n^{k}\langle n^{k+1}\rangle}
  f_{k}\left(\frac {n\langle n^{k}\rangle }{\langle n^{k+1}\rangle }\right),
\label{hegy}
\end{equation}
which reduces to KNO result for~$k$=0.

Another proposal is based on the scaling properties of systems 
undergoing a second order phase transition,
and yields~\cite{plos, bopl}
\begin{equation}
  P(n)=\frac {1}{\langle n\rangle ^{\delta }}\phi \left( 
\frac {n-\langle n\rangle }{\langle n\rangle ^{\delta }}\right), \label{plos}
\end{equation}
which reduces to KNO case for $\delta$=1. 
It has been argued that solutions of this type appear for 
QCD equations when they are modified by non-perturbative terms 
(see Section~\ref{sec-nonperturbative}).
It has also been proposed to consider so-called
log-KNO scaling~\cite{hegyi2},
which has the general form
\begin{equation}
  P(n)=\frac {1}{\lambda (s)}\varphi \left(
   \frac {\ln n+c(s)}{\lambda (s)}\right), \label{lkno}
\end{equation}
which amounts to a scale- and power-transform of multiplicity.

In eq.~(\ref{20}),
a continuous function $f$ is used to approximate
the discrete distribution~$P_n$. 
This procedure is justified for $\langle n\rangle \gg 1$. 
The procedure to restore the discrete distribution $P_n$ from 
the continuous function $f$ is~\cite{golo}
\begin{equation}
  P_{n} =\int _{n}^{n+1}\frac {dm}{\langle m\rangle } f\left( \frac {m}{\langle
  m\rangle }\right).   \label{disc}
\end{equation}
Note that $\langle m\rangle$ is slightly shifted compared 
to $\langle n\rangle$,
i.e.~$\langle m\rangle \approx \langle n\rangle +0.5$.


\subsection{Entropy scaling}

A convenient means to study the energy dependence
of multiplicity in an integrated form is
provided by the so-called information entropy~\cite{wehr},
defined by
\begin{equation}
  S=-\sum _{n=0}^{\infty } P_{n}\ln P_n . \label{entr}
\end{equation}
Information entropy possesses some peculiar properties
(see Refs.~\cite{ssz1, ssz2}), viz.:
\begin{itemize}
  \item The entropy of $k$ statistically independent sources equals the sum
    of the entropies of the individual sources: $S=\sum _{m=1}^{k}S_m$.
  \item The entropy is invariant under an arbitrary distortion of the 
    multiplicity scale, in particular 
    if a subsample of particles is chosen such as 
    the charged particles.
  \item There exists a relationship between the entropy $S$, 
    the average multiplicity $\langle n\rangle $,
    and the KNO function $f(x)$,
    as long as the KNO form eq.~(\ref{20}) is valid:
    \begin{equation}
      S-\ln \langle n\rangle =\int _{0}^{\infty }f(x)\ln \frac {1}{f(x)}dx.
       \label{Snf}
     \end{equation}
  \item For functions satisfying the normalization conditions~(\ref{21}),
    the right hand side of eq. (\ref{Snf}) does not exceed unity. 
    Therefore the following inequality is valid:
    \begin{equation}
      \frac {S}{\ln \langle n\rangle }\leq 1+\frac {1}{\ln \langle n\rangle }.
      \label{esc}
    \end{equation}
\end{itemize}
From eq.~(\ref{esc}) we see that KNO scaling is equivalent to
scaling of the ratio $S/\ln \langle n\rangle$
in the limit of asymptotics,
and that the approach to this limit is very slow,
of order $\cal{O}$($1/\ln\langle n\rangle$). 
In contrast,
the ratio of the entropy $S$ to the total available
rapidity range, 
namely $S/\ln (\sqrt {s}/m_{\pi})$,
has been shown to exhibit scaling at experimentally
accessible scales.
The observed constancy of $S/\ln (\sqrt {s}/m_{\pi})$ 
is called entropy scaling~\cite{ssz1}. 
If correct, 
entropy scaling implies a violation of KNO scaling 
unless the average multiplicity increases as a power of energy.

If subsamples of particles are chosen,
it is necessary to employ the proper
mean values of the subset multiplicities and 
the correspondingly normalized KNO
functions in eqs.~(\ref{Snf}) and~(\ref{esc}) 
(for more details, see~\cite{ssz2}). 
The values of entropy are the same 
for the subsamples as for the full sample,
as noted above.

The above treatment can be generalized~\cite{pssz}
using the so-called
R\'{e}nyi information entropy of $q$th order:
\begin{equation}
  I_{q}=\frac {1}{1-q}\ln \sum _{n=0}^{\infty } (P_{n})^{q}.   \label{rent}
\end{equation}
Entropy $S$ is obtained from~eq. (\ref{rent})
by setting $q$=1. 
Such a generalization is reminiscent of the moments approach 
to multiplicity described in Section~\ref{sec-definitions}.
The relation between $I_q$ and the KNO function is
\begin{equation}
  I_q -\ln \langle n\rangle =\frac {1}{1-q}\ln \int _{0}^{\infty }f^{q}(z)dz.
\label{iqf}
\end{equation}
Asymptotically,
KNO scaling therefore corresponds to scaling of the ratio
$I_q /\ln \langle n\rangle$,
analogous to the result discussed above for 
the ratio $S/\ln \langle n\rangle$.

The entropy approach is also valid for smaller regions of 
phase space~\mbox{\cite{pssz,ftak}} and can be
formulated in terms of the fractals discussed 
in Section~\ref{sec-intermittency}.
Here we merely note that a multi-fractal character of the
multiparticle production process requires violation of KNO scaling,
and predicts a widening of the KNO function $f(x)$ 
and an increase of its maximum as energy increases~\cite{pssz}.

\subsection{Conventional distributions}
\label{sec-conventional}

The purely phenomenological study of multiplicity
is based on the use of well known functions from probability theory.
Here we consider three functions for which analytic expressions for 
the generating functions and all moments can be derived~\cite{20, 21}. 
These will serve as a starting point for our
discussion of the QCD results presented in 
Sections~\ref{sec-equations}-\ref{sec-exactsolutions}.

\subsubsection{The Poisson distribution}

The presence of correlations in a process is conventionally described 
by the difference between its typical distribution and the 
Poisson distribution, which is written as
\begin{equation}
  P_{n} = \frac {\langle n \rangle ^n}{n!} \exp (-\langle n \rangle ). \label{25}
\end{equation}
The generating function is  (see eq.~(\ref{3}))
\begin{equation}
  G^{(P)}(z) = \exp (\langle n \rangle z),           \label{26}
\end{equation}
where, from eqs. (\ref{4}) and (\ref{5}),
\begin{equation}
  F_{q} = 1 , \;\;\; K_{q} = H_{q} = \delta _{q1} .   \label{27}
\end{equation}
Therefore the measure of correlations can be defined as the 
difference between $F_q$ and 1, 
or, equivalently,
between $K_q$ (or $H_q$) and~0. 
The Poisson distribution exhibits
exact $F$ scaling and asymptotic KNO scaling.

\subsubsection{The negative binomial distribution and its generalizations}
\label{sec-nbd}

The negative binomial distribution (NBD) deserves special attention. 
The NBD~\cite{g} has been used to describe experimental 
measurements of multiplicity in a wide variety of processes
and over a large energy range.
In particular, the NBD qualitatively describes the distribution of
multiplicity in almost all inelastic, high energy processes,
except for data at the highest available energies.
The NBD is given by
\begin{equation}
  P_{n} = \frac {\Gamma (n+k)}{\Gamma (n+1)\Gamma (k)} \left ( \frac {\langle n
  \rangle }{k}\right )^{n} \left (1+\frac {\langle n \rangle }{k}\right )^{-n-k} ,
  \label{31}
\end{equation}
where $k$ is an adjustable parameter with the physical meaning 
of the number of independent sources. 
The Bose-Einstein distribution is a special
case of the NBD with $k$=1. 
The Poisson distribution is obtained from eq.~(\ref{31})
in the limit $k$$\,\rightarrow\,$$\infty$. 

The generating function is
\begin{equation}
  G^{(NBD)}(z) = \left ( 1-\frac {z\langle n \rangle }{k}\right ) ^{-k} .  
  \label{32}
\end{equation}
The integer rank moments are
\begin{equation}
  F_{q} = \frac {\Gamma (k+q)}{\Gamma (k) k^q} ,   \label{33}
\end{equation}
\begin{equation}
  K_{q} = \frac {\Gamma (q)}{k^{q-1}} ,   \label{34}
\end{equation}
\begin{equation}
  H_{q} = \frac {\Gamma (q)\Gamma (k+1)}{\Gamma (k+q)} = kB(q, k) . \label{35}
\end{equation}
For a fixed value of $k$, 
the rate of increase of the factorial moments with $q$ 
is larger than exponential. 
As $q$ increases,
the cumulants at first decrease,
reaching a minimum at $q\approx k$,
after which they increase.
The cumulants are all positive. 
The product of several generating functions 
of negative binomial distributions with different parameters 
also yields positive cumulants since the unnormalized
total cumulant is just the sum of unnormalized individual cumulants.
Similarly, the ratio $H_q$ is always positive.
The $H_q$ moments decreases monotonically 
and are proportional to $q^{-k}$ at large $q$.

\begin{figure}[tphb]
\vspace*{-1cm}
\begin{center}
  \epsfxsize=12cm
  \hspace*{.2cm}\epsffile{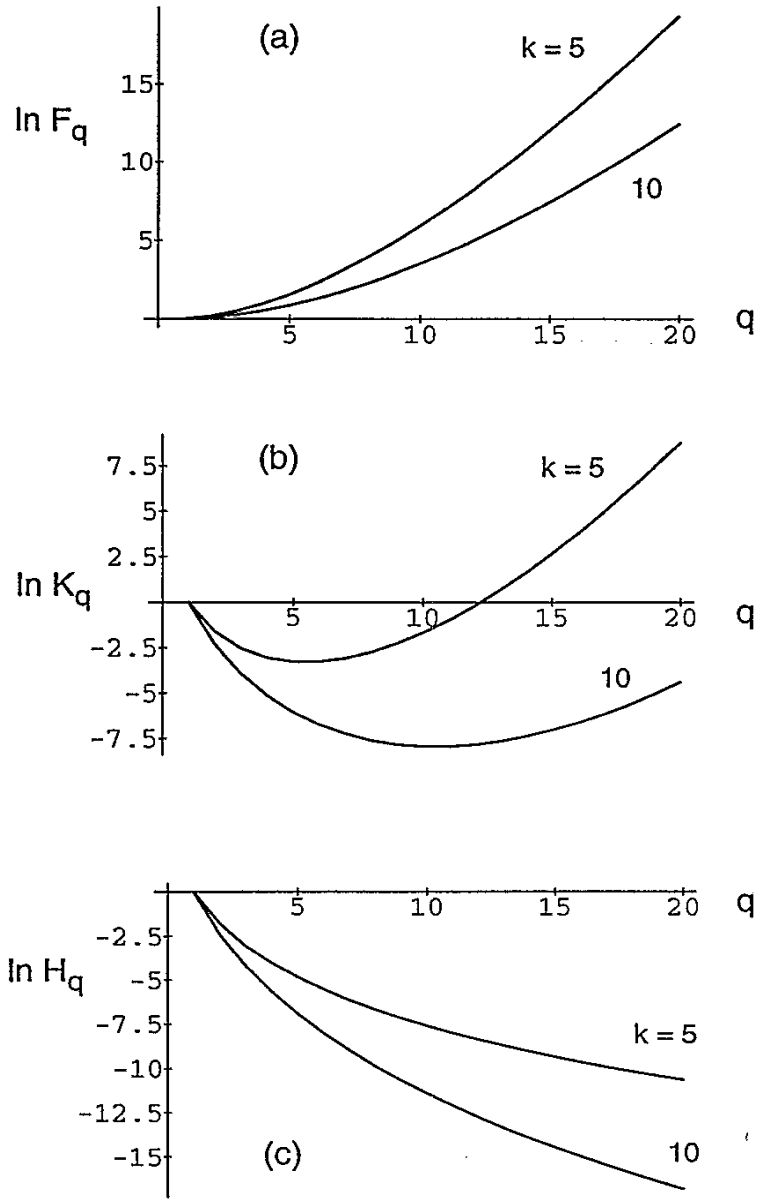}
\end{center}
\vspace*{-.5cm}
\caption{Moments of the negative binomial distribution~\cite{21} 
for $k$=5 and 10,
calculated for integer values of rank $q$:
(a) $\ln F_{q}$ , (b) $\ln K_{q}$ , (c) $\ln H_{q}$.
The curves are drawn to guide the eye.
} 
\label{fig-one}
\end{figure}

Fig.~\ref{fig-one} shows the behavior of 
$\ln F_q$, $\ln K_q$ and $\ln H_q$ as a function of
$q$ for \mbox{$k$=5 and 10.} 
The function $P_n$ becomes narrower as $k$ increases.
This explains the slower rise of $F_q$ for $k$=10 
compared to $k$=5. 
The dependence of $K_q$ on $k$ is
seen to be more pronounced than for~$H_q$. 
These properties are characteristic of the NBD
and not of QCD
(see e.g. Section~\ref{sec-hoapprox}).

The negative binomial distribution with fixed
parameter $k$ exhibits $F$ scaling
and asymptotic KNO scaling. 
For large $n$,
the KNO function with fixed $k$ behaves as
\begin{equation}
  f(x) = \frac {k^k}{(k-1)!}x^{k-1}e^{-kx} ,   \label{36}
\end{equation}
where $x$=$n/\langle n\rangle $. 
This distribution is wider than the 
Poisson distribution for all~$k$.
The generating function eq.~(\ref{32}) is singular for
$z$=$k/\langle n \rangle$$\,\rightarrow\,$0 if
$\langle n\rangle\rightarrow\infty$ and $k$ is constant. 
Therefore,
it is necessary to approach the singularity when calculating 
the derivatives of the generating function at $z$=0,
i.e. the factorial moments (see eq.~(\ref{4})). 
The singularity moves closer to $z$=0 as the energy increases.




The generalized negative binomial distribution can be represented as the 
Poisson transform of the generalized gamma distribution~\cite{heg1},
given by the special scaling function (see eq.~(\ref{20}))
\begin{equation}
  f(x)=\frac {\vert \mu \vert }{\Gamma (k)}\lambda ^{\mu k}x^{\mu k-1}
  \exp (-[\lambda x]^{\mu })   \label{hnbd}
\end{equation}
with three adjustable parameters. 
The Poisson transform of eq.~(\ref{hnbd})
leads to a multiplicity distribution 
which can be expressed in terms of different Fox functions in 
various intervals of the parameter~$\mu$,
as was studied in Ref.~\cite{biya}. 
The notion of fractional moments is quite
useful when considering the generalized NBD because its integer
order moments are formally equivalent to the fractional moments of the 
ordinary NBD (see Ref.~\cite{heg1}).

The so-called modified negative binomial distribution 
with the generating function
\begin{equation}
  G^{(MNBD)}(z)=\left( 
    \frac {1-\Delta z}{1-\chi z}\right)^k   \label{mnbd}
\end{equation}
has also been used to describe data~\cite{ct}. 
The number of adjustable parameters is again three.
They are related to the average multiplicity through
the relation $\langle n\rangle$=$k(\chi -\Delta )$. 
The negative binomial distribution corresponds to $\Delta$=0. 
Such fits of multiplicity usually yield 
better results than the~NBD,
which is no surprise given the larger number of parameters.
On the other hand,
it is difficult to give a physical interpretation to these parameters. 
The cumulants derived from eq.~(\ref{mnbd}) are
\begin{equation}
  K_{q}^{MNBD} = k^{1-q}(q-1)!(\chi ^{q} - \Delta ^{q})/(\chi -\Delta )^q ,
  \label{kmnb}
\end{equation}
which change sign at each subsequent integer $q$ 
for the range of parameters $\chi$ and $\Delta$ given in~\cite{ct}.
This differs from QCD results described below. 
Moreover negative probabilities appear at large $n$ which is
forbidden.

A three parameter fit also has been attempted in the framework 
of the Saleh-Teich distribution~\cite{rzk},
with the claim that it perfectly describes 
the observed shape of the distribution at high multiplicities,
including the shoulder (see Section~\ref{sec-expphen}). 
Moments of the multiplicity distribution
have not been studied using this approach, however.

A combination of two NBDs has also been used in some fits.
This also enlarges the number of adjustable parameters.

We will also mention the log-normal distribution which describes 
the Gaussian distribution of the logarithms of the variable. 
The discrete probability distribution for charged particle 
multiplicity $n$ is defined by
\begin{equation}
  P_{n}(\mu ,\sigma ,c)=N\int _{n}^{n+\delta n} \frac {dn'}{n'+c} \exp (-\frac
  {[\ln (n'+c)-\mu ]^{2}}{2\sigma ^2}),      \label{logn}
\end{equation}
where $\mu$, $\sigma$ and $c$ are adjustable parameters
and $N$ is a normalization factor.
The integration parameter is $\delta n$=2 for full phase space.
In restricted rapidity intervals,
where both even and odd multiplicities contribute, 
\mbox{$\delta n$=1.}
This distribution corresponds to a scale invariant stochastic 
branching process (for details see Refs.~\cite{ci, sww}).

\subsubsection{The fixed multiplicity and Gaussian distributions}
\label{sec-fixed}

We next consider the example of fixed multiplicity.
Fixed multiplicity demonstrates how severely experimental 
selection criteria can influence moments.
Events with a given multiplicity have often
been studied in the past, the so-called semi-inclusive events.
In this case one deals with the distribution
\begin{equation}
  P_{n} = \delta _{nn_{0}} \;\;\;\;\;\;\; (n_{0} = const ) .  \label{40}
\end{equation}
The generating function is
\begin{equation}
  G^{(F)}(z) = (1+z)^{n_0} .   \label{41}
\end{equation}
Since $\langle n \rangle$=$n_0 $, one obtains
\begin{equation}
  F_{q} = \frac {n_{0}!}{n_{0}^{q} (n_{0}-q)!} = \frac {\Gamma (n_{0}) n_{0}^{1-q}}
  {\Gamma (n_{0}-q+1)} ,    \;\;\;\;\;  1<q\leq n_{0} ,   \label{42}
\end{equation}
\begin{equation}
  F_{q} = 0 ,     \;\;\;\;\;\;\;\;  q>n_{0} ,    \label{43}
\end{equation}
\begin{equation}
  K_{q} = (-n_{0})^{1-q}(q-1)! = (-n_{0})^{1-q}\Gamma (q) ,   \label{44}
\end{equation}
\begin{equation}
  H_{q} = (-1)^{1-q}n_{0}B(q, n_{0}-q+1) .   \label{45}
\end{equation}
The factorial moments decrease monotonically with $q$ 
until $q$=$n_{0}$.
All factorial moments with rank larger than $n_{0}$ are zero.
Thus the $H_q$ ratios are defined for $q\leq n_{0}$ only. 
The integer order cumulants alternate in sign for successive ranks, 
being positive at odd values of $q$ and negative at even values:
therefore the cumulants ``oscillate.''
As $q$ increases,
the amplitude of the oscillations decreases until $q$=$n_{0}$,
after which the amplitude increases.
 
Oscillations of the cumulants are also observed in QCD 
but with a different periodicity
(see Section~\ref{sec-hoapprox}).
Furthermore,
in QCD, 
the factorial moments increase rapidly with~$q$.

\begin{figure}[tphb]
\vspace*{-1cm}
\begin{center}
  \epsfxsize=12cm
  \hspace*{.2cm}\epsffile{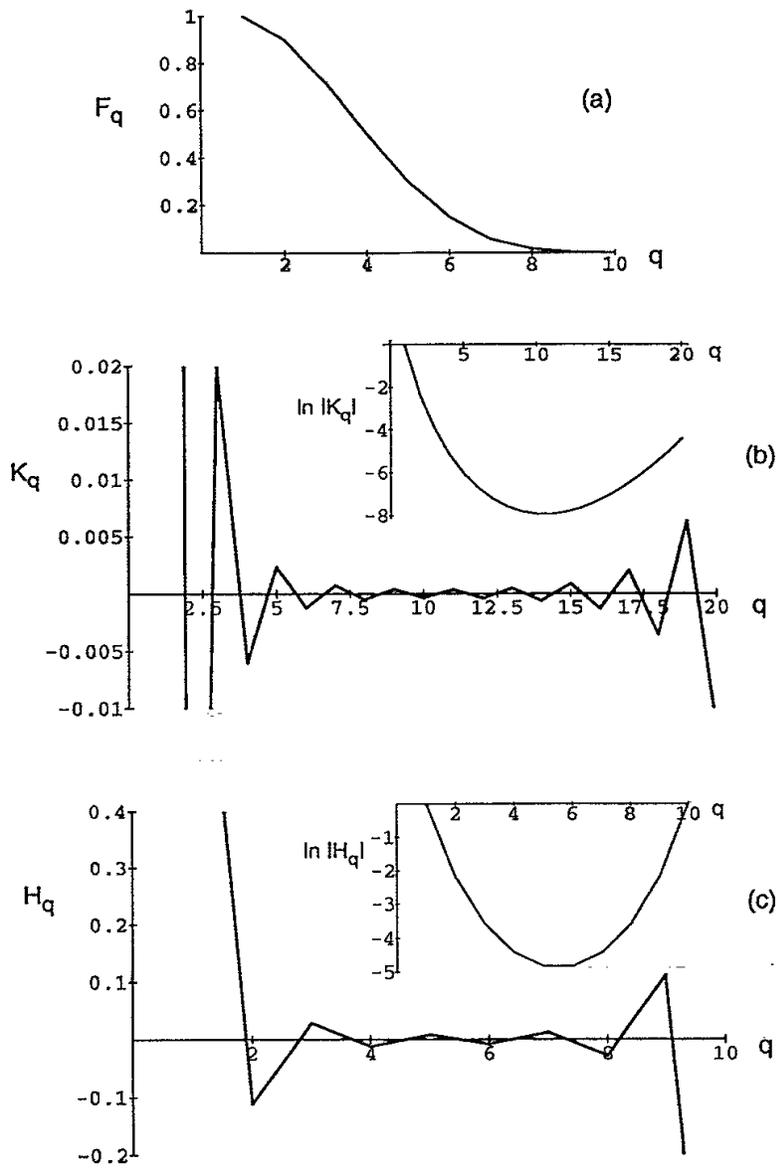}
\end{center}
\vspace*{-.5cm}
\caption{Moments of the
fixed multiplicity distribution~\cite{21} for $n_{0}$=10,
calculated for integer values of rank $q$: 
(a)~$F_q$ , (b)~$K_q$ , $\ln \vert K_q\vert $ , 
(c)~$H_q$ , $\ln \vert H_{q}\vert $.
The curves are drawn to guide the eye.
} 
\label{fig-two}
\end{figure}

Fig.~\ref{fig-two} shows the $F_{q}$, $K_{q}$ and $H_{q}$ 
moments of the fixed multiplicity distribution for $n_{0}$=10.
The insets show $\ln\vert K_q\vert$ and $\ln\vert H_q\vert$.

For fixed multiplicity,
the oscillations of the cumulant moments
are entirely a consequence of the event selection procedure.
This is easy to recognize when one selects, 
for example, 10-particle events from a set of events which
are distributed according to the Poisson distribution. 
Then we obtain alternating sign cumulants at integer ranks 
instead of cumulants equal to zero. 
The amplitude of the oscillations preserves information about 
the original distribution as long as the normalization is not changed. 

A similar behavior of cumulants is observed for the Gaussian distribution
\begin{equation}
  P_n=\frac {1}{\sqrt {2\pi \sigma ^2}}\exp [-\frac {(n-\langle n\rangle
  )^2}{2\sigma ^2}].     \label{pgau}
\end{equation}
If $\sigma \propto \langle n\rangle $, the Gaussian exhibits KNO scaling. 
Its generating function is
\begin{equation}
  G(z)=\exp [\langle n\rangle \ln (z+1)+\frac {\sigma ^2}{2}\ln ^{2}(z+1)].
\label{ggau}
\end{equation}
For $\sigma$$\,\rightarrow\,$0, 
the generating function tends to the fixed multiplicity case, 
i.e.~the cumulants are positive for odd
ranks and negative for even ranks. 
For $\sigma ^2>\langle n\rangle $,
the cumulants are positive for even ranks and negative for odd ranks.

\subsection{Some models}
\label{sec-models}

The distributions discussed above are 
well known from probability theory.
In particle physics, 
they have been used in the context of cluster or clan~\cite{6} models
and in the multiperipheral cluster model~\cite{28} 
with one particle exchange.

At first sight, the graph-theoretic description of multiparticle
production seems entirely different between {\epem} 
and hadron initiated processes. 
In the former case, the main graphs are tree-like,
with a highly virtual initial time-like parton.
In the latter case, 
one used to consider a sequence of multiperipheral-type graphs 
with low space-like virtualities and rather complicated topologies. 
A more general and unifying picture emerges by considering
strings to be stretched between the color charges 
(cf. the Lund model~\cite{22, 23}, dual topological model~\cite{24, 25}, 
and the quark-gluon string model~\cite{26, 27}) 
and by introducing models for final particle clusterization 
(the multiperipheral cluster model~\cite{28}, clans~\cite{6, 29}, etc.). 
The multiplicity distributions in these models are not usually 
described by a single analytic formula but are formed from
a combination of distributions. 
For example, the multiperipheral model with a single ladder 
(chain, string) is based on the Poisson distribution  
for the particle emission centers 
(resonances, fireballs, clusters, clans, etc.).
In general, the resulting distribution is obtained by convolution of
the Poisson distribution,
which specifies the number of sources,
with the decay distribution of the sources.
If one chooses a logarithmic distribution to describe the decay multiplicity,
the distribution of final particle multiplicity is given by the NBD.
This model does not incorporate energy-momentum conservation,
however,
which would almost certainly modify the simplicity of this result.

The simultaneous creation of several ladders yields a 
more complicated distribution. 
Particles produced in different chains are treated as uncorrelated. 
For each configuration,
the multiplicity distributions of different ladders are convoluted. 
Averaging over all multiple scattering processes yields 
the final multiplicity distribution.

The multiplicity is sometimes approximated by a
sum of negative binomials with different parameters,
resulting in a distribution with ``shoulders'' or ``quasi-oscillations'' 
imposed on a smooth curve. 
(The possible relation between these oscillations and
those of the $H_q$ moments discussed below in
Sections~\ref{sec-hoapprox} and~\ref{sec-oscillating}
is considered in Refs.~\cite{30, 30a}.)
In particular, the multiplicity distribution within a single jet in {\epem} 
annihilation is sometimes approximated by a single NBD,
while the superposition of several jets yields a
distribution with a shoulder~\cite{31}
in qualitative agreement with experiment
(see Section~\ref{sec-expphen}). 
However,
the solutions of QCD equations do not support 
the hypothesis of superimposed NBDs,
as we discuss below. 
The detailed study of
semi-phenomenological models is usually performed 
using Monte Carlo computations,
involving assumptions about hadronization.
It is possible that the discrepancy between the QCD solutions and
the results of the phenomenological approaches
is related to these assumptions.

In hadron-induced reactions, one generally describes
multiparticle production as proceeding through a
set of clusters created through the exchange of some mean
transferred momentum between the colliding constituents. 
The general kinematic relation between the primary energy~$s$, 
the cluster masses~$s_i$,
and the effective transferred momentum 
$p_t^2$ is easily obtained by iteration of the formula
$s\,$=$\,s_1s_2/p_t^2$~\cite{drch}. 
For equal cluster masses and transferred momenta, 
this relation can be written as 
$s/p_t^2\,$$\approx$$\,(s_i/p_t^2)^N$ or, 
in terms of mean multiplicities, as
\begin{equation}
  \langle n(y)\rangle =\frac {y-L_p}{y_i-L_p}\langle
  n(y_i)\rangle \approx \frac {y}{y_i-L_p}\langle
  n(y_i)\rangle ,    \label{yilp}
\end{equation}
where $y\,$=$\,\ln s$, $y_i\,$=$\,\ln s_i$, and $L_p\,$=$\,\ln p_t^2$ 
(with $s$, $s_i$ and $p_t^2$ appropriately normalized by,
say, 1~GeV$^2$),
and $N\,$=$\,\langle n(y)\rangle /\langle n(y_i)\rangle$ 
is the number of clusters. 
The traditional multiperipheral approach assumes
constant values of $y_i$ and $L_p$ and therefore constant 
$\langle n(y_i)\rangle $. 
Thus $\langle n(y)\rangle \propto y$,
i.e.~the mean multiplicity is predicted to
increase logarithmically with energy. 
A faster than logarithmic rise 
is observed experimentally, however.
According to eq.~(\ref{yilp}),
this implies that the ratio of $s_i$ to $p_t^2$
tends to unity at high energies,
an effect which is also observed by experiment.
The model of classical fields as applied to 
nucleus-nucleus collisions (see Refs.~\cite{mcve,klm}) 
should yield similar features
if interpreted in terms of clusters.

Another important consideration in the comparison of 
theory with hadron-hadron data is that
the phase space is relatively unoccupied in typical events,
i.e. for events with multiplicities close to the mean,
because the phase space volume 
is proportional to the primary energy $\sqrt s$
whereas the mean multiplicity increases more slowly,
as $\exp (c\sqrt {\log s})$
according to perturbative QCD or as a power of $\log s$ 
according to some phenomenological models. 
At present, we describe these data using peripheral-like
ladder type processes with reggeons or reggeized gluon exchange. 
The phase space is elongated along the axis of
the impinging hadrons and fixed in the transverse direction. 
For the tail of the multiplicity distribution at 
very high multiplicities,
the phase space becomes saturated. 
Which dynamics governs this region is not yet clear.
Surely, the longitudinal momenta will decrease 
compared to typical events
because of conservation laws.
However it is unclear whether the transverse momenta 
are the same or larger than in than in typical events.
This latter possibility has been
suggested by some theoretical speculations. 
In the former case, 
one should employ non-perturbative approaches while
in the latter case perturbative solutions can be found. 
Additional experimental and theoretical effort is required in this area,
which could result in the observation of new physical effects
or at least a deeper understanding of the dynamics. 
The Fermi statistical model~\cite{ferm, pome} and the 
Landau hydrodynamical model~\cite{land} (or their modifications),
popular in the 1950s, 
could apply here. 
A statistical approach to the description of very high multiplicity 
processes based on the generating 
function technique is proposed in~\cite{kmsi}.

The treatment of nucleus-nucleus interactions is even more complicated,
involving an extremely large number of exchanged ladders between 
numerous colliding nucleons,
yielding a much wider multiplicity 
distribution than hadron-hadron collisions.
This feature is in qualitative agreement with experiment.
In central heavy ion collisions, particle production is described 
by multi-ladder exchange in the reggeon approach. 
This implies that the number
of partons emitted by colliding nuclei is very large, 
and that their combined action can be treated 
as a source of some classical field which creates
the final-state particles~\cite{mcve}. 
The multiplicity distribution is a
Gaussian type~\cite{klm} and exhibits KNO scaling,
i.e.~$\sigma$$\,\propto\,$$\langle n\rangle $ in eq.~(\ref{pgau}). 
The cumulants change sign at each subsequent rank. 
However, such a classical field model is found to be
applicable only to the tail of the observed multiplicity 
distribution.

Despite its greater complexity,
collisions involving nuclei share some features
with the simpler hadron-hadron and {\epem} collisions,
as we discuss in the following sections.

\subsection{Experimental phenomenology}
\label{sec-expphen}

When KNO scaling was first proposed~\cite{7},
experimental data were available
at comparatively low energies only. 
The available data from hadron collisions
approximately satisfied the KNO scaling condition. 
A slight deficiency of the fits at low multiplicities was cured
by use of the correct relation between discrete and continuous 
distributions~\cite{golo}.
It was only when UA5 data became available in 1985~\cite{ua5},
with energies of several hundred GeV,
that violation of KNO scaling was clearly observed.
The negative binomial distribution was then used to fit 
the UA5 data. 
However, a shoulder structure was present in the data
which could not be reproduced using a single NBD.

The situation in {\epem} annihilations followed a similar path,
with suggestions at low energies that KNO scaling was valid
followed by observations at higher energies that it was not.
A shoulder structure in the multiplicity distribution,
similar to that observed in hadron collisions,
was also observed here.
The shoulder is ascribed to the
multi-source nature of the processes, 
either from additional ladders in hadron reactions 
or to multi-jet emission in the {\epem} events. 

A universal energy dependence of average multiplicities 
in reactions initiated by different particles
was proposed~\cite{bib-zich,bib-bas}.
It was argued that the scale for hadron-initiated reactions
should be the so-called effective energy,
obtained by subtracting the energy of leading particles
from the total event energy.
The effective energy corresponds to the notion of
inelasticity coefficients in cosmic ray studies.
The mean multiplicities of {\epem} and hadron reactions
were found to coincide up to ISR energies 
($\sqrt{s}$$\,\sim\,$50~GeV)
if the effective energy was used in the latter case.
This correspondance has not yet been checked
for higher energies, however,
where it is known that multiplicity sensitive
quantities such as rapidity differ between the two processes.

\begin{figure}[thb]
\begin{center}
  \epsfxsize=14cm
  \epsffile{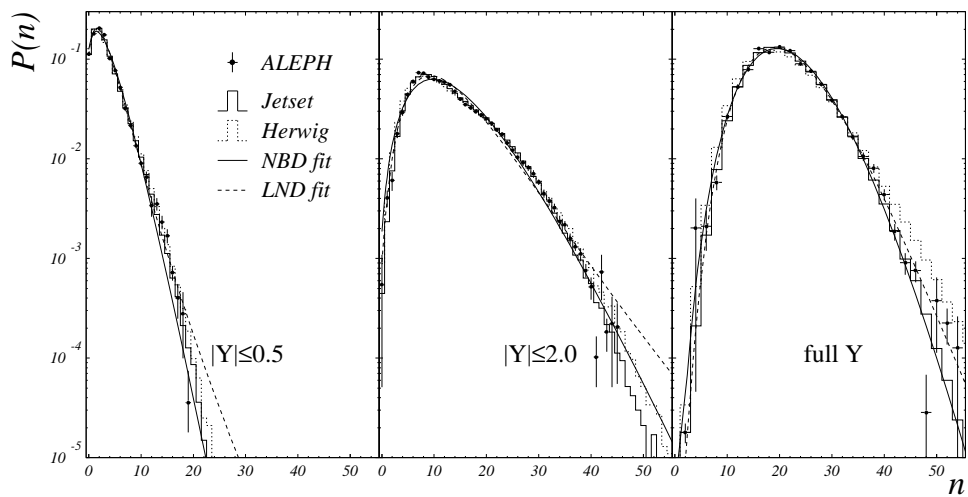}
\end{center}
\vspace*{-.5cm}
\caption{
The measured charged particle multiplicity distribution
in {\epem} hadronic Z$^0$ decays~\cite{alp},
for small ($|$Y$|\leq 0.5$, left) 
and intermediate ($|$Y$|\leq 2.0$, center)
rapidity windows,
and in full phase space (right),
in comparison to various models.
} 
\label{fig-three}
\end{figure}

With the advent of high statistics
{\epem} experiments at the Z$^0$ energy, 
it became possible to analyze multiplicity not only 
in full phase space but in limited phase space regions. 
As an example,
Fig.~\ref{fig-three} shows results from the ALEPH Collaboration
at LEP for the distribution of charged particle multiplicity 
in hadronic Z$^0$ decays~\cite{alp}.
The data are shown for full phase space
and in limited phase space intervals defined by
rapidity $|$Y$|$$\,\leq\,$2.0 and $|$Y$|$$\,\leq\,$0.5.
The data are compared to the results of the
negative binomial and log-normal distributions.
Overall, the log-normal result is in somewhat better agreement 
with data than the NBD.
The shoulder structure,
especially pronounced in the data for multiplicities $n$$\,\approx\,$30 
in the $|$Y$|$$\,\leq\,$2.0 rapidity bin,
is not well described by either model, however.
Concerning factorial moments, 
the NBD underestimates the experimental results
whereas the log-normal distribution overestimates them~\cite{opi,sar}. 
Further analysis of factorial moments
reveals disagreement between the data and 
the log-normal distribution both in small
intervals and full phase space~\cite{sar}. 
This is an example of how moment analysis can
reveal small differences between theory and data 
in a more clear manner than a direct fit of multiplicity.
Neither the modified negative binomial distribution 
nor the so-called pure birth stochastic model~\cite{biy}
improve the situation. 

The predictions of the JETSET~\cite{bib-jetset} and 
HERWIG~\cite{bib-herwig} parton shower Monte Carlo 
event generators are included in Fig.~\ref{fig-three}.
These two programs,
along with the ARIADNE parton shower Monte Carlo~\cite{bib-ariadne},
are widely used to interpret
{\epem} data (see, for example, Ref.~\cite{bib-physicsatlep}).
The Monte Carlo results are seen to describe the data quite well,
presumably because they account for hard gluon emission 
relevant for the shoulder. 
Monte Carlo studies based on the quark-gluon string model and
the dual topological model similarly provide a rather successful guide
to experimental distributions
from hadron- and nucleus-initiated reactions.

As a final remark we
note that other phenomenological distributions
have been used to describe experimental data on multiplicity,
such as the so-called modified gamma distribution~\cite{bib-goul}.

\section{Equations of Quantum Chromodynamics}
\label{sec-equations}

Multiparticle production in Quantum Chromodynamics 
arises from the interactions of quarks and gluons.
The interactions lead to the creation of additional quarks and gluons
and eventually to the formation of hadrons.
The most characteristic features of QCD processes are determined 
by the vector nature of the massless gluons and 
by the dimensionless coupling constant. 
Gluons carry color charge and therefore
emit gluons in addition to quark-antiquark pairs. 
The development of quark and gluon jets
is described by evolution equations. 
The main parameter of the evolution is the 
angle of divergence of the jet or, more precisely, 
its transverse momentum. 
The essential difference between Quantum Electrodynamics
and Quantum Chromodynamics 
--~due to the triple gluon coupling~--
can be traced to the lowest order graphs
in which two colliding electrons 
(or quarks) emit a single photon (or gluon). 
In the latter case there exists an additional graph 
with a gluon emitted by the exchanged gluon. 
The interference of this graph with the others 
leads to transverse momentum ordering,
in contrast to pure angular ordering as in Quantum Electrodynamics 
(see e.g. Ref.~\cite{gpw}).

The subsequent emission of gluons and quarks within the jet fills in
the internal regions of cones defined by the previous emissions
(transverse momentum ordering). 
This remarkable property of ``angular ordering''
can be exploited to formulate 
a probabilistic scheme for the development of the jet as a whole. 
In this case the evolution equations are reminiscent of the
well known classical Markovian equations for ``birth-death'' 
or ``mother-daughter'' processes.
(For a detailed discussion of this approach, 
based on the coherence phenomenon, 
see Refs.~\cite{5,bib-ermolaev}).

The system of two equations for the generating functions 
$G_F$ and $G_G$ of quark and gluon jets, 
respectively, is 
(with A, B, C = F, G)~\cite{1,5,bib-konishi}
\begin{eqnarray}
  G_{A}(y,z)& = & e^{-w_{A}(y)}z \nonumber \\ 
  & + &\frac {1}{2}\sum _{B,C}\int ^{y}dy^\prime 
  \int _{0}^{1}dx \exp [-w_{A}(y)+w_{A}(y^\prime )] \nonumber \\ &\times &
  \frac {\alpha_{S}}
  {2\pi }K_{A}^{BC}(x)G_{B}[y^\prime+\ln(x),z]
        G_{C}[y^\prime+\ln(1-x),z] , 
  \label{49}
\end{eqnarray}
where
\begin{equation}
  y=\ln (p\Theta /Q_0 )=\ln (2Q/Q_{0})=\ln(Q/\Lambda), 
\end{equation}
with $p$ the initial momentum, 
$\Theta$ the angle of divergence of the jet 
(the initial jet opening angle), 
assumed here to be small, 
$Q$ the jet virtuality, 
$Q_{0}$ a constant which
defines the limit of perturbative evolution, 
and $\alpha _{S}$ the strong interaction 
coupling strength or ``constant.'' 
$\Lambda$ is the so-called scale parameter of QCD,
related to $Q_0$ by $\Lambda$=$Q_0/2$.\footnote{
Note that the relationship between $Q_0$ and $\Lambda$
is not unique;
other relationships have been chosen 
(see e.g. Refs.~\cite{bib-klo98,lo1})
and they are often treated as independent variables.}
The scale parameter $\Lambda$ is strongly related to
$\lms$~\cite{bib-bardeen,pdg}
but is not, in general, the same.
We shall consider both running and fixed $\alpha_S$. 
An effective infrared safe coupling constant 
is sometimes used as a substitute for the  
phenomenological parameter $Q_0$ (see~\cite{khoc,dokt}). 
Here, we use the more traditional approach, however. 

The first term on the right hand side of eq.~(\ref{49})
corresponds to the propagation of the primary
parton without interaction,
and is described by the form factor $\exp [-w_{A}(y)]$. 
The second term describes the creation of two jets $B$ and $C$
which carry proportions $x$ and $1-x$ of the primary energy
after their production at the vertex $K _{A}^{BC}$,
which is reached by the primary parton after it evolves
to scale $y^\prime$ without splitting,
as dictated by the factor $\exp [-w_{A}(y)+w_{A}(y^\prime )]$. 
Iteration of eq.~(\ref{49}) 
generates all tree graphs of the perturbation series,
ordered by transverse momentum. 

Considering just terms of order $(\alpha_{S}\log^{2}s)^n$, 
where $s$$\,\approx\,$$4p^2$ is the cms energy squared, 
and then summing,
leads to the so-called DGLAP evolution equation 
(see Ref.~\cite{bib-dglap})
with kernels explicitly written below. 
The approximation in which only these terms are taken into 
account is the double-logarithmic approximation (DLA).
Adding further terms with lower powers of $\log s$
yields the modified leading-logarithmic
approximation (MLLA). 
Adding yet lower powers of $\log s$,
one obtains the next-to-next-to-leading order
approximation (NNLO), and so on. 
Even the lowest order approximation includes processes 
with an arbitrary number of partons and is described 
by a sub-series of the perturbative expansion with different powers 
of the coupling constant. 
Thus the classification of approximations is 
not based simply on powers of $\alpha_S$ but rather on 
the product of $\alpha_S^n$
with the corresponding power of $\log s$. 
This explains why terms from different sub-series are combined
when the conservation laws (e.g.~energy conservation) are included,
as discussed in more detail below
(see Section~\ref{sec-hoapprox}).

Multiplying both sides of eq.~(\ref{49}) by $\exp [w_{A}(y)]$ 
and then differentiating by $y$, 
we eliminate the form factors and obtain the final system of 
equations~\mbox{\cite{1, 5,bib-konishi}}\footnote{The system of 
eq.~(6.29) in Ref.~\cite{1} is the same as the
system of eqs.~(\ref{50}) and~(\ref{51}) in the present work,
using a slightly different notation;
the notion of the generating function in this context was 
first introduced in Ref.~\cite{bib-konishi}
as is also mentioned in Ref.~\cite{5}.}:
\begin{eqnarray}
  &G_{G}^{\prime }&= \int_{0}^{1}dxK_{G}^{G}(x)\gamma _{0}^{2}
    [G_{G}(y+\ln x)G_{G}
  (y+\ln (1-x)) - G_{G}(y)] \nonumber \\ 
  &+&n_{f}\int _{0}^{1}dxK_{G}^{F}(x)\gamma _{0}^{2}
  [G_{F}(y+\ln x)G_{F}(y+\ln (1-x)) - G_{G}(y)] ,   \label{50}
\end{eqnarray}
\begin{equation}
  G_{F}^{\prime } = \int _{0}^{1}dxK_{F}^{G}(x)\gamma _{0}^{2}[G_{G}(y+\ln x)
  G_{F}(y+\ln (1-x)) - G_{F}(y)] ,     \label{51}
\end{equation}
where $G^{\prime }(y)=dG/dy$, 
$n_f$ is the number of active quark flavors,
and
\begin{equation}
  \gamma _{0}^{2} =\frac {2N_{c}\alpha _S}{\pi } .     \label{52}
\end{equation}
The kernels of the equations are
\begin{equation}
  K_{G}^{G}(x) = \frac {1}{x} - (1-x)[2-x(1-x)] ,    \label{53}
\end{equation}
\begin{equation}
  K_{G}^{F}(x) = \frac {1}{4N_c}[x^{2}+(1-x)^{2}] ,  \label{54}
\end{equation}
\begin{equation}
  K_{F}^{G}(x) = \frac {C_F}{N_c}\left[ \frac {1}{x}-1+\frac {x}{2}\right] ,   
\label{55}
\end{equation}
with $N_c$=3 the number of colors
and $C_{F}$=$(N_{c}^{2}-1)/2N_{c}$=4/3 in QCD. 

The variable $z$ has been omitted in eqs.~(\ref{50}) and~(\ref{51}).
It should be remembered, however,
that derivation of the equations for moments
relies on the expansions (\ref{7}) and (\ref{8}) 
of the above equations.

A typical feature of any field theory with a dimensionless 
coupling constant,
and of QCD in particular,
is the presence of singular terms at $x\rightarrow 0$ 
in the kernels of the equations. 
These singularities imply an uneven sharing of energy between 
the newly created jets and play an important role
in the jet evolution,
giving rise to a larger average multiplicity
compared to the case of equal sharing of energy
(the non-singular case).

The system of equations (\ref{50}) and (\ref{51}) is 
physically appealing but not absolutely exact. 
This last point is clear since, for example, 
there is no four-gluon interaction term in 
eqs.~(\ref{50}) and~(\ref{51}).
Such a term does not contribute a singularity to the kernels,
justifying its omission from the lowest order approximation. 
Despite this lack of exactness,
the modified series of perturbation theory
(i.e.~with three parton vertices only),
is well reproduced by eqs.~(\ref{50}) and~(\ref{51})
up to terms including higher order logarithmic corrections. 
As shown in Ref.~\cite{5}, 
the neglected terms contribute at the level of the product of, 
at least, five generating functions. 
The physical interpretation of the corresponding graphs leads 
to treatment of the ``color polarizability'' of jets. 

There are complications associated with
the definition of the evolution parameter,
with preasymptotic corrections,~etc. 
(see, e.g., Ref.~\cite{32}). 
For example,  
the limits of integration in eqs.~(\ref{50}) and (\ref{51})
are in principle constrained by a restriction 
on the transverse momentum:
\begin{equation}
  k_t = x(1-x)p\Theta^{\prime } > Q_0/2.   \label{ktli}
\end{equation}
This condition originates from the requirement that 
the formation time of a gluon ($t_{form}\sim k/k_{t}^{2}$) 
be less than its hadronization time
($t_{had}\sim kR^{2}\sim k/Q_{0}^{2}$). 
This leads to the requirement that the arguments of the generating
functions in eqs.~(\ref{50}) and (\ref{51}) be positive,
which in turn implies that the limits of integration in
eqs. (\ref{50}) and (\ref{51}) should be
$e^{-y}$ and $1-e^{-y}$ rather than 0 and~1.
These limits tend to 0 and 1 at high energies.
Therefore it seems reasonable to use integration limits 
of 0 and 1 in eqs.~(\ref{50}) and~(\ref{51}),
to study their analytic solutions with high accuracy,
and then to account for the neglected terms
by considering them to be 
preasymptotic corrections to these solutions.

This issue of the limits of integration has physical significance. 
With limits of $e^{-y}$ and $1-e^{-y}$, the partonic cascade 
terminates at the virtuality~$Q_0$,
as seen from the arguments 
of the multiplicities in the integrals. 
With limits of 0 and 1, the cascade extends into the
non-perturbative region with virtualities smaller than $Q_{0}/2$, 
i.e.~$Q_{1}\approx xp\Theta /2$ and $Q_{2}\approx (1-x)p\Theta /2$. 
This region contributes
terms of the order of $e^{-y}$, power-suppressed in energy. 
It is not known whether the equations and the LPHD hypothesis
are valid below the cutoff $Q_0$.

\section{Gluodynamics}
\label{sec-gluodynamics}

It is natural to begin our study of QCD
with the case of gluodynamics,
in which there are no quarks and only the 
interactions of gluons are considered.
Gluodynamics exhibits all the qualitative features 
of QCD evolution while being more transparent.
In gluodynamics,
the system of equations (\ref{50}) and (\ref{51}) 
reduces to the single equation
\begin{equation}
  G^{\prime }(y) = \int _{0}^{1}dxK(x)
     \gamma _{0}^{2}[G(y+\ln x)G(y+\ln (1-x)) -
  G(y)] ,   \label{56}
\end{equation}
with $G(y)$$\,\equiv\,$$G_{G}(y)$ and $K(x)$$\,\equiv\,$$K_{G}^{G}(x)$. 
Eq.~(\ref{56}) is a non-linear, integro-differential equation 
with shifted arguments in the nonlinear term which account 
for energy conservation in the triple QCD vertex,
when one gluon splits into two. 
As mentioned above,
$k_t$~ordering influences the limits of integration, 
and instead of 0 and 1 in eq. (\ref{56}) we should insert 
$e^{-y}$ and $1-e^{-y}$.
Use of these latter limits leads to
preasymptotic power corrections which we
neglect for the moment and estimate later. 
Even though momentum conservation is not, 
strictly speaking, included, 
it is taken into account in an approximate manner 
through a combined action of energy conservation 
and transverse momentum ordering. 

In the lowest order (double-logarithmic) approximation, 
only the most singular terms in the kernel $K(x)$ 
and within the square brackets are retained.
Thus $K\rightarrow 1/x$ and $\ln (1-x)\rightarrow 0$
(therefore one neglects energy
conservation in the asymptotic limit), 
while $\gamma _{0}^{2}$ is chosen to be constant.

\subsection{Approximate solutions of equations with fixed coupling 
constant and the shape of the KNO function}
\label{sec-approximatesolutions}

In Section~\ref{sec-exactsolutions} we describe 
the exact solutions of the QCD equations
for the moments of multiplicity,
assuming a fixed coupling strength.
We show that the properties of gluon jets do not 
change appreciably if quarks are included. 
Therefore it is instructive to examine the more transparent 
approximate solutions of eq.~(\ref{56}).

Formally, the assumptions of the double-logarithmic approximation 
are equivalent for the three factors in the integrand 
of eq.~(\ref{56}) since non-leading terms are neglected.
In Quantum Chromodynamics with a fixed coupling strength,
$F$ scaling is favored over KNO scaling 
(see Section~\ref{sec-exactsolutions}).
The difference between $F$ and KNO scaling 
is usually neglected, however,
since the calculations are often performed 
for asymptotically high energies. 
Preasymptotic corrections for the second moment 
of the multiplicity distribution are discussed in Ref.~\cite{18},
and for higher moments in Ref.~\cite{dln}.

The generating function for the lowest order solution 
of eq.~(\ref{56}) is independent of both the energy 
and coupling constant (see Refs.~\cite{5,19}).
In this case eq.~(\ref{56}) reduces to the differential equation
(for more details see eq.~(\ref{63}) below) 
\begin{equation}
  [\ln G(y)]''=\gamma _{0}^{2}(G(y)-1).     \label{edla}
\end{equation}
The corresponding KNO function $f(x)$ defined by eq.~(\ref{20})
decreases exponentially at large values of $x$:
\begin{equation}
  f(x) \sim 2C(Cx-1+\frac {1}{3Cx}+...)\exp (-Cx) ;
   \;\;\;\;\; Cx\gg 1, \label{23}
\end{equation}
where $C\approx 2.552$. 
For small values of $x$ its behavior is
\begin{equation}
  f(x) \sim x^{-1}\exp (-\ln ^{2}x/2) .    \label{24}
\end{equation}
The appearance of asymptotic KNO scaling and its 
independence of the coupling constant in the lowest order
approximation are by themselves a success of perturbative Quantum 
Chromodynamics~\cite{11}.
The shape of the scaling function~(\ref{23}) 
does not fit experiment, however.
Experiment favors shapes which are much narrower than 
predicted by eqs.~(\ref{23}) and~(\ref{24}). 
The corrections of the modified leading-logarithmic approximation 
yield a function that is less wide~\cite{19},
reducing the width of $f(x)$
and introducing a dependence on~$\alpha_S$.

Detailed studies of the solutions of gluodynamics
have been presented in many
papers~\cite{9}-\cite{11}, \cite{12,13}
\cite{21}, \cite{32}, \cite{33}-\cite{cdfw}.
In most cases only moments of low rank are considered,
i.e.~the average multiplicity and its dispersion. 
The approximations in these papers do not always allow
higher rank moments to be treated,
leading to unphysical results
such as negative factorial moments of the fifth rank 
(see Ref.~\cite{33}) 
which are forbidden by the definition, eq.~(\ref{4}). 
This illustrates the importance of consistency when treating
terms of the same order in QCD equations.
The role of conservation laws 
in the shifted arguments of the generating
functions is extremely important,
providing large corrections. 
It was shown in Ref.~\cite{12} that 
conservation laws can be taken into account. 
However, in Ref.~\cite{12} the running property of the 
coupling constant was disregarded, 
the non-singular terms in the kernel were neglected 
(as were some other terms),
and the difference between the coupling constant 
$\gamma _{0}$ (eq.~(\ref{52}))
and the QCD anomalous dimension $\gamma $, defined as
\begin{equation}
  \langle n \rangle = \exp (\int ^{y}\gamma (y^\prime )dy^\prime )  ,
    \label{57}
\end{equation}
was neglected as well. 

Eqs.~(\ref{50}) and (\ref{51})
possess exact solutions for a fixed coupling constant,
as discussed in Section~\ref{sec-exactsolutions}.
To find these solutions requires the numerical solution
of some algebraic equations 
(see Section~\ref{sec-exactsolutions}).
Therefore we first consider the approximate approach of 
Ref.~\cite{12} because it yields an analytic expression 
for the KNO function 
which reveals the importance of the conservation laws,
while differing from eq.~(\ref{23}) by predicting a narrower width, 
in much better agreement with experiment.

\begin{figure}[tphb]
\vspace*{-2cm}
\begin{center}
  \begin{turn}{.99}
  \epsfxsize=16cm
  \hspace*{-.8cm}\epsffile{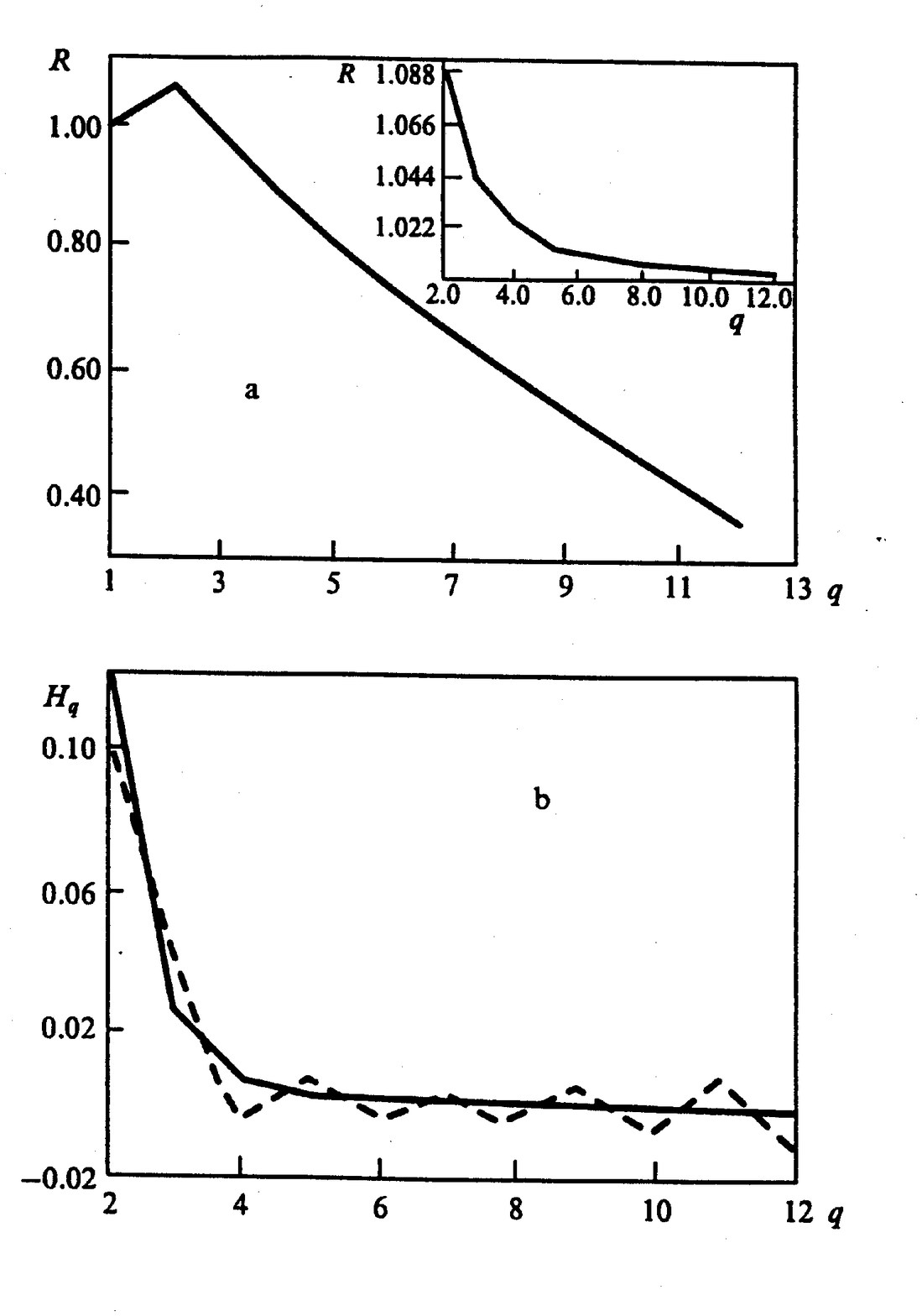}
  \end{turn}
\end{center}
\vspace*{-4cm}
\caption{(a) The ratio of factorial moments derived 
from the KNO curve in Fig.~\ref{fig-five} to the 
asymptotic result of eq.~(\ref{59});
the inset shows the corresponding ratio
between the results of eqs.~(\ref{58}) and~(\ref{59}). 
(b)~The ratio $H_q$ obtained from the modified
KNO curve in Fig.~\ref{fig-five} (solid line) 
compared to its NBD counterpart with $k$=7 (dashed line).
} 
\label{fig-four}
\end{figure}

First, one obtains a recurrence relation for the factorial moments
by substituting expression~(\ref{7}) into eq.~(\ref{56}) and
equating the coefficients of $z^q$ on both sides:
\begin{equation}
  (q-q^{-1})F_{q} = \gamma \sum _{l=1}^{q-1} 
    C_{q}^{l}B(\gamma l,\gamma (q-l)+1)
  F_{q-l}F_{l} .   \label{58}
\end{equation}
This system of equations can be solved\footnote{The exact solution 
of the system of equations for quark and gluon jets in case of 
the fixed coupling constant is given in Section~\ref{sec-widthshigher}.} 
using the initial conditions $F_{0}$=$F_{1}$=1.
The inset in Fig.~\ref{fig-four}(a) shows the ratio of the
resulting moments to their values~\cite{12} at large ranks~$q$:
\begin{equation}
  F_{q}^{as} = \frac {[\Gamma (1+\gamma )]^{q}}{\Gamma (1+\gamma q)}\cdot 
  \frac {2q\Gamma (q+1)}{C^{q}} .   \label{59}
\end{equation}
For large values of $q\gamma $ one obtains
\begin{equation}
  F_{q}\approx \frac {2\mu D^{-q}}{\sqrt {2\pi \gamma }}\Gamma
   \left ( \frac {3}{2}
  +\frac {q}{\mu }\right ) ,    \label{60}
\end{equation}
where
\begin{equation}
  \mu =(1-\gamma )^{-1},\,\,\,\,\,\, 
  D=C\frac {\gamma ^{\gamma }(1-\gamma )^{1-\gamma }}
  {\Gamma (1+\gamma )}.   \nonumber
\end{equation}
The asymptotic results for $F_q$ at large ranks
determine the asymptotics of the KNO function $f(x)$ 
at high multiplicities:
\begin{equation}
  f(x)\approx \frac {2\mu ^{2} (Dx)^{3\mu /2}}{x\sqrt {2\pi \gamma }} 
  \exp [-(Dx)^{\mu }] ,   \;\;\;\;\;\;  (\mu -1)(Dx)^{\mu }\gg 1 .  \label{61}
\end{equation}
Thus the tail of the distribution at large multiplicities 
is far more suppressed 
than in the double-logarithmic approximation,
eq.~(\ref{23}). 
Using the value of $\gamma$$\,\approx\,$0.4 
at presently accessible energies
for which $\mu$=$(1-\gamma )^{-1}$$\,\approx\,$1.6,
one obtains a Gaussian-like suppression
rather than an exponential falloff as in eq.~(\ref{23}).
Thus, we conclude that the conservation laws drastically 
reduce the width of the multiplicity distribution. 

\begin{figure}[tphb]
\vspace*{-2.5cm}
\begin{center}
  \epsfxsize=15cm
  \hspace*{-.3cm}\epsffile{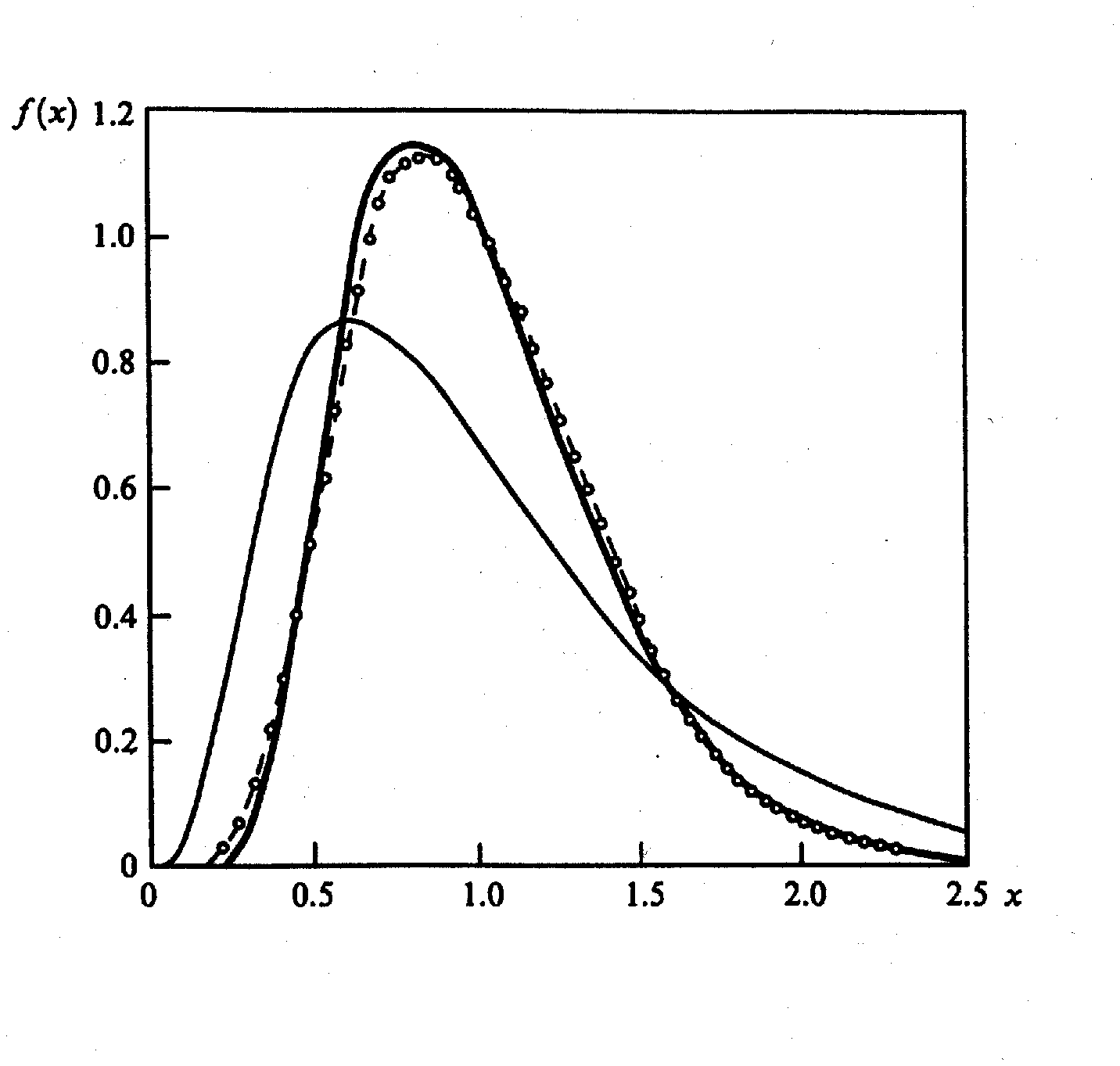}
\end{center}
\vspace*{-10cm}
\caption{The modified KNO function~\cite{12} 
(thick solid curve) for $\gamma $=0.4 
is much narrower than the lowest order distribution (thin solid curve). 
The negative binomial distribution with $k$=7 is shown for comparison 
(dashed curve).
} 
\label{fig-five}
\end{figure}

This is demonstrated in Fig.~\ref{fig-five}, 
where the modified QCD distribution,
i.e.~the distribution including energy conservation, 
is compared with the lowest order QCD result.
The modified distribution also accounts for some 
corrections at low multiplicities~\cite{12}.
The modified distribution (thick solid line)
is seen to be much narrower than the DLA distribution
(thin solid line).
Also shown in Fig.~\ref{fig-five}
is the result of a fit of the modified distribution by the NBD,
with parameters $k$=7 and \mbox{$\langle n \rangle $=30.} 

Making use of the modified QCD curve in Fig.~\ref{fig-five}, 
the genuine factorial moments can be calculated,
i.e.~the moments
including the corrections at low multiplicities. 
Their ratio to the asymptotic solution, eq.~(\ref{59}), 
is shown in the main part of Fig.~\ref{fig-four}(a).
Comparison of the two curves in Fig.~\ref{fig-four}(a) reveals the 
importance of the corrections at low multiplicities. 
Using the genuine factorial moments,
one can compute the cumulants and the $H_q$ ratios. 
The latter are shown in Fig.~\ref{fig-four}(b)
and compared with the $H_q$ moments obtained from the
fitted NBD curve in Fig.~\ref{fig-five}.
A dramatic difference is seen between the behavior of
the modified QCD solution and the NBD.
The $H_q$ ratios in the former case are seen to
oscillate as the rank increases.
In contrast, the $H_q$ ratios of the NBD decrease monotonically,
as was already seen in Fig.~\ref{fig-one}(c).
These oscillations of the $H_q$ ratios are reminiscent of 
the oscillations observed for the fixed multiplicity distribution,
Fig.~\ref{fig-two}(c). 
Oscillations with a different periodicity are obtained from 
the solutions to the full QCD generating equations,
as discussed below in
Sections~\ref{sec-computer} and~\ref{sec-exactsolutions}.

\subsection{Higher order approximations with a running coupling constant}
\label{sec-hoapprox}

Eq.~(\ref{56}) for the generating function in gluodynamics can be 
solved in a more general approximation~\cite{13},
by accounting for all terms of the kernel $K$
including the non-singular ones,
by incorporating the running coupling constant, 
and by using a Taylor series expansion of the generating 
functions in the nonlinear term at large $y$:
\begin{equation}
  G(y+\epsilon )\approx G(y) + G^{\prime }(y)\epsilon +\frac {1}{2} G^{\prime
  \prime }(y)\epsilon ^{2} + ...     \label{62}
\end{equation}
This approach illustrates the distinction 
between the various assumptions 
and their relative importance.

Each derivative in eq.~(\ref{62}) produces a derivative of 
$\langle n\rangle$ which is proportional to the anomalous dimension 
$\gamma$ according to eq.~(\ref{57}). 
In turn, the anomalous dimension $\gamma$ can be represented 
by a perturbative series with $\gamma_0$ as the leading term. 
Therefore, higher derivatives in the Taylor series give rise 
to terms of a higher order approximation in the perturbative expansion.
The Taylor series leads to a recombination 
of different order terms in this expansion. 
It is important to note that energy conservation,
accounted for by $x$ and $1-x$ in eqs.~(\ref{50}) and~(\ref{51}),
is an essential and integral aspect of
the higher order perturbative corrections. 
At first sight, it may seem odd that
the order of the approximation can be changed
when energy conservation is included.
However, this becomes clear if one considers that energy sharing 
among partons results in terms with lower 
powers of $\log s$ at the same power of $\alpha_S$, 
i.e.~in a shift to a higher approximation as seen from 
eq.~(\ref{62}).

The running coupling strength $\alpha_S$ should be 
considered in the two-loop approximation when working in NNLO. 
It is given by the formula
\begin{equation}
  \alpha _{S}(y)=\frac {2\pi }{\beta _{0}y}\left( 1-\frac {\beta _1}
  {\beta _{0}^{2}}\cdot \frac {\ln 2y}{y}\right)
   + {\cal{O}}(y^{-3}), \label{al}
\end{equation}
with
\begin{equation}
 \beta _{0}=\frac {11N_{c}-2n_f}{3}, \;\; \beta _1 =
    \frac {17N_c^2-n_f(5N_c+3C_F)}{3},
 \label{be}
\end{equation}
where $n_f$ is the number of active quark flavors 
(in gluodynamics $n_{f}$=0), 
and the difference between the QCD scale $\Lambda$ and 
$Q_0/2$ is neglected in asymptotics. 
In MLLA, one can use the one-loop expression, 
i.e.~the first factor of the product in eq.~(\ref{al}) only.

Note that --~in contrast to QED~--
the QCD series is an expansion in the product of the 
coupling strength and energy dependent ($\ln s$) terms,
rather than an expansion in the coupling strength alone.
Therefore higher order corrections appearing in the 
QCD expansion arise not only from higher loop graphs 
but also from a proper accounting for conservation laws,
as mentioned above.
In this review,
we show that the effects of the latter
--~resulting in a limitation of the available phase space~--
are much more important than the two-loop terms of
the coupling strength.
This explains e.g. why NNLO calculations of the multiplicity
ratio $r$ between gluon and quark jets yield rather different 
results depending on the accuracy with which
energy conservation is included,
as we discuss below in
Sections~\ref{sec-rtheory} and~\ref{sec-computer}.

Inserting expression~(\ref{62}) for the generating function
into the nonlinear term of eq.~(\ref{56}), 
dividing both sides of (\ref{62}) by $G(y)$ and differentiating
with respect to $y$, we obtain
\begin{eqnarray}
  [\ln G(y)]^ {\prime \prime }= \gamma _{0}^{2}(y)[(1+4h_{1}^{2}
  \gamma _{0}^{2}(y))(G(y)-1)-2h_{1}G^{\prime }
  (y) \nonumber \\ +\sum _{n=2}^{\infty }(-1)^{n}h_{n}G^{(n)}(y) 
   + \sum _{m,n=1}^{\infty }(-1)^{m+n}
  h_{nm}\left( \frac {G^{(m)}G^{(n)}}{G}\right) ^{\prime }] ,  \label{63}
\end{eqnarray}
where
\begin{eqnarray}
  h_{1}&=&11/24, \nonumber \\
  h_{2}&=&(67-6\pi ^{2})/36, \nonumber \\
  h_{3}&=&(4\pi ^{2}-15)/24, h_{4}=13/3, \nonumber \\
  h_{n}&=&\vert 2-2^{-n}-3^{-n}-\zeta (n)\vert , \nonumber \\
  \zeta (n)&=&\sum _{m=1}^{\infty }m^{-n}, \;\;  n\geq 2,  \label{64}
\end{eqnarray}
\begin{equation}
  h_{mn}=\vline \frac {1}{m!n!}\int _{0}^{1}dxK(x)
    \ln ^{n}x\ln ^{m}(1-x)\vline \, .
  \label{65}
\end{equation}
The first term on the right hand side $(G(y)-1)$ without its 
pre-factor is the well known~\cite{5} expression of the 
double-logarithmic approximation, eq.~(\ref{edla}).
The additional higher order contribution 
$4h_{1}^{2}\gamma _{0}^{2}$ is due to the derivative of~$\gamma_{0}^{2}$, 
and is considered in~\cite {39} but not in~\cite{13}.
Similar corrections appear in the subsequent terms of eq.~(\ref{63}).
In particular, the term proportional to $h_1$
corresponds to the modified leading-logarithmic approximation, 
while the term proportional to $h_2$ together with the 
$4h_{1}^{2}\gamma _{0}^{2}$ correction to the first term,
mentioned above,
corresponds to the next-to-next-to-leading order correction. 
Actually, the transverse momentum $k_t$ given by eq.~(\ref{ktli}) 
should be used as the argument of $\alpha_S$ in place of $y$, 
but it can be shown that such a replacement leads 
to yet higher order corrections than those considered here. 
Note that the use of the preasymptotic limits of integration,
i.e.~$e^{-y}$ and $1-e^{-y}$,
do not lead to a differential equation which is as 
simple as eq.~(\ref{63})
but result in power-like corrections.

A straightforward solution of eq.~(\ref{63}) appears problematic 
even if only the terms with $h_1$ and $h_2$ are included 
in addition to the double-logarithmic terms. 
The solution is very simple for the moments of the
distributions~\cite{13}, however,
since $G(z)$ and $\ln G(z)$ are the 
generating functions for factorial 
moments and cumulants, respectively. 
Using formulas~(\ref{7}) and~(\ref{8}) 
and assuming asymptotic $F$ scaling, 
one obtains the product $q\gamma$ (and its derivatives) 
at each differentiation in eq.~(\ref{63})
because the average multiplicity 
is the only $y$-dependent term which remains. 
The coefficients of $z^q$ on both sides should be equal. 
Hence, one obtains in NNLO:
\begin{equation}
  H_{q} = \frac {K_q}{F_q} = \frac {\gamma _{0}^{2}[1-2h_{1}q\gamma 
  +4h_{1}^{2}\gamma _{0}^{2} +h_{2}(q^{2}\gamma ^{2}
  +q\gamma ^{\prime })]}{q^{2}\gamma ^{2}+q\gamma ^{\prime }} . \label{66}
\end{equation}
The anomalous dimension $\gamma$ is defined by eq.~(\ref{57}). 
The condition $F_{1}$=$K_{1}$=1 determines the relation between 
$\gamma$ and $\gamma_{0}$:
\begin{equation}
  \gamma \approx \gamma _{0} - \frac {1}{2}h_{1}\gamma _{0}^{2} + \frac {1}{8}
  (4h_{2}+15h_{1}^{2})\gamma _{0}^{3} 
  + {\cal{O}}(\gamma _{0}^{4}) ,  \label{67}
\end{equation}
which shows that the increase of the average multiplicity with energy is
slower in the modified leading-logarithmic approximation than in the 
double-logarithmic approximation,
since the term with $h_{1}$ is negative 
(see eq.~(\ref{57})).
However, the higher orders reverse this situation.
The MLLA and NNLO terms in eq.~(\ref{67}) practically cancel 
each other for present values of $\gamma_{0}$. 
Therefore the series~(\ref{67}) essentially yields the asymptotic result.
Only at very large energies when $\gamma_{0}$$\,\approx\,$0 
do subsequent terms in the series
become smaller than the lower order ones
(if only the first term in the expression for $\alpha_S$ is considered).
However, one should keep in mind that in NNLO the value of 
$\gamma_0$ itself acquires a correction,
as given by the second term in the brackets in eq.~(\ref{al}).
This correction is negative and drastically reduces the value of 
the coefficient in front of $\gamma_{0}^{3}$ in eq.~(\ref{67}),
resulting in a much better convergence of the series. 
This is discussed below in connection with the NNLO and 3NLO
corrections to the expression for the energy dependence of
mean multiplicities.

The running property of $\gamma _{0}$ was taken into account 
in expression~(\ref{67}) according to eq.~(\ref{al}):
\begin{equation}
  \gamma _{0}^{\prime }\approx -h_{1}\gamma _{0}^{3} + 
   {\cal{O}}(\gamma _{0}^{5}) ,
  \label{68}
\end{equation}
which leads to
\begin{equation}
  \gamma ^{\prime }\approx -h_{1}\gamma _{0}^{3}(1-h_{1}\gamma _{0}) +
  {\cal{O}}(\gamma_{0}^{5}) .   \label{69}
\end{equation}
The lesson we learn from eq.~(\ref{66}) is that in all correction terms
(which contain $h_{1}$, $h_{2}, \ldots$),
the expansion parameter $\gamma$ appears 
in a product with the rank,
namely the product~$q\gamma$, 
which becomes large at high ranks, i.e.~at high multiplicities. 
Therefore, for high multiplicity events one should 
take into account ever higher order terms in $\gamma$. 
This problem was mentioned a long time ago~\cite{5} and 
is discussed in some detail in Ref.~\cite{38}, 
but has only recently been analyzed.

As already mentioned, 
the double-logarithmic formulas are obtained from eq.~(\ref{63}) 
by setting $h_{1}$=$h_{2}$=0, $\gamma$=$\gamma_{0}$,
and $\gamma^{\prime}$=0. 
In this case,
\begin{equation}
  H_{q}= q^{-2} ,  \label{70}
\end{equation}
which is similar to the asymptotic form of the negative binomial 
distribution with $k$$\,\approx\,$2,
corresponding to an extremely wide multiplicity distribution 
(see eq.~(\ref{23})).
In contrast, experimental data typically yield
values of $k$ in the range from 3.5 to~100. 
We stress, however, that the NBD result for $k$=2 is
\begin{equation}
  H_q=\frac {2}{q(q+1)}, \label{70a}
\end{equation}
which shows that the ``genuine'' correlations 
are about twice as large in the NBD as in QCD.
Therefore the QCD predictions for a single jet do not support 
the hypothesis of the negative binomial distribution even in asymptotics,
not to mention other distinctive features
of eq.~(\ref{66}) and its generalization. 

It is easy to recognize that eq.~(\ref{70}) is just an asymptotic form
(for $y$$\,\rightarrow\,$$\infty $) of eq.~(\ref{66}) 
for any fixed rank~$q$. 
The value of $\gamma$ tends to zero as $y^{-1/2}$ at large~$y$. 
In this limit one obtains eq.~(\ref{70}) from eq.~(\ref{66}). 
The approach to the asymptotic limit is very slow,
however.
The omitted terms are of order ${\cal{O}}(\ln^{-1/2}s)$. 
Therefore they should be taken into account at present energies.

Note that the preasymptotic difference between assumed $F$ scaling 
and KNO scaling does not alter our conclusions because it is of order
${\cal{O}}(\langle n\rangle^{-1})$, 
i.e.~it decreases with energy faster than
the terms ${\cal{O}}(y^{-1/2})$ and ${\cal{O}}(y^{-1})$ considered.

A more interesting feature of the solutions
is the qualitative behavior of $H_q$ 
as a function of rank at a given energy,
i.e.~when $\gamma_0$ is kept constant in eq.~(\ref{66}). 
According to eq.~(\ref{70}),
the $H_q$ moments are a smoothly decreasing function of $q$ 
in the double-logarithmic approximation. 
In the modified leading-logarithmic approximation in which the linear 
$h_{1}$ term is kept but the $h_{2}$ term 
and higher order terms are neglected in eqs.~(\ref{63}) and~(\ref{66}), 
$H_{q}$ acquires a minimum at
\begin{equation}
  q_{min}\approx \frac {1}{h_{1}\gamma _{0}} 
   + \frac {1}{2}+{\cal{O}}(\gamma _0). \label{71}
\end{equation}
Note that the minimum position is shifted to larger rank values at 
higher energies due to the decrease of $\gamma_{0}$,
and to smaller ranks for lower energies or for smaller phase space 
windows~\cite{taiw}
if higher order corrections are neglected 
(see Section~\ref{sec-intermittency}).
At present energies,
with a value of $\gamma _{0}$ of about 0.45--0.5,
we obtain
\begin{equation}
  q_{min}\approx 5 \; (\pm 1) .  \label{71a}
\end{equation}
The uncertainty of $\pm 1$ for this result is
due to the ${\cal{O}}(\gamma_{0})$ correction in eq.~(\ref{71}).

In MLLA, the ratio $H_q$ crosses the abscissa axis, 
is negative at the minimum,
and tends to zero from below as $\sim -q^{-1}$ as
the rank increases. 
If one includes the NNLO terms (proportional to $h_{2}$), 
the ratio $H_{q}$ is shifted up by an amount 
$h_{2}\gamma _{0}^{2}$ independent of~$q$. 
The ratio then exhibits a second zero,
crosses the abscissa axis again,
and tends asymptotically 
to a positive constant $h_{2}\gamma_{0}^{2}$. 
The location of the minimum shifts to slightly 
larger values of $q$ because of the term with $h_{1}^{2}$ 
and the two-loop correction to $\alpha_S$.
If the the exact limits of integration 
$e^{-y}$ and $1-e^{-y}$ are used instead of 0 and~1,
the minimum moves back to smaller ranks, however.
Thus we conclude that the location of the minimum as approximately 
given by eqs.~(\ref{71}) and~(\ref{71a}) is a rather stable 
prediction of gluodynamics.
Furthermore we see that eq.~(\ref{66}) predicts an oscillation
of the $H_q$ ratio analogous to that found using the
more approximate methods of Section~\ref{sec-approximatesolutions}.

One could object that the expansion parameter 
$q\gamma$ entering eq.~(\ref{66}) is large at present energies. 
This, however, is similar to the situation with the expansion of, 
for example, $\cos x$ in a Taylor series,
for which the higher order terms approximate the function 
quite well even for large values of~$x$. 
As more terms of the Taylor expansion of $\cos x$ are included, 
the approximation improves at yet larger values of~$x$. 

\begin{figure}[tphb]
\begin{center}
  \epsfxsize=15cm
  \epsffile[100 100 600 700]{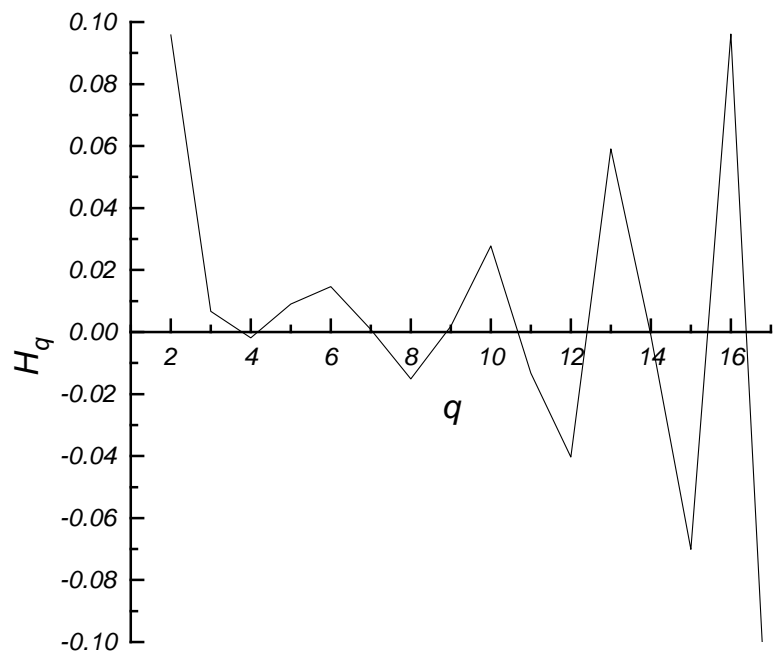}
\end{center}
\vspace*{-10cm}
\caption{
The ratio $H_q$ as a function of $q$ reveals 
oscillations in higher order perturbative QCD~\cite{41}.
The results shown here correspond to the Z$^{0}$ energy. 
} 
\label{fig-hqanalytic}
\end{figure}

A computer solution of eq.~(\ref{63})~\cite{41} 
also yields an oscillating behavior for $H_q$ 
when account is taken of the higher order terms 
with $h_3$ and $h_{11}$,
see Fig.~\ref{fig-hqanalytic}.
This is consistent with the discussion in the
previous paragraph.
However, 
the position of the first minimum
is shifted to $q$$\,\approx\,$4 
in the approximation of Ref.~\cite{41}, 
illustrating the sensitivity of $H_q$ to the various assumptions. 
In particular, the term with $h_4$, 
important for the proper limit in supersymmetric (SUSY) 
QCD as shown below, 
is not included in the results of Ref.~\cite{41}. 

Although the assumptions of the calculation
influence the quantitative details of the oscillations, 
the main qualitative features are stable. 
An implicit cutoff of the multiplicity distribution
at some maximum value due to 
the restrictions of finite energy 
produces similar oscillations,
although with a smaller amplitude~\cite{30,ugl}. 
Our present goal is to discuss the qualitative properties of~$H_q$.
More precise expressions are presented in the following sections.
We emphasize that the amplitudes and periodicity
of the oscillations are different in the more precise
treatment than in the results shown in 
Figs.~\ref{fig-two} and~\ref{fig-four}.

In this section,
we have demonstrated that conservation laws and 
higher order perturbative corrections lead, 
in the framework of gluodynamics, 
to a substantial reduction in the width of the multiplicity distribution 
and to a qualitative change in the behavior of 
the cumulant and factorial moments,
namely to oscillations of their ratio $H_q$ as a function of $q$ 
which disappear only at extremely high energies.

\section{Perturbative solutions of QCD equations 
with a running coupling constant}
\label{sec-psolutions}

A discussion of Quantum Chromodynamics,
in which quarks are created as well as gluons, 
leads back to the system of coupled equations 
(\ref{50}) and~(\ref{51}) for the generating functions.
The structure of these equations does not differ, 
however, from the equation for gluodynamics, eq.~(\ref{56}).
Therefore we shall not present the results for the full solutions
(see e.g. Refs.~\cite{dln,39}, \cite{cdnt8}-\cite{dg}), 
but will merely describe them. 

In analogy to gluodynamics, 
one obtains a system of coupled recurrent equations 
for factorial moments and cumulants when the Taylor series 
expansion~(\ref{62}) and the formulas (\ref{7}) and~(\ref{8}) are used. 
This system of equations has been solved analytically 
for average multiplicities,
accounting for corrections up to ${\cal{O}}(\gamma _{0}^{3})$, 
i.e.~the 3NLO approximation~\cite{cdnt8}.
Analytic calculations of the second and third factorial moments 
have been performed in the same approximation~\cite{dln}.
Numerical solutions are available
for yet higher moments~\cite{39, 41, 40}. 
The properties of gluon jets do not change appreciably
compared to Section~\ref{sec-gluodynamics}, 
i.e.~the gluon jet cumulants and factorial moments are very similar
to those determined in gluodynamics. 
In particular,
the ratio $H_q$ of gluon jets exhibits a minimum at the 
same value $q$$\,\approx\,$5 found in Section~\ref{sec-gluodynamics}.
The factorial moments of quark jets are determined to be larger 
than those of gluon jets, 
i.e.~the parton multiplicity distribution of quark jets is wider than 
that of gluon jets even though the average multiplicity is smaller.
This fact was known long ago~\cite{shty}-\cite{azdk}. 
The quark jet cumulants and $H_q$ ratios are also
found to oscillate.
The first minimum of the quark jet oscillation is 
predicted by the analytic solutions to be
located at $q$$\,\approx\,$7-8,
but it moves to $q$$\,\approx\,$5 if the non-asymptotic 
limits of integration are incorporated~\cite{lo2}. 

To apply these results to the two jet process of 
electron-positron annihilations, 
it is necessary to relate the {\epem} generating function 
to those of quarks and gluons. 
Bearing in mind the Feynman diagram for the production of two
quark jets in {\epem} annihilations at a very early stage, 
one obtains
\begin{equation}
  G_{e^{+}e^{-}}\approx \; G_{F}^{2} ,     \label{72}
\end{equation}
with further corrections (see, e.g., Ref.~\cite{32}). 
Thus the zeros of the quark jet and {\epem} cumulants
coincide since the logarithms of their generating functions 
are proportional to each other (see eq.~(\ref{5})).

\subsection{Average multiplicities and their slopes}
\label{sec-rtheory}

We now describe the QCD results for average 
multiplicities~\cite{cdnt8, dg} in more detail.
The equations for the average multiplicities in jets are obtained 
from the system of equations~(\ref{50}) and~(\ref{51}) 
by expanding the generating functions in $z$ and 
keeping the terms with $q$=0 and~1 in the 
series (\ref{7}) and~(\ref{8}). 
They read
\begin{eqnarray}
  \langle n_G(y)\rangle^{'} =\int dx\gamma _{0}^{2}[K_{G}^{G}(x)
  (\langle n_G(y+\ln x)\rangle +\langle n_G(y+\ln (1-x)\rangle -\langle n_G(y)
  \rangle ) \nonumber  \\
  + n_{f}K_{G}^{F}(x)(\langle n_F(y+\ln x)\rangle +\langle n_F(y+
  \ln (1-x)\rangle -\langle n_G(y)\rangle )], \;\;\;\;\; \label{ng}
\end{eqnarray}
\begin{equation}
  \langle n_F(y)\rangle ^{'} =\int dx\gamma _{0}^{2}K_{F}^{G}(x)
  (\langle n_G(y+\ln x)\rangle +\langle n_F(y+\ln (1-x)\rangle -\langle n_F(y)
  \rangle ).   \label{nq}
\end{equation}
From these equations one can predict the energy evolution of the ratio 
of multiplicities $r$ and the QCD anomalous dimension $\gamma$
(the slope of the logarithm of the average multiplicity in a gluon jet),
defined by
\begin{equation}
  r=\frac {\langle n_G\rangle }{\langle n_F\rangle } ,\;\;\;\;\; \;\;\;
  \gamma =\frac {\langle n_G\rangle ^{'}}{\langle n_G\rangle } 
  =(\ln \langle n_G\rangle )^{'}.  \label{def}
\end{equation}
The perturbative expansions of $\gamma $ and $r$ are
\begin{equation}
  \gamma = \gamma _{0}(1-a_{1}\gamma _{0}-a_{2}\gamma _{0}^{2}-
  a_3\gamma _0^3)+{\cal{O}}(\gamma _{0}^{5}), \label{X}
\end{equation}
\begin{equation}
  r = r_0 (1-r_{1}\gamma _{0}-r_{2}\gamma _{0}^{2}-
  r_3\gamma _0^3)
  + {\cal{O}}(\gamma _{0}^{4}).  \label{Y}
\end{equation}
The asymptotic value of $r$ is $r_0$=$N_{c}/C_{F}$=9/4,
as was first obtained in~\cite{brgu}. 

Using a Taylor series expansion of $\langle n\rangle$ at large $y$ in
eqs.~(\ref{ng}) and (\ref{nq}),
in conjunction with (\ref{X}) and (\ref{Y}),
the coefficients $a_i$ and~$r_i$ can be determined analytically.
Their numerical values are given in Table~\ref{tab-randa} 
for various numbers of active quark flavors
and for SUSY QCD.
In SUSY QCD where $n_f$=$N_{c}$=$C_{F}$,
all $r_i$ are identically zero.


\begin{table}[t]
\begin{center}
\begin{tabular}{|c|c|c|c|c|c|c|}
\hline
$n_f$ & $r_1$ & $r_2$ & $r_3$ & $a_1$ & $a_2$ & $a_3$\\
\hline
3 &  0.185 & 0.426 & 0.189   & 0.280 & - 0.379 & 0.209\\
\hline
4  &  0.191 & 0.468 & 0.080  & 0.297 & - 0.339 & 0.162 \\
\hline
5 &    0.198 & 0.510 &  -0.041  & 0.314 & - 0.301 & 0.112\\
\hline
S &  0     &  0    &    0    & 0.188 & -0.190  & -0.130  \\
\hline
\end{tabular}
\end{center}
\caption{Numerical values
of the perturbative corrections up to order $\gamma _{0}^{3}$
for the multiplicity ratio $r$ and the QCD anomalous dimension~$\gamma$,
based on integration limits of the generating functions
from 0 to~1 (see text).
$n_f$ is the number of active quark flavors while S refers
to supersymmetric (SUSY) QCD.
}
\label{tab-randa}
\end{table}

The value of the coefficient $r_{1}$ was first reported in
Ref.~\cite{43} and later in Ref.~\cite{33}. 
The coefficient $r_2$ was first presented in Ref.~\cite{34}. 
The $r_2$ results in Ref.~\cite{34} do not account
for energy conservation in the equations 
for the generating functions, however,
and as a consequence are much smaller 
than the $r_2$ values given in Table~\ref{tab-randa}.
The larger values of $r_2$,
i.e.~including energy conservation,
were first presented in Ref.~\cite{39}.
Similar results to those in Ref.~\cite{39} were subsequently
obtained in the framework of the
Lund dipole model~\cite{ed} with cutoff fractal triangles. 
Large NNLO corrections are also advocated in Ref.~\cite{37} 
using somewhat different reasoning.
The energy conservation effects responsible for the
large values of $r_2$ are closely related 
to the oscillations of higher rank moments discussed below. 
The 3NLO values of $a_3$ and $r_3$ were first 
presented in~\cite{cdnt8} where one can also find analytic 
expressions for all $a_i$ and $r_i$ with $i\leq 3$.

According to eq.~(\ref{Y}) and the values of 
$r_i$ in Table~\ref{tab-randa},
$r$ at higher orders is much smaller~\cite{39} than 
in the double-logarithmic approximation. 
On average, it is smaller by about~20\%. 
At the Z$^0$ resonance, 
the subsequent terms in eq.~(\ref{Y})
diminish the theoretical value of $r$ compared 
with its asymptotic value by
approximately 10\%, 13\%, and 1\% for $n_f$=4,
see Section~\ref{sec-rexp} and Fig.~\ref{fig-ratio}.

Inserting eq.~(\ref{X}) into eq.~(\ref{57}),
the energy behavior of the average multiplicity 
in gluon jets at 3NLO is determined to be~\cite{dg}
\begin{eqnarray}
  \langle n_{G}\rangle=Ky^{-a_{1}c^2 }\exp ( 2c\sqrt y+
  \frac {c}{\sqrt y}[2a_2c^2+\frac {\beta _1}{\beta _{0}^{2}}(\ln 2y+2)]
  \nonumber \\
   +\frac {c^2}{y}[a_3c^2-\frac {a_1\beta _1}{\beta _{0}^{2}}(\ln 2y +1)]),
 \label{eq-ng}
\end{eqnarray}
where $c$=$(4N_c/\beta _0)^{1/2}$=$(2B)^{-1/2}$
and $K$ is a normalization constant.
For $n_f$=0 and $B$=$h_1$ this reduces to the corresponding
result of gluodynamics.
The pre-exponential term and the first term in the exponent correspond 
to the MLLA expression~\cite{web1, dktr} 
(for a coupled system of quark and gluon jets,
these terms were first found in~\cite{cdfw}). 
The second term in the exponent,
proportional to $c/\sqrt y$,
is the NNLO correction.
This term can be also written as a small negative 
correction to the pre-exponent
(see Table~\ref{tab-randa}). 
The third term in the exponent,
proportional to $c^2/y$,
is the 3NLO result.
The role of the 3NLO term in the gluon jet multiplicity
is not important compared to the lower order terms
because of the smallness of~$a_3$. 
Note that the NNLO and 3NLO corrections are almost constant and 
somewhat compensate each other at currently accessible energies.
As a consequence,
the MLLA expression for gluon jets is a good approximation
to the higher order result, eq.~(\ref{eq-ng}).

\begin{figure}[thb]
\begin{center}
  \begin{turn}{.9}
  \epsfxsize=13cm
  \hspace*{-2.5cm}\epsffile[5 100 505 700]{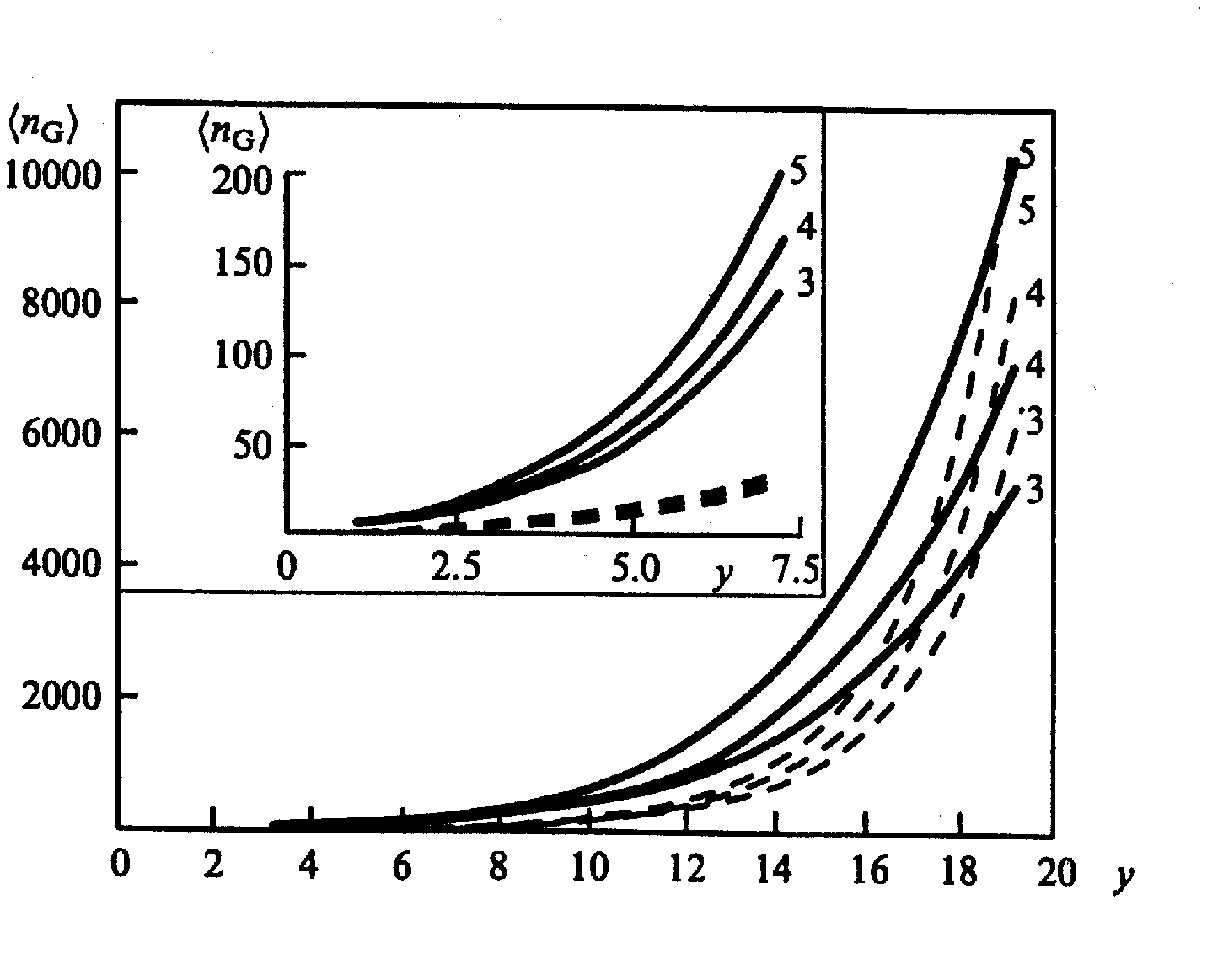}
  \end{turn}
\end{center}
\vspace*{-8.5cm}
\caption{The energy ($y$=$\ln(Q/Q_0)$) dependence of the average 
multiplicity of gluon jets,
for a running (solid curves) or fixed (dashed curves) 
coupling constant~\cite{39}.
The results are shown for different numbers of active flavors,
$n_f$= 3, 4, and~5.
} 
\label{fig-seven}
\end{figure}

The solid curves in Fig.~\ref{fig-seven} show
the behavior of the average gluon jet multiplicity as a
function of energy (i.e.~$y$)
as calculated in NNLO using a running coupling constant. 
The different curves show the results for $n_f$=3, 4 and~5.
The parameter $Q_{0}$=1.3$\lms$,
with $\lms$=175~MeV for \mbox{$n_f$=5},
and taken in the proportions 63:100:130 for $n_f$=5:4:3,
has been chosen as in~\cite{45}.
For purposes of comparison, 
we also show the energy dependence of the mean multiplicity
for a fixed coupling constant.
This is indicated by the dashed curves. 
At low energies, the multiplicity for fixed coupling increases rather slowly.
At higher energies, the rate of increase exceeds 
that found using the running coupling constant. 
This is reasonable since the coupling strength in the case
of fixed coupling has been evaluated at a rather high scale,
namely~$y_{Z^0}$$\,\approx\,$6.6 corresponding to the mass of Z$^0$,
yielding a relatively small (fixed) coupling strength. 
In actuality, the coupling strength should increase
during the course of evolution of the jet,
while the number of active flavors should decrease. 
The two trends somewhat compensate each other.

The slope of $r$,
namely $r'\equiv{\mathrm{d}}r/{\mathrm{d}}y$,
is extremely sensitive to higher order perturbative corrections.
The role of higher order corrections is
increased here compared to $r$ because each $n$th order term,
proportional to $\gamma_{0}^{n}$,
gets an additional factor of $n$ in front of it when
differentiated, the main constant term disappears, 
and the large value of the ratio $r_2/r_1$ 
becomes crucial~\cite{cdnt8}:
\begin{equation}
  r^{'} =Br_{0}r_{1}\gamma _{0}^{3}\left [1+\frac {2r_{2}\gamma _{0}}
  {r_1}+\left (\frac {3r_3}{r_1}+B_{1}\right )\gamma _{0}^{2}
  + {\cal{O}}(\gamma _{0}^{3})
  \right ],       \label{rpri}
\end{equation}
where the relation $\gamma _{0}^{'}\approx -B\gamma _{0}^{3}(1+B_{1}
\gamma_{0}^{2})$ has been used,
and where $B$=$\beta _{0}/8N_c$ and 
$B_{1}$=$\beta_{1}/4N_c\beta _0$. 
The factor in front of the brackets is very small even
at present energies: $Br_0r_1$$\,\approx\,$0.156 
and $\gamma _0$$\,\approx\,$0.5.
Nonetheless, the numerical estimate of 
$r^{'}$ is unreliable because of the expression inside the brackets.
Each differentiation leads to a factor $\alpha_S$ (or $\gamma_{0}^{2}$), 
i.e.~to terms of higher order.
The values of $r_1$, $r_2$ and $r_3$ given in Table~\ref{tab-randa} 
for $n_f$=4 yield
$2r_{2}/r_{1}$$\,\approx\,$4.9 and 
$(3r_3/r_1)+B_1$$\,\approx\,$1.5.
Thus the first correction in eq.~(\ref{rpri})
(proportional to $\gamma_0$)
is more than twice as large as unity at present energies,
while the next correction
(proportional to $\gamma _0^2$) is about~0.4. 
Therefore even higher order terms are necessary 
before the perturbative result for $r'$ can be considered
to be reliable.

The slope $r'$ is 0 for a fixed coupling constant.
In the case of the running coupling, 
$r'$ evolves rapidly with $y$ in
the lowest perturbative approximation,
ranging according to eq.~(\ref{rpri}) from 0.06
at the Z$^0$ to 0.25 at the~$\Upsilon$. 

Next we consider the ratios of slopes $r^{(1)}$ 
and curvatures $r^{(2)}$, defined as
\begin{equation}
  r^{(1)}=\frac {\langle n_G\rangle ^{'}}{\langle n_F\rangle ^{'}}, \;\;\;\;\;\;\;
  r^{(2)}=\frac {\langle n_G\rangle ^{''}}{\langle n_F\rangle ^{''}}. \label{r12}
\end{equation}
Their ratios to $r$ can be written as
\begin{equation}
  \rho _1 =\frac {r}{r^{(1)}}=1-\frac {r^{'}}{\gamma r},   \label{rh1}
\end{equation}
\begin{equation}
  \rho _2 =\frac {r}{r^{(2)}}=1-\frac {2\gamma rr^{'}+rr^{''}-2r^{'2}}
  {(\gamma ^{2}+\gamma ^{'})r^2}.   \label{rh2}
\end{equation}
Since $r^{'}$$\,\propto\,$$\gamma _{0}^{3}$, 
the asymptotic values of $r$, $r^{(1)}$ and $r^{(2)}$ coincide,
equaling~2.25.
Moreover, the values of $r$, $r^{(1)}$ and $r^{(2)}$ coincide in MLLA.  
The first preasymptotic corrections are very small. 
They are of order ${\cal{O}}(\gamma _{0}^{2})$ (NNLO) with a small
factor in front and contribute about 2-4$\%$ at  present energies:
\begin{equation}
  \rho _1 =1-Br_{1}\gamma _{0}^{2}\approx 1-0.07\gamma _{0}^{2},  
  \label{cor1}
\end{equation}
\begin{equation}
  \rho _2 =1-2Br_{1}\gamma _{0}^{2}\approx 1-0.14\gamma _{0}^{2}.
  \label{cor2}
\end{equation}
However, this favorable situation 
does not persist when the next terms are considered.
For $\rho _1$ the numerical values of the higher order terms are
\begin{equation}
  \rho _1 =1-0.07\gamma _{0}^{2}(1+5.38\gamma _{0}+4.21\gamma _{0}^{2}) \;\; 
  (n_f =4),   \label{cor3}
\end{equation}
and similarly for $\rho _2$ (see Ref.~\cite{cdnt8}).
All the correction terms are much smaller than~unity.
However,
the numerical factors inside the brackets are so large that
at present values of $\gamma _{0}$$\,\approx\,$0.5 the 
subsequent terms are larger than the 
first one and the sum of the series is unknown. 
The series can be summed using the simplest Pad\'{e} approximant
(i.e.~by assuming the steady decrease of the unaccounted terms 
in proportions determined by the ratio of two last calculated terms), 
resulting in
\begin{equation}
  \rho_1=1-0.07\gamma _0^2\left 
   [1+\frac {5.38\gamma _0}{1-0.78\gamma _0}\right ].
\label{rpad}
\end{equation}

In contrast to $\rho_1$,
the perturbative corrections to
the ratio of slopes $r^{(1)}$ are small.
The lowest order ${\cal{O}}(\gamma_0)$ correction to $r^{(1)}$ 
is the same as for $r$ but higher order corrections are smaller 
because they are negative both in $r$ and $\rho_1$
which define $r^{(1)}=r/\rho_1$. 
This explains why experimental values of $r^{(1)}$
are similar to values of $r$ calculated in the MLLA approximation 
(see Section~\ref{sec-experiment}),
whereas experimental values
of $r$ are about 25\% lower than the MLLA prediction.
Yet more substantial cancelations of higher order terms 
should occur for~$r^{(2)}$.
It would be interesting to test this prediction experimentally.

An interesting feature of eq.~(\ref{cor3}) 
is that $\rho _1$ contains
terms up to ${\cal{O}}(\gamma _0^4)$ whereas $r$ in the 
numerator of eq.~(\ref{rh1}) is known to ${\cal{O}}(\gamma _0^3)$ only. 
This situation is related
to the fact that $r'/\gamma \sim O(\gamma _0^2)$.

The value of $\rho _1$ also determines the ratio of the slope 
of the logarithm of the average multiplicity of a quark jet 
($\gamma _F$) to that of a gluon jet ($\gamma $):
\begin{equation}
  \gamma _F =(\ln \langle n_F\rangle )'= \rho _1 \gamma 
    =\gamma -\frac {r^{'}}{r}. 
\label{gf}
\end{equation}
Once again we see that the logarithmic slopes 
of quark and gluon jets are equal in MLLA.
They differ in higher orders in such a manner that 
$\gamma _F$$\,<\,$$\gamma$ since both $r$ and $r^{'}$ are positive.
Our failure to obtain a precise estimate of $\rho _1$ implies that
we cannot reliably evaluate $\gamma _F$ either.
Thus the ratio of logarithmic slopes $\rho_1$
is more sensitive to higher order perturbative terms
than the ratio of slopes $r^{(1)}$.

A high sensitivity of multiplicities in quark jets to higher order terms 
is especially noticeable if one writes down the expression for their 
energy dependence,
analogous to eq.~(\ref{eq-ng}) for gluon jets. 
Taking into account perturbatively the ratio $r(y)$ 
one obtains~\cite{cdnt8,dg}
\begin{eqnarray}
  \langle n_F\rangle=\frac {K}{r_0}y^{-a_1c^2}\exp (2c\sqrt y+
     \frac {c}{\sqrt y}[r_1+2a_2c^2+\frac{\beta _1}{\beta _{0}^{2}}(\ln 2y+2)]
         \nonumber \\ 
     +\frac {c^2}{y}[ a_3c^2 + \frac{r_1^2}{2} + r_2 
       -\frac {a_1\beta _1}{\beta _{0}^{2}}(\ln 2y+1)] ).  \label{eq-nq}
\end{eqnarray}
Especially striking is the role of the last term which 
cannot be neglected even at the energy of the~Z$^0$. 

Another representation of this dependence is the direct 
use of formula~(\ref{def}):
\begin{equation}
  \langle n_F\rangle =\frac {\langle n_G\rangle }{r_0(1-r_1\gamma _0-
  r_2\gamma _{0}^{2}-r_3\gamma _{0}^{3})}.   \label{nus}
\end{equation}
However one easily observes that formulas~(\ref{eq-nq})
and~(\ref{nus}) differ since there is no term with $r_3$ 
in eq.~(\ref{eq-nq}). 
This is an important point which emphasizes that the notion
of the order of approximations is different in the definitions 
of $\gamma $ and~$r$. 
The MLLA term of $r$ (proportional to $\gamma _0$),
given by $r_1$,
appears in the NNLO term of $\gamma_F$ 
(proportional to $\gamma _0^2$)
in combination with~$a_2$. 
This situation is a consequence of the fact that
$r^{'}$$\,\sim\,$${\cal{O}}(\gamma _{0}^{3})$. 
Therefore the notion of orders is somewhat a question of convention. 
We conclude that  it is improper to use the term with $r_3$ in 
eq.~(\ref{nus}) until the $a_4$ contribution to $\gamma$ is known. 
By extension this implies that if the
MLLA formula is used to describe multiplicities,
as is common practice, 
the DLA result $r$$\,=\,$$r_0$ should be inserted into all formulas
to be self consistent within the perturbative approach,
leading to the same energy dependence for quark and gluon jets.
The NNLO term with $r_1$ may be included in $\gamma _F$
only if the $a_2$ contribution has been incorporated in $\gamma$, 
and so on.

Therefore within the present accuracy of ${\cal{O}}(\gamma_0^3)$ corrections,
the perturbative QCD approach fails in the precise determination of the
logarithmic slope of the quark jet multiplicity $\gamma _F$,
even at the Z$^0$,
and can be trusted only at much higher energies.

\subsection{Widths of the distributions}
\label{sec-widths}

Next we consider higher order corrections for the moments of multiplicity. 
The normalized factorial moments of any rank $q$ can be obtained by
differentiating the generating functions according to eq.~(\ref{4})
or, equivalently, by using the series~(\ref{7}) and collecting the terms
with equal powers of $z$ on both sides of eqs.~(\ref{50}) and~(\ref{51}).
The QCD equations for the moments of multiplicity 
are given in Refs.~\cite{32, dln, 39, 41, 40}. 

The normalized second factorial moments of gluon and quark jets,
$F_2$, are given by
\begin{equation}
  F_2^G =\frac {\langle n_G(n_G-1)\rangle }{\langle n_G\rangle ^2}, \;\;\;\;
  F_2^F =\frac {\langle n_F(n_F-1)\rangle }{\langle n_F\rangle ^2}.
   \label{fgff}
\end{equation}
The $F_2$ moments define the widths of the multiplicity distributions,
being related to the dispersion 
$D^{2}=\langle n^2\rangle -\langle n\rangle ^2$ by
\begin{equation}
  D^{2}=(F_{2}-1)\langle n\rangle ^{2}+\langle n\rangle =
  K_{2}\langle n\rangle^{2} + \langle n\rangle .     \label{disp}
\end{equation}
The perturbative expansions of $F_2$ up to $\gamma _0^3$ are
\begin{equation}
  F^G_2=\frac{4}{3}(1-f_1\gamma_0-f_2\gamma_0^2-f_3\gamma _0^3), 
  \label{f2g}
\end{equation}
\begin{equation}
  F^F_2=(1+\frac{r_0}{3})(1-\phi_1\gamma_0-
  \phi_2\gamma_0^2-\phi _3\gamma _0^3). \label{f2f}
\end{equation}

Inserting expressions~(\ref{f2g}) and~(\ref{f2f})
into the QCD equations yields 
predictions for the coefficients $f_i$ and~$\phi_i$. 
The asymptotic ($\gamma_0$$\,\rightarrow\,$0) 
values of $F_2^G$ and $F_2^F$
can also be determined by equating the leading terms 
in $\gamma_0$ on both sides of the equations. 
We have presented their explicit expressions in eqs.~(\ref{f2g})
and~(\ref{f2f}) to simplify later notation.
Analytic formulas for $f_i$ and $\phi_i$ are given in Ref.~\cite{dln}.
Their numerical values for different numbers of active flavors 
and for SUSY QCD
are listed in Table~\ref{tab-fandphi}.
From these values it is seen that
the series~(\ref{f2g}) and~(\ref{f2f}) change in sign at each term
and that higher order terms are more important for 
the width of a quark jet than for a gluon jet.

\begin{table}[t]
\begin{center}
\begin{tabular}{|c|c|c|c|c|c|c|}
\hline
$n_f$ & $f_1$ & $f_2$&$f_3$ & $\phi_1$ & $\phi_2$&$\phi _3$\\
\hline
3 & 0.364 & -0.0279 &  0.795 & 0.637 & -0.276 &  2.12      \\
\hline
4 & 0.358 & -0.0457 & 0.740 & 0.631 & -0.286 &  2.04  \\
\hline
5 & 0.352 & -0.0629 & 0.689 & 0.625 & -0.295 &  1.95        \\
\hline
S & 0.313 & 0.310 &  -0.120    & 0.313 & 0.310 & -0.120  \\
\hline
\end{tabular}
\end{center}
\caption{Numerical values of the perturbative corrections 
up to order $\gamma _{0}^{3}$ for the normalized second
factorial moments of gluon ($f_i$) and quark ($\phi_i$) jets.
$n_f$ is the number of active quark flavors while S refers
to supersymmetric (SUSY) QCD.
}
\label{tab-fandphi}
\end{table}

The asymptotic values 
of $F_{2}^{G}$ and $F_{2}^{F}$ are~\cite{1}:
\begin{equation}
  F_{2, as}^{G}=\frac {4}{3}, \;\;\;\; F_{2, as}^{F}=
    1+\frac {r_0}{3}=\frac {7}{4}.  \label{fas}
\end{equation}
At lower energies the widths are slightly smaller 
because of the increase of~$\alpha _S$,
leading to larger values of the first corrections 
(with coefficients $f_1$ and~$\phi_1$)
in eqs.~(\ref{f2g}) and~(\ref{f2f}). 

\begin{figure}[tphb]
\vspace*{-4cm}
\begin{center}
  \epsfxsize=18cm
  \hspace*{-1.5cm}\epsffile{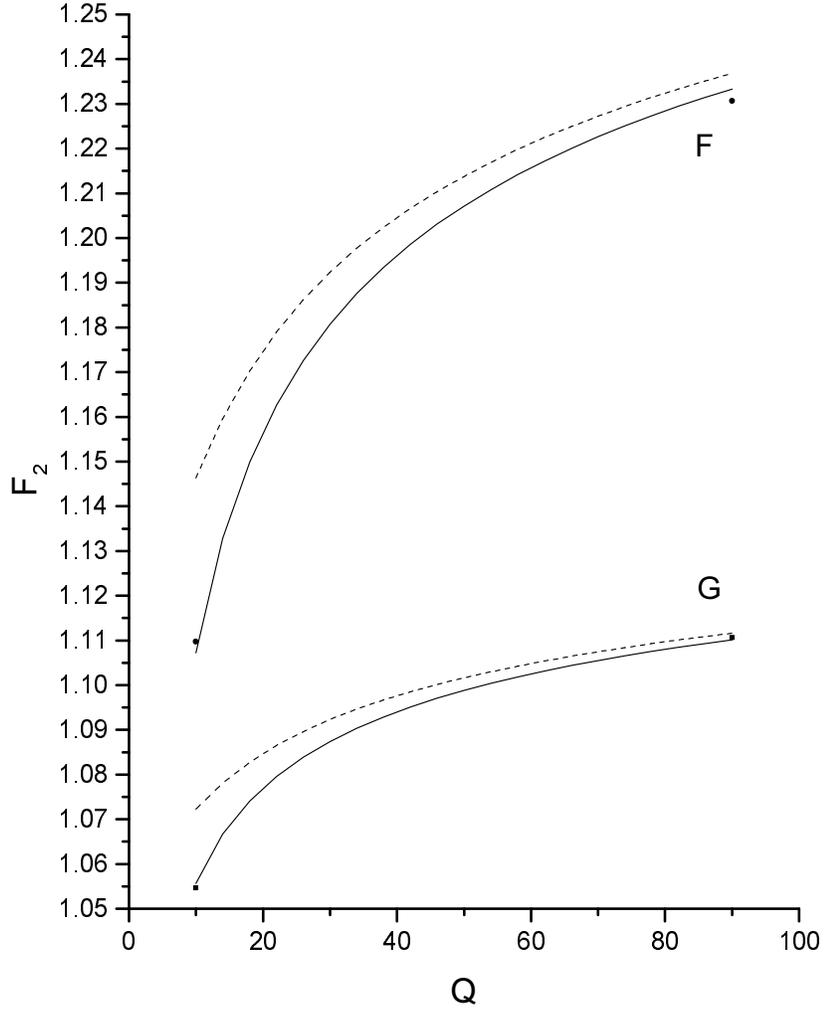}
\end{center}
\vspace*{-5cm}
\caption{The energy behavior of the second factorial moments of
quark (F) and gluon (G) jets~\cite{dln}.
The scale of the energy $Q$ is~GeV.
The limits of integration are chosen to be either 0 and 1 (solid curves)
or $\varepsilon$ and 1$\,-\,$$\varepsilon$ (dashed curves).
The dots at the ends of the curves demonstrate that the results are
insensitive to variation of the effective number of flavors (see text).
} 
\label{fig-lam}
\end{figure}

The MLLA prediction for the energy dependence of the 
second factorial moments is shown in Fig.~\ref{fig-lam},
using $n_f$=4 and the values of $\alpha_S$ in Ref.~\cite{pdg}.
The curves demonstrate that the widths are almost
Poissonian at low energies,
i.e.~they approach unity as the scale decreases.
The small dots at the ends of the curves show the effects
of a change in $n_f$ to 3 at $Q$=10~GeV and to 5 at $Q$=90~GeV:
it is seen that the results are almost insensitive
to the number of effective flavors.
Note that $n_f$ is the only free parameter in these results.
Cutoff of the integration at 
$\varepsilon$=$e^{-y}$$\,\approx\,$$e^{-2\pi /\beta _0\alpha _S}$ 
from below and at $1-\varepsilon$ from above as a consequence
of preasymptotic effects is also not particularly important 
at present energies,
as shown by the dashed curves in Fig.~\ref{fig-lam}.
The preasymptotic corrections slightly increase the widths 
and reduce the slope with respect to~$Q$. 

Unfortunately, higher order corrections do not improve the
theoretical estimates. 
The NNLO terms are positive,
while the 3NLO corrections are negative 
and so large they lead to sub-Poissonian 
widths ($F_2$$\,<\,$1) for~\mbox{$\alpha_S$=0.2.}
Indeed, the 3NLO corrections are just as large as
those in MLLA. 
The origin of the large values of $f_3$ and $\phi_3$
can be traced to rather large contributions of integrals 
of the form $\int _0^1\ln ^nxdx \propto n!$,
i.e.~to the region of very soft gluons. 
Thus the cutoff at $e^{-y}$ and $1-e^{-y}$ should be
quite important for the higher order terms.
This is analogous to the situation with renormalons 
(see, e.g., Ref.~\cite{zak}).

The 3NLO corrections are overestimated in the method of
the Taylor series expansion because of the assumption 
$y$$\,\gg\,$$\vert \ln x\vert $ which is invalid for soft gluons. 
For example, the $k_t$ dependence of the coupling strength is
transformed so that
\begin{equation}
  \alpha _S\propto \frac {1}{y+\ln x(1-x)}\approx \frac {1}{y}\left (1-
  \frac {\ln x(1-x)}{y}\right ).  \label{yx}
\end{equation}
The second term becomes infinitely large 
for~$x$$\,\rightarrow\,$0. 
The cutoff at $x$=$e^{-y}$ yields a factor of 2 only. 
Thus the above expansion implies a presumption about 
the behavior of the coupling strength in the non-perturbative region 
as well as its modification at the limits of the perturbative one.

The slopes of the widths are especially sensitive to 
higher order terms because each of them is multiplied 
by the factor $n$ when differentiating $\gamma_0^n$. 
Thus the 3NLO term in the perturbative expression for the
slopes of the widths is about 3 times larger than the MLLA term.
This suggests that a precise quantitative
estimate of these slopes is not possible. 
At present energies,
$F_2^{G'}(MLLA)$$\,\approx\,$0.04 and
$F_2^{F'}(MLLA)$$\,\approx\,$0.092 
for~\mbox{$\alpha _S$=0.2.}
The asymptotic value of the ratio of slopes can be
estimated to be
\begin{equation}
  \frac {(F_{2}^{G})^{'}_{as}}{(F_{2}^{F})^{'}_{as}}=\frac {16f_1}{21\phi _1}
  \approx 0.43,   \label{rfpr}
\end{equation}
which surely coincides with the MLLA prediction for this ratio. 
Thus the second factorial moment of quark jets 
approaches its asymptotic value faster than the
corresponding quantity for gluon jets.

We stress that all slopes and curvatures in pQCD are related to the running
property of the QCD coupling constant since they are proportional to its
derivatives which are zero for a fixed coupling strength.

One can easily check (see Ref.~\cite{dln})
that the relations of SUSY QCD,
where $n_f$=$N_c$=$C_F$,
are valid for the coefficients given above 
(e.g.,~$F_{2}^{G}$=$F_{2}^{F}$ etc.).
The asymptotic SUSY values of $F_2$ equal~4/3
for both gluon and quark jets.

Third moments of the multiplicity distributions
have been calculated~\cite{dln} in a manner analogous
to that described above for the second moments.
The results are
\begin{equation}
  F_3^G=h_0(1-\sum _{i=1}^{3}h_i\gamma _0^i); \;\;\;\;
  F_3^F=g_0(1-\sum _{i=1}^{3}g_i\gamma _0^i),  \label{f32}
\end{equation}
where $h_0$=9/4, $g_0$=$1+r_0+r_0^2/4$,
and the values of the coefficients $h_i$ and $g_i$ 
are listed in Table~\ref{tab-handg}. 
In the asymptotic limit,
the third moment of quark jets is
about twice as large as the third moment of gluon jets.

\begin{table}[t]
\begin{center}
\begin{tabular}{|c|c|c|c|c|c|c|}
\hline
$n_f$ & $h_1$ & $h_2$ & $h_3$ & $g_1$ & $g_2$ & $g_3$ \\
\hline
3     & 0.986  & -0.342 & 2.49 & 1.61 & -1.58 & 7.74   \\
\hline
4     & 0.972 & -0.380 & 2.36 & 1.60 & -1.59 & 7.54 \\
\hline
5     & 0.957 & -0.417 & 2.25 & 1.59 & -1.60 & 7.34 \\
\hline
S     & 0.844 & 0.722 & -1.09 & 0.844 & 0.722 & -1.09 \\
\hline
\end{tabular}
\end{center}
\caption{Numerical values of the perturbative corrections 
up to order $\gamma _{0}^{3}$ for the normalized third
factorial moments of gluon ($h_i$) and quark ($g_i$) jets.
$n_f$ is the number of active quark flavors while S refers
to supersymmetric (SUSY) QCD.
}
\label{tab-handg}
\end{table}

Comparing $f_i$ and $\phi_i$ to $h_i$ and $g_i$,
one concludes that the corrections
increase for higher moments,
even in MLLA. 
Moreover, at present values of $\gamma _0$$\,\approx\,$0.5, 
the corrections are rather large.
Note the similarity in the structure of the corrections 
for the second and third moments. 
The corrections alternate in sign, 
and the third coefficients are larger than the first ones. 
This is an indication of a sign-alternating asymptotic series,
and Borel summation could be effective. 

In SUSY QCD the asymptotic values of 
the third moments equal~9/4. 
The first correction,
given by $h_1$(SUSY)=$g_1$(SUSY)=0.844,
is about the same size as for ordinary gluon jets and
is similar to the correction for second moments.
However, the NNLO and 3NLO terms for moments in SUSY QCD 
differ drastically from those for ordinary jets,
both in absolute value and sign,
as seen from
Tables~\ref{tab-fandphi} and~\ref{tab-handg}. 
This demonstrates the sensitivity of the 
results to the value of~$r_0$,
which is different in the two cases.

\begin{figure}[tphb]
\begin{center}
  \epsfxsize=14cm
  \epsffile{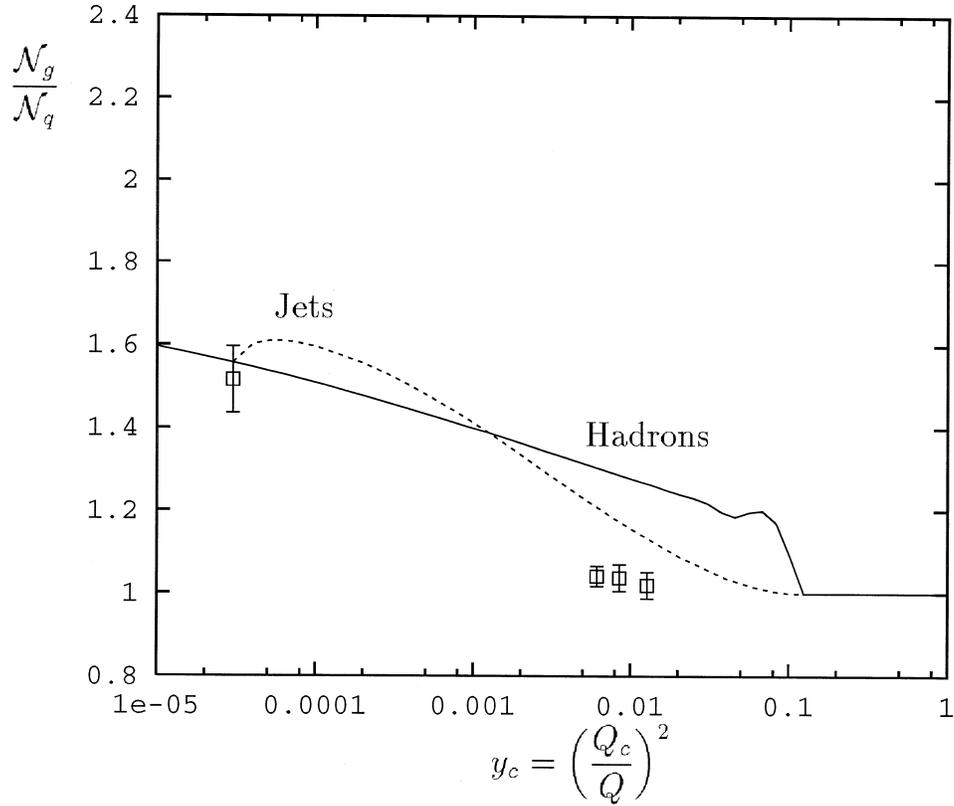}
\end{center}
\caption{Computer solution~\cite{lo1}
for the ratio of mean hadron multiplicities
between gluon and quark jets (solid curve)
in comparison to data.
The experimental result at $y_c$$\,\approx\,$$3\times 10^{-5}$
is based on {\epem} Z$^0$ events from 
the OPAL Collaboration~\cite{bib-opal96,bib-opal99}.
The results at $y_c$$\,\approx\,$0.01 are from
$\Upsilon$ decays recorded by the 
CLEO Collaboration~\cite{bib-cleo97}.
The dashed curve shows the analogous prediction~\cite{lo1}
for subjet multiplicities.
} 
\label{fig-eight}
\end{figure}

A similar procedure can be used to determine 
yet higher rank moments.
However,
the series will probably be badly convergent.

\section{Computer solutions}
\label{sec-computer}

The analytic approach described in Sections~\ref{sec-gluodynamics}
and~\ref{sec-psolutions} accounts for 
energy conservation in an approximate manner.
Energy conservation can be included more accurately
by solving eqs.~(\ref{50}) and~(\ref{51}) numerically,
i.e.~by implementing a computer solution.
The computer solution also allows the preasymptotic 
limits of integration to be directly incorporated,
i.e.~the limits $e^{-y}$ and $1-e^{-y}$ rather than 0 and~1. 
Thus the non-perturbative region of $x$ values
near 0 and~1 can be avoided.

A computer solution of eqs.~(\ref{50}) and~(\ref{51})
is presented in \mbox{Refs.~\cite{lo2,lo1}.}
Thresholds for new quark production are
approximated using a smoothed function.
The coupling constant is considered 
in the one-loop approximation. 
A deficiency,
shared by the analytic calculations,
is the neglect of strict conservation of transverse momentum
and a corresponding limitation on the available phase space.
This deficiency effectively corresponds to a modification of
eqs.~(\ref{50}) and~(\ref{51}).

The principal result of the computer solution
is that the value of $r$ is reduced by more than 10\% compared 
with the 3NLO prediction.
The computer solution for $r$ versus energy scale $Q$ 
is shown in Fig.~\ref{fig-eight}.
The experimental results shown in this figure
are discussed in Section~\ref{sec-experiment}.

\begin{figure}[tphb]
\begin{center}
  \epsfxsize=15cm
  \hspace*{-1cm}\epsffile{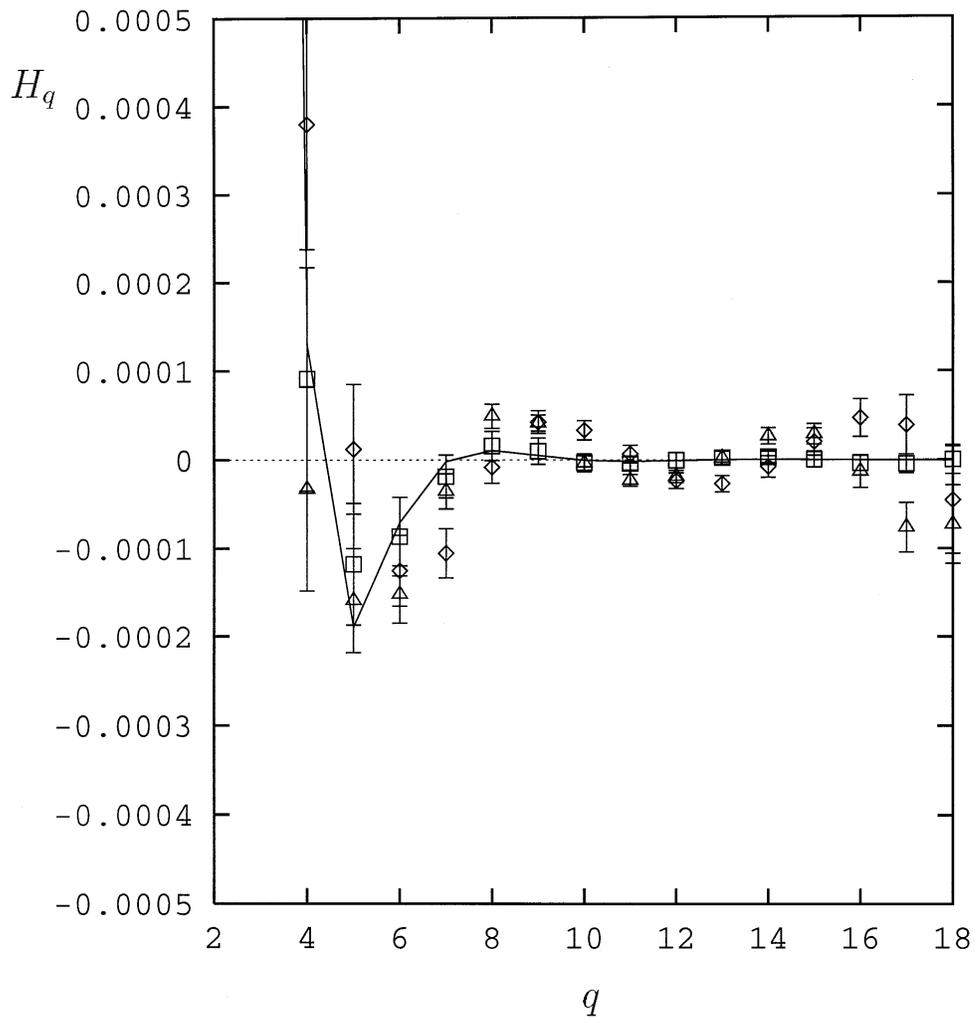}
\end{center}
\caption{Computer solution for the ratio $H_q$ of cumulant 
over factorial moments as a function of rank $q$ (solid curve)~\cite{lo2}
in comparison to experimental results from 
{\epem} Z$^0$ decays from the DELPHI Collaboration~\cite{bib-delphi}
as analyzed in~\cite{bib-alberto}.
} 
\label{fig-nine}
\end{figure}

A perturbative cutoff $Q_0$=507~MeV is found from a fit
to the measured mean particle multiplicity in quark jets~\cite{lo1}. 
With increasing resolution,
more and more jets are resolved, 
and predictions for the evolution of the subjet multiplicity 
(see Section~\ref{sec-subjets})
can be obtained as well. 
The subjet multiplicities are found to agree quite well 
with experiment using this same value of~$Q_0$~\cite{lo1}.
Correlation functions of high ranks up to $q$=18 in quark
and gluon jets were also determined,
fully accounting for
energy conservation~\cite{lo2}. 
A quantitative description of the
existing data on factorial moments, 
factorial cumulants,
and their ratio $H_q$ was obtained
for single quark and gluon jets 
and for two-jet {\epem} events,
using the same parameters determined for the average 
subjet and particle multiplicities, 
see Fig.~\ref{fig-nine} and Section~\ref{sec-highermoments}.

\section{Exact solutions of QCD equations with a fixed coupling constant}
\label{sec-exactsolutions}

The QCD results described above demonstrate that conservation laws 
and the non-singular terms of the kernels play a more important role 
up to MLLA than the running coupling constant,
which becomes essential in NNLO only. 
Thus the running property of the coupling constant does not
alter the qualitative predictions of QCD for multiplicity. 
It is possible to obtain an exact analytic solution of
eqs.~(\ref{50}) and~(\ref{51}),
i.e.~without resorting to a perturbative approximation,
if the coupling constant is fixed~\cite{21,42}.
No other assumption is necessary. 
One can obtain the general solution for the moments of any rank
in this approximation.
The effect of the running coupling constant can be assessed
by varying the value chosen for~$\gamma_0$.
This analytic approach provides further insight into the behavior
of the solutions.

In the following we first discuss the lowest rank moments
obtained in the fixed coupling approximation,
and compare them with the results of the previous sections.
Following this we discuss the solutions for higher rank moments.

\subsection{First moments and the ratio of average multiplicities in gluon jets
to those in quark jets}

The equations for average multiplicities,
i.e.~unnormalized moments of rank $q$=1,
are obtained by substituting the series~(\ref{7}) 
for the generating functions
into eqs.~(\ref{50}) and~(\ref{51}) and equating the terms linear in $z$,
with the conditions $F_{0}$=$F_{1}$=$\Phi_{0}$=$\Phi_{1}$=1. 
The factorial moments of quark jets are denoted by $\Phi_q$ 
and their cumulants by~$\Psi_q$.
If the coupling constant is fixed, the average 
multiplicities behave like
\begin{equation}
  \langle n_{G}(y)\rangle \propto \exp (\gamma y) , \;\;\;\;\;\;\;
  \langle n_{F}(y)\rangle \propto r^{-1}\exp (\gamma y) ,   \label{73}
\end{equation}
where the anomalous dimension $\gamma $ and the ratio $r$ are constants. 
This behavior follows directly from eqs.~(\ref{50}) and~(\ref{51})
as can be seen using the relations
\begin{equation}
  \langle n_{G,F}(y+\ln x)\rangle /\langle n_{G,F}(y)\rangle = x^{\gamma } ,
  \label{74}
\end{equation}
\begin{equation}
  \langle n_{G,F}(y)\rangle ^{\prime } = \gamma \langle n_{G,F}\rangle . \label{75}
\end{equation}
Then eqs.~(\ref{50}) and~(\ref{51}) can be rewritten as a system 
of algebraic equations for the variables $\gamma $ and~$r$:
\begin{equation}
  \gamma = \gamma _{0}^{2} [M_{1}^{G} + n_{f}r(M_{1}^{F} - M_{0}^{F})] , \label{76}
\end{equation}
\begin{equation}
  \gamma = \gamma _{0}^{2} [L_{2} - L_{0} + rL_{1}] ,  \label{77}
\end{equation}
where
\begin{eqnarray*}
  M_{1}^{G}&=&\int _{0}^{1} dxK_{G}^{G}[x^{\gamma }+(1-x)^{\gamma }-1] , \\
  M_{1}^{F}&=&\int _{0}^{1} dxK_{G}^{F}[x^{\gamma }+(1-x)^{\gamma }] ,  \\
  M_{0}^{F}&=&\int _{0}^{1} dxK_{G}^{F} = M_{1}^{F}(\gamma =0)/2 ,  \\
  L_{1}&=&\int _{0}^{1} dxK_{F}^{G}x^{\gamma } ,   \\
  L_{2}&=&\int _{0}^{1} dxK_{F}^{G}(1-x)^{\gamma } ,  \\
  L_{0}&=&\int _{0}^{1} dxK_{F}^{G} = L_{1}(\gamma =0) . \\
\end{eqnarray*}
The coefficients $M_{i}$ and $L_{i}$  can be expressed 
in terms of Euler beta functions and psi functions and 
depend only on~$\gamma$. 

From this, one can estimate the corrections induced 
by the preasymptotic limits of integration
(see e.g.~Section~\ref{sec-equations}).
For example, the leading term in $M_{1}^{G}$ gives rise 
to a correction of order ${\cal{O}}(-\exp (-\gamma y)/\gamma)$,
about 10$\%$ of the main term at the Z$^0$ energy.
Computer calculations support this estimate~\cite{lo1}. 
The corrections decrease with energy in a power-like 
manner similar to higher twist effects. 
It is possible that other effects of the same order,
e.g.~due to instantons,
could influence the multiplicity distributions. 
No definite answer to this question has yet been given.

It should be stressed that $\gamma$ does not equal
$\gamma_{0}$ even in gluodynamics ($n_{f}$=$0$) because 
$M_{1}^{G}$ differs from~$\gamma^{-1}$. 
An approximate equality between $\gamma$ and $\gamma_{0}$ 
is valid for $\gamma_{0}$$\,\ll\,$1
but the perturbative expansion for $\gamma$ 
in the fixed coupling approximation,
\begin{equation}
  \gamma \approx \gamma _{0} - h_{1}\gamma _{0}^{2} 
     + \frac {1}{2}(h_{1}^{2}+h_{2})
  \gamma _{0}^{3}+ {\cal{O}}(\gamma _{0}^{4}) ,        \label{78}
\end{equation}
differs from the corresponding formula~(\ref{67}) 
for the case of the running coupling constant,
from which one sees that the first correction 
is twice as large in the former case as in the latter.
This explains why the average multiplicity
at low energies increases more slowly for fixed coupling than
for running coupling (see Fig.~\ref{fig-seven}),
and why the increase in multiplicity for fixed coupling 
follows a power law at extremely high energies only.

The ratio $r$ appears linearly
in eqs.~(\ref{76}) and~(\ref{77}), 
allowing these two equations to be written as
\begin{equation}
  r(\gamma ) = b(\gamma ) \left [\frac {\gamma }{\gamma _{0}^{2}} - a(\gamma )
  \right ]^{-1} ,    \label{79}
\end{equation}
\begin{equation}
  r(\gamma ) = \left [\frac {\gamma }{\gamma _{0}^{2}} - d(\gamma )\right ]
  \frac {1}{c(\gamma )} ,  \label{80}
\end{equation}
where
\begin{eqnarray*}
  a(\gamma )&=&\psi (1)-\psi (\gamma +1)+B(\gamma ,1)
  -2B(\gamma +1,2) \nonumber \\
  &-&2B(\gamma +2,1)+B(\gamma +2,3)+B(\gamma +3,2)+\frac {11}{12}-
  \frac {n_{f}}{6N_{c}},\nonumber   \\
  b(\gamma )&=&\frac{n_{f}}{2N_{c}}[B(\gamma +3,1)+B(\gamma +1,3)] , 
     \nonumber \\
  c(\gamma )&=&\frac {C_{F}}{N_{c}}[B(\gamma ,1)-B(\gamma +1,1)+
  \frac {1}{2}B(\gamma +2,1)] ,\nonumber  \\
  d(\gamma )&=&\frac {C_{F}}{N_{c}}[\psi (1)-\psi (\gamma +1)-B(\gamma +1,1)+
  \frac{1}{2}B(\gamma +1,2)+\frac {3}{4}] .  \nonumber \\
\end{eqnarray*}
The beta functions are just inverse polynomials of $\gamma$ but the
above notation is less cumbersome. 

Thus the non-linear, integro-differential 
equations~(\ref{50}) and~(\ref{51}) can be reduced 
to the algebraic expressions~(\ref{79}) and~(\ref{80}),
which yield $\gamma$ and $r$ as a function of $\gamma_{0}$ and~$n_{f}$. 
The dependence of the solution on $n_{f}$ is very small.

The anomalous coupling
$\gamma$ can be related to $\gamma_{0}$ 
by the simple fitted formula
\begin{equation}
  \gamma  = 0.077 + 0.62\gamma _{0} ,     \label{81}
\end{equation}
or by a more theoretically motivated one
based on eq.~(\ref{78}),
\begin{equation}
  \gamma = 0.97\gamma _{0} - 0.48\gamma _{0}^{2} + 0.2\gamma _{0}^{3} ,
    \label{82}
\end{equation}
fitted in the range of $\gamma _{0}$ from 0.48 to~0.6. 
Thus $\gamma$ changes very slowly with~$\gamma_{0}$. 
The alternating signs of the correction terms correspond to the
tendencies observed for the running coupling constant,
see Table~\ref{tab-randa}. 
In itself, the value of $\gamma$ is not of much interest. 
Even though it is related to the energy dependence 
of the average multiplicity, 
it is known that the power law increase of mean multiplicity 
for fixed coupling is replaced asymptotically by a slower
dependence $\sim\exp (\ln ^{1/2}s)$ for running coupling
(see Fig.~\ref{fig-seven}). 
The more realistic behavior provided by the 
running coupling constant was discussed in the previous sections.

\begin{figure}[tp]
\begin{center}
  \epsfxsize=17cm
  \epsffile[90 80 590 680]{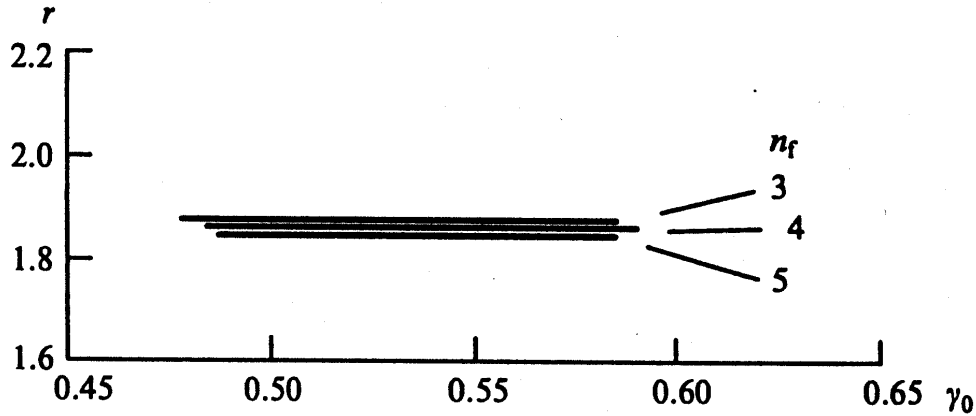}
\end{center}
\vspace*{-15.5cm}
\caption{$r$ versus $\gamma _{0}$ in fixed coupling
QCD for $n_f$= 3, 4 and 5~\cite{42}.
} 
\label{fig-ten}
\end{figure}

The ratio $r$ is of more interest because the energy dependence 
of the average multiplicities in gluon and quark jets cancels. 
Thus the fixed coupling prediction for $r$ can be expected to
be rather general.
A computer solution of eqs. (\ref{79}) and~(\ref{80})
yields the results shown in~Fig.~\ref{fig-ten}.
The dependence on $n_{f}$ is seen to be very mild. 
More important, the dependence of $r$ on $\gamma_{0}$ 
is even weaker than that of $\gamma$, 
and its average effective value is
\begin{equation}
  r = 1.84 \pm 0.02 .    \label{83}
\end{equation}
This value of $r$ is much lower than the prediction from the 
double-logarithmic approximation and corresponds well with the 
estimate of $r$ from the higher order approximate solution of 
QCD equations with a running coupling constant,
discussed in Section~\ref{sec-psolutions}.
This suggests that yet higher order corrections to eq.~(\ref{Y})
will be small.
Introduction of the exact (i.e.~preasymptotic)
limits of integration would reduce the value of $r$ 
in eq.~(\ref{83}) yet further,
as mentioned in Section~\ref{sec-computer}.

In a realistic process, the virtuality in a jet degrades as the 
partons evolve toward hadrons, 
presumably with an associated change in the value of~$n_f$.
In the framework of the calculations with a fixed coupling strength, 
the dependence on $n_f$ can be addressed with 
the aid of formulas (\ref{52}) and (\ref{al}).
The dependence on $n_f$ is found to be very small.
Thus, as a jet of partons evolves toward lower $Q$, 
we need not be concerned with a change in the number 
of active flavors.
The parton multiplicity will depend on the evolving 
virtuality through a mild variation of~$\gamma$, 
but not enough to invalidate the fixed coupling approximation. 
Certainly, in the ratio of the multiplicities, 
such dependences cancel, yielding a stable result for $r$. 
This conclusion is supported by the perturbative
solutions of the equations with the running coupling~\cite{39,cdnt8},
as has already been discussed (see Table~\ref{tab-randa}).

\subsection{Widths and higher rank moments of 
multiplicity distributions in gluon and quark jets}
\label{sec-widthshigher}

Relations~(\ref{74}) and~(\ref{75}) allow the solution 
for higher moments of multiplicity 
to be obtained in the fixed coupling approximation
for any rank~$q$.
A system of coupled recurrent equations~\cite{21} 
is obtained by substituting eq.~(\ref{7})
into eqs.~(\ref{50}) and (\ref{51}) 
and by comparing the coefficients of $z^q$ on both sides. 
These equations can be solved by iteration. 
The results will not be given here (see~\cite{21}). 
Only the final analytic expressions for the moments 
of rank $q$ as they relate to the lower rank moments will be presented. 
For this purpose, let us introduce
\begin{equation}
  f_{q} =\frac {F_{q}}{q!} ,  \;\;\;\;\;  
  \hat{\phi }_{q} = \frac {\Phi _{q}}{r^{q}q!} . \label{84a}
\end{equation}
The solution of the equations is~\cite{21}
\begin{equation}
  f_{q} = [a_{q}S_{q}(f,\hat{\phi }) + b_{q}T_{q}(f,\hat{\phi })]
  \Delta _{q}^{-1} , \label{85}
\end{equation}
\begin{equation}
  \hat{\phi }_{q} = [c_{q}S_{q}(f,\hat{\phi }) + d_{q}T_{q}(f,\hat{\phi })]
  \Delta _{q}^{-1} , \label{86}
\end{equation}
where
\begin{equation}
  S_{q} = \sum _{l=1}^{q-1}[N_{q,l}^{G}f_{l}f_{q-l} + n_{f}N_{q,l}^{F}
  \hat{\phi }_{l}\hat{\phi }_{q-l}] ,  \label{87}
\end{equation}
\begin{equation}
  T_{q} = \sum _{l=1}^{q-1}L_{q,l}\hat{\phi }_{l}f_{q-l} ,   \label{88}
\end{equation}
\begin{equation}
  a_{q} = \frac {q\gamma }{\gamma _{0}^{2}} + L_{0,0} - L_{q,q} ,  \label{89}
\end{equation}
\begin{equation}
  b_{q} = n_{f}M_{q}^{F} ,   \label{90}
\end{equation}
\begin{equation}
  c_{q} = L_{q,0} ,  \label{91}
\end{equation}
\begin{equation}
  d_{q} = \frac {q\gamma }{\gamma _{0}^{2}} - M_{q}^{G} + n_{f}N_{0,0}^{F} ,
\label{92}
\end{equation}
\begin{equation}
  \Delta _{q} = a_{q}d_{q} - b_{q}c_{q}  ,  \label{93}
\end{equation}
\begin{eqnarray*}
  M_{q}^{G}&=&\psi (1) - \psi (q\gamma +1) 
  + B(q\gamma ,1) - 2B(q\gamma +1,2) \nonumber \\
  &-&2B(q\gamma +2,1) + B(q\gamma +2,3) + B(q\gamma +3,2) 
  +\frac {11}{12} , \nonumber \\
  M_{q}^{F}&=&\frac {1}{2N_{c}}[B(q\gamma +3,1) + B(q\gamma +1,3)] , \\
  N_{q,l}^{G}&=&B[l\gamma ,(q-l)\gamma +1] 
   - 2B[l\gamma +1,(q-l)\gamma +2]\nonumber \\
  &+&B[l\gamma +2,(q-l)\gamma +3] , \nonumber   \\
  N_{q,l}^{F}&=&\frac {1}{4N_{c}}(B[l\gamma +3,(q-l)\gamma +1] + B[l\gamma +1,
  (q-l)\gamma +3]) , \\
  L_{q,l}&=&\frac {C_{F}}{N_{c}}[B[l\gamma +1,(q-l)\gamma ] - B[l\gamma +1,
  (q-l)\gamma +1]\nonumber \\&+&\frac {1}{2} B[l\gamma +1,(q-l)\gamma +2)] .\\
\end{eqnarray*}
The above expressions may seem cumbersome 
but their structure is simple.
They generalize the formulas of the preceding section to any~$q$. 
The formulas of gluodynamics are obtained by setting 
$n_{f}$=$C_{F}$=0 if one retains, 
in $M_{q}^{G}$ and $N_{q,l}^{G}$, 
the leading terms $B(q\gamma ,1)\equiv 1/q\gamma$
and $B[l\gamma ,(q-l)\gamma +1]$, respectively. 
Using the results for $\gamma$ and $r$ from the
preceding section at a given value of $\gamma _{0}$ and~$n_{f}$, 
one can determine $F_2$ and $\Phi_2$, 
then increase $q$ by 1 to determine $F_3$ and $\Phi_3$,~etc.

The evolution parameter $y$ disappears from the formulas. 
A posteriori, this means that $F$ scaling,
with all dependence on $y$ hidden in the average 
multiplicities $\langle n_{G,F}(y)\rangle$,
is valid for fixed coupling.
This leads to a self consistent system 
of algebraic equations where all quantities, 
including $F_{q}$ and $\Phi _q$, are independent of energy.
It should be stressed that $F$ scaling is only as accurate
as the main equations of the fixed coupling approximation. 
In fact, one should speak about asymptotic $F$ scaling since
the limits of \mbox{$x$-integration}
in eqs.~(\ref{50}) and (\ref{51}) are asymptotic. 

\begin{figure}[tphb]
\begin{center}
  \epsfxsize=12cm
  \hspace*{.1cm}\epsffile{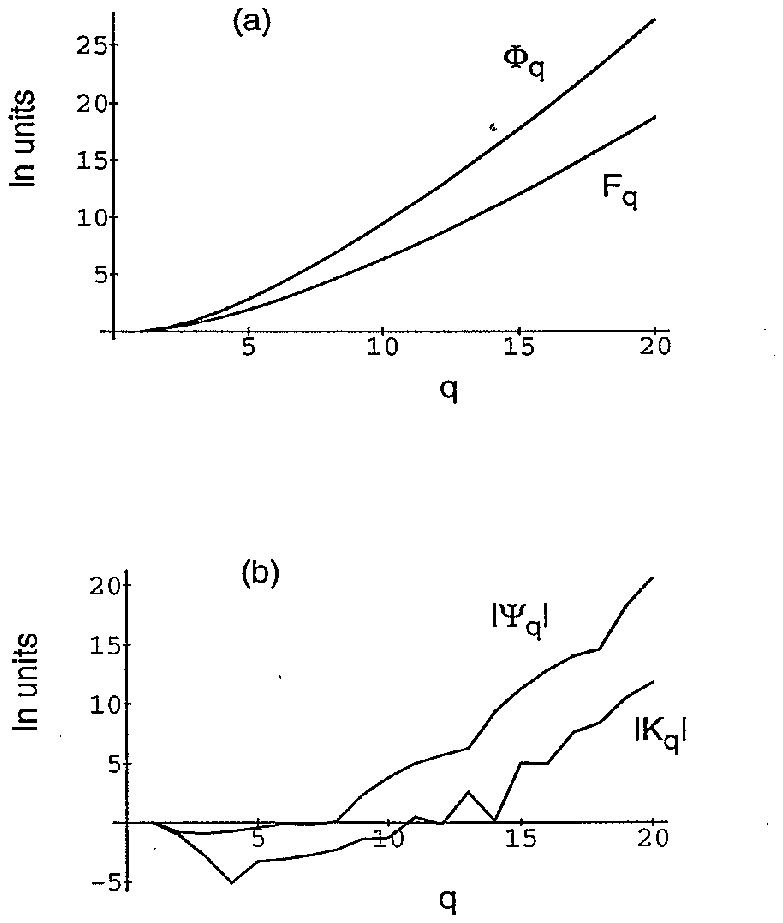}
\end{center}
\caption{Moments of the multiplicity distribution 
in fixed coupling QCD for
$\gamma_{0}$=0.48 and $n_f$=5~\cite{21}; 
(a)~$\ln F_{q}$ and $\ln \Phi _{q}$,
(b)~$\ln \vert K_{q}\vert$ and $\ln \vert \Psi _{q}\vert $.
} 
\label{fig-eleven}
\end{figure}

The results of the calculation, 
when expressed in terms of $F_q$ and~$\Phi _q$,
are shown in Fig.~\ref{fig-eleven}(a)
for $\gamma_{0}$=0.48 and $n_{f}$=5. 
The solutions are seen to increase rapidly with $q$, 
more so for $\Phi_q$ than for~$F_q$. 
Since these are normalized factorial moments, 
this once again demonstrates that the multiplicity distribution 
of quark jets is wider than that of gluon jets. 
The results are insensitive to the value of~$n_{f}$. 
The dependence on the coupling strength is also very mild.
Indeed, the results are insensitive to whether
the coupling strength is fixed or running.
The corresponding results for cumulants are shown
in Fig.~\ref{fig-eleven}(b).

\begin{figure}[tphb]
\begin{center}
  \epsfxsize=13cm
  \epsffile{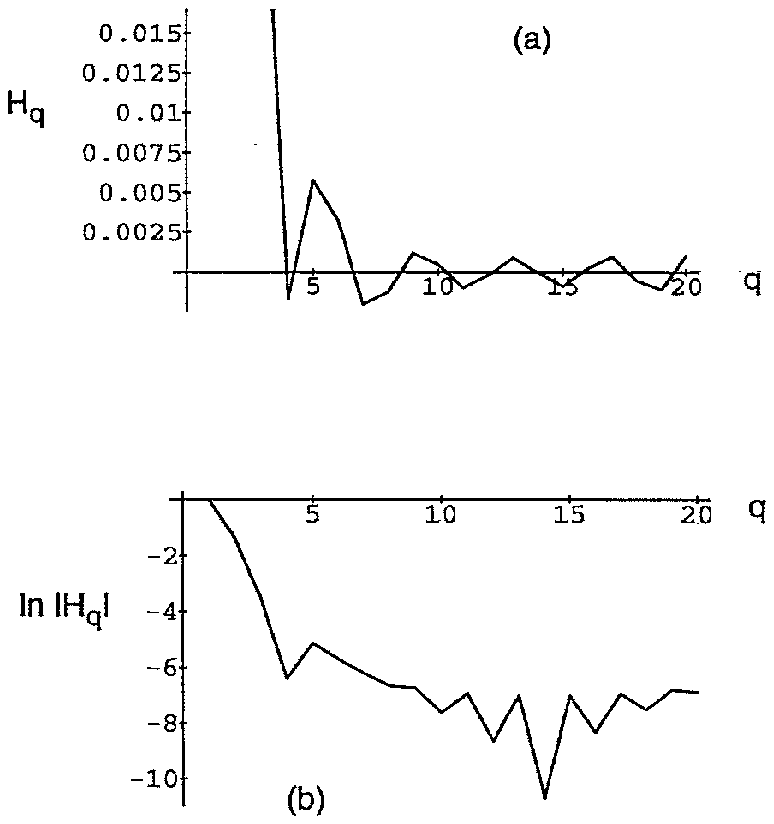}
\end{center}
\caption{Ratio $H_q$ of the
gluon jet distribution in fixed coupling QCD for
$\gamma _{0}$=0.48 and $n_f$=5~\cite{21};
(a)~$H_q$, (b)~$\ln \vert H_{q}\vert$.
} 
\label{fig-twelve}
\end{figure}

\begin{figure}[tphb]
\begin{center}
  \epsfxsize=13cm
  \hspace*{.3cm}\epsffile{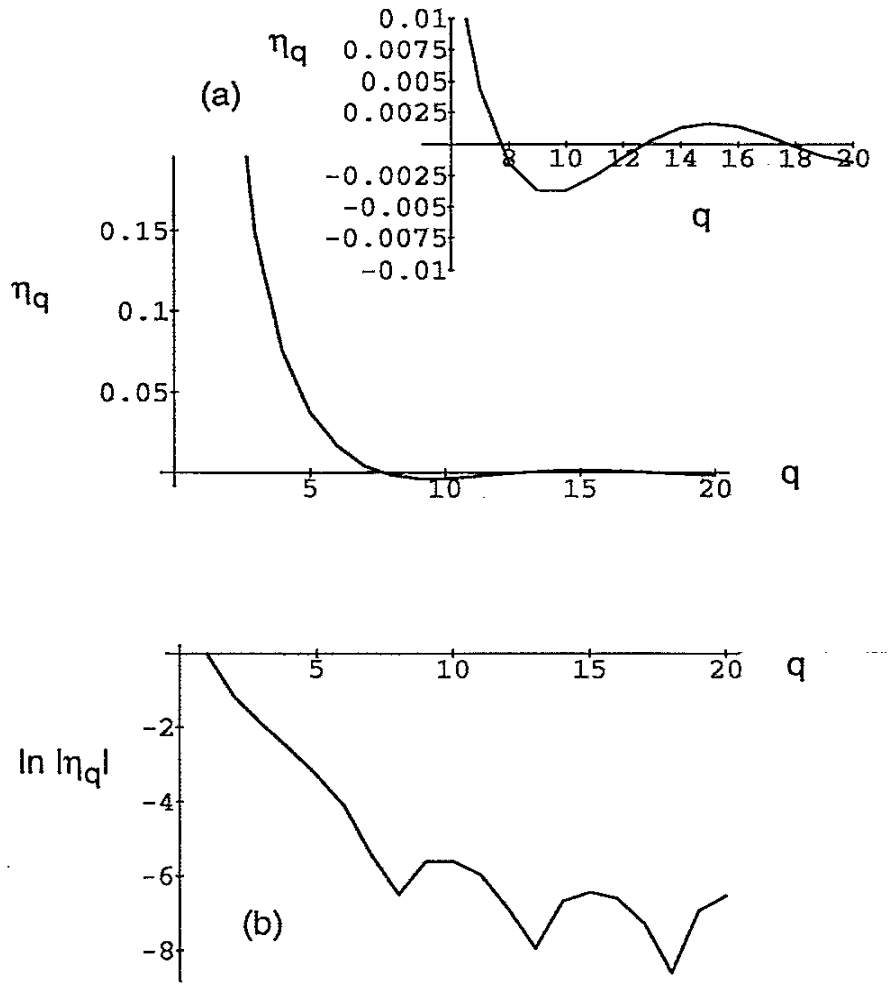}
\end{center}
\caption{Ratio $\eta_{q}$ of the
quark jet distribution in fixed coupling QCD for
$\gamma _{0}$=0.48 and $n_f$=5~\cite{21};
(a) $\eta _{q}$, (b) $\ln \vert \eta _{q} \vert$.
} 
\label{fig-thirteen}
\end{figure}

The ratios of factorial moments to cumulants
for gluon and quark jets are defined here as:
\begin{equation}
  H_{q} = \frac {K_{q}}{F_{q}} ,  \label{94}
\end{equation}
\begin{equation}
  \eta _{q} = \frac {\Psi _{q}}{\Phi _{q}} . \label{95}
\end{equation}
The results of the calculations for $H_{q}$ and $\eta _{q}$
in fixed coupling QCD are presented in
Figs.~\ref{fig-twelve} and~\ref{fig-thirteen},
respectively.

It is of interest to compare the QCD results 
in Figs.~\ref{fig-eleven}-\ref{fig-thirteen}
to those of the phenomenological models discussed in 
Section~\ref{sec-phenomenology}.
By comparison of Fig.~\ref{fig-eleven} with 
Figs.~\ref{fig-one} and~\ref{fig-two}, 
the QCD results are seen to be more similar
to the NBD than to fixed multiplicity. 
In fact, $F_q$ in Fig.~\ref{fig-eleven}(a) can be approximated 
by the corresponding NBD result with~$k$=5. 
While this characterization is convenient, 
fits of higher rank moments by the NBD are not appropriate. 
For example,
the ratio $H_q$ of the NBD decreases monotonically with $q$,
as shown in Fig.~\ref{fig-one}(c).
In QCD at present energies,
the $H_q$ moments become negative and then oscillate. 
These oscillations become even more pronounced if
the preasymptotic limits of integration are used.
Thus the QCD and NBD predictions 
for higher moments differ drastically
from each other.

The negative binomial distribution fits
the second and third moments well
and provides a good overall description of experiment.
It does not describe features of the data such as the shoulders 
of the distributions or the oscillations of the $H_q$ ratios, however.
The fixed multiplicity distribution,
in contrast to the NBD,
{\it does} yield oscillations.
However the $H_q$ moments change sign at each
subsequent integer value of $q$ in this case, 
yielding results for the moments and for the period of oscillation
in disagreement with experiment
(see Section~\ref{sec-oscillating}).

The sensitivity of $H_q$ to the detailed shape of the 
multiplicity distribution is clearly demonstrated by its various 
qualitative forms in the DLA, MLLA, NNLO,
and higher order solutions of 
eqs.~(\ref{50}) and~(\ref{51}).
At large ranks the moments behave in radically different manners
in these different approximations,
as discussed in Section~\ref{sec-hoapprox}.
Moreover, the behavior of $H_q$ depends strongly on slight variations 
of the particular shape of the factorial moments at low values of $q$. 
One can demonstrate how easy it is to obtain oscillations 
of the fixed multiplicity type imposed on the double-logarithmic 
behavior from the following exercise. 
It is known~\cite{19, 46} that the large $q$ behavior of the factorial 
moments $F_q$ in the double-logarithmic approximation is
(see eq.~(\ref{23}))
\begin{equation}
  F_{q} = \frac {2q\Gamma (q+1)}{C^{q}} .    \label{96}
\end{equation}
If one adds a preasymptotic term by replacing the factor 
$2q$ by $2q+1$ in the numerator 
(which restores the condition $F_{0}$=1 but not $F_{1}$=1), 
one obtains an additional term in $H_{q}$ which imposes 
oscillations of the fixed multiplicity type on the monotonous 
decrease of the form $q^{-2}$, 
and the ratio $H_q$ becomes
\begin{equation}
  H_{q} = \frac {2+(-1)^{q-1}}{q(2q+1)} ,  \label{97}
\end{equation}
where the second term in the numerator appears because 
of the newly added preasymptotic term.

The above examples illustrate the sensitivity of $H_q$ to the
approximations made in solving the set of equations 
for the generating functions. 
The different qualitative behavior obtained for different
approximations in QCD is illustrated in 
Figs.~\ref{fig-four}, \ref{fig-hqanalytic}, 
\ref{fig-twelve}, and~\ref{fig-thirteen},
while the difference between QCD and the
phenomenological distributions is illustrated by comparing
these figures with Figs.~\ref{fig-one} and~\ref{fig-two}. 
The experimental cutoff of the tail of multiplicity 
at finite energies induces additional oscillations which 
have been shown to be rather small~\cite{ugl},
as already mentioned in Section~\ref{sec-hoapprox}.
The numerical solution of equations which incorporate
energy conservation both in the generating functions 
and in the limits of integration yields an extremely good fit 
of experimental data even for high moments~\cite{lo2}
(Section~\ref{sec-oscillating}
and Fig.~\ref{fig-nine}),
but also poses some questions about the potential importance
of neglected power corrections,
as mentioned above.

Unfortunately, there is still no clear understanding 
of the physical origin of the oscillations, 
i.e.~of their periodicity, amplitude, 
and dependence on rank $q$ 
(it seems that the amplitude increases and the 
period decreases with increasing~$q$). 
It has been suggested that the oscillations are related to the 
experimental fact of the shoulder structure 
of the multiplicity distribution,
interpreted as originating from the superposition of jets with
different flavors and topologies~\cite{30a, giug}.
However, the insight provided by the solution 
of fixed coupling QCD suggests otherwise. 
In QCD, the jet initiated by a 
single light flavored parton 
exhibits $H_q$ oscillations,
as long as account is taken of conservation laws.
In the NBD framework
one needs a superposition of at least two NBDs,
interpreted as the contributions of both
light- and $b$-quark jets,
to describe the oscillations~\cite{giolu}.
An analysis of the light quark jet data in Ref.~\cite{bib-opal98}
demonstrates that the $H_q$ moments of
single light quark jets do oscillate, however,
with no admixture of $b$-quarks~\cite{bib-jwg2}.
This supports the QCD prediction but argues against
the NBD conjecture.

The fixed coupling solution demonstrates that the running property 
of the coupling constant is not important for the oscillations.
More significant is the vector nature of massless gluons 
which controls the singularities of the kernels in the QCD equations. 
This has been demonstrated~\cite{dnb} by comparing the QCD results 
with those of the $\lambda \phi ^{3}_{6}$ theory 
(i.e.~for scalar mesons in six dimensions). 
This model exhibits asymptotic freedom
and contains a non-singular kernel in the evolution equation. 
There is no minimum at $q$=5 in this theory.
Oscillations appear only at much larger ranks than in QCD. 
Perhaps the behavior of $H_q$ will provide some insight into
how the equations for the generating functions can be generalized, 
including the fine effects of the interaction of 
``color monsters''~\cite{5, 19, 46}. 
However, this is a difficult problem to solve.

Note that the above oscillations occur for integer values of $q$
and are not related to oscillations of fractional moments. 
The latter would impose lower period harmonics 
on the oscillations from integer ranks.

\section{Non-perturbative modifications of QCD equations}
\label{sec-nonperturbative}

The non-perturbative aspects of jet evolution 
are less clear at the moment.
They are usually hidden within the hypothesis of local 
parton-hadron duality or in Monte Carlo hadronization models. 
The influence of the QCD vacuum condensate on jets cannot be 
estimated within the perturbative approach. 
The first proposal to phenomenologically account for its effects
within the equations describing QCD jet evolution 
is given in Refs.~\cite{dnon, dlno}. 

The action of the vacuum condensate on partons in jets 
was assumed to be similar to the influence of the medium
on electron-positron pairs in electromagnetic cascades. 
As is well known, 
electromagnetic cascades evolve to a large number of particles,
of the order of the ratio of the initial energy to the electron mass,
if the ionization processes in the medium are neglected. 
Because of ionization,
low energy electrons stop and exit the cascade. 
The ionization process is described by a non-scaling 
term in the evolution equations. 

A similar term was added to the QCD equations~\cite{dnon, dlno}. 
Later on, when angular ordering was properly incorporated, 
it was shown that this non-scaling term could also be included
in the equations for the generating functionals~\cite{leos}. 
This implies that some energy is spent on the 
formation of non-perturbative strings,
diminishing the number of partons in the cascade. 
The ratio of partons with and without account of 
the non-scaling (non-perturbative) term is,
in the lowest order approximation~\cite{leos},
\begin{equation}
  \frac {n_{np}}{n_{p}}=1-\gamma _{0}\frac {\kappa }{p}
  -2\gamma _{0}^{3}\frac
  {\kappa \Theta}{Q_{0}},     \label{nonp}
\end{equation}
where $\kappa$ denotes the typical energy spent for soft 
non-perturbative interactions,
so that $\kappa/p$ is a small parameter. 
It is also possible to estimate the energy distribution of 
partons in the jet~\cite{leos} and the behavior of moments of the 
multiplicity distribution~\cite{leos1}. 
The most interesting property of eq.~(\ref{nonp})
is its non-scaling behavior,
namely it does not depend on the product $p\Theta$ (i.e. on~$y$),
but on $p$ and $\Theta$ separately. 
Also, it depends on the product of the perturbative 
and non-perturbative scales.

A similar approach is advocated in Refs.~\cite{plos,bopl}. 
A non-perturbative term is added to the QCD equation for the 
generating function,
and is interpreted as an analogue of 
an ``inactivation process'' implying hadronization. 
This leads to modifications of KNO scaling 
(see eq.~(\ref{plos})).

A somewhat different approach to the same problem 
is formulated in Refs.~\cite{elge, elge1} where 
non-perturbative effects are included in the effective Lagrangian, 
and a Monte Carlo scheme is developed.
The general idea is to construct an effective field theory that 
embodies both partonic and hadronic degrees of freedom. 
The effective action is composed of three parts,
with quark and gluon fields at small distance scales,
an effective low energy Lagrangian with vacuum condensate fields
representing hadronic states at large scales, 
and a term which couples these different degrees of freedom 
at intermediate scales. 
The final goal is to describe the transition from the quark-gluon interaction 
stage to hadrons within a unified scheme of strong interactions. 

All such proposals aim at the description of the 
hadronization stage and, in particular, 
of the relation between the numbers of partons 
and final-state hadrons.
Unfortunately, it is still necessary to introduce ad hoc 
phenomenological assumptions to achieve this goal.

\section{Singularities of the generating functions}
\label{sec-singularities}

To this point,
we have considered $z$ to be a subsidiary variable in the 
generating function $G(y,z)$,
which is set to a constant value 
after calculation of the moments. 
It is of interest to study the behavior
of $G(y,z)$ in the complex $z$ plane. 
This interest stems from the fact that the singularities
of the generating function are located close to the
point $z$=0 where the moments are calculated. 
For example, the singularity of the generating function of the 
negative binomial distribution is located at $z$=$k/\langle n\rangle$,
as mentioned in Section~\ref{sec-nbd}.
For $k$=2 this is just a pole of second order at 
$z$=$2/\langle n\rangle$,
which tends to $z$=0 at high energies. 

In QCD the singularities of the generating function are known
in DLA only (see~\cite{5}, p.~137). 
In this approximation,
eq.~(\ref{63}) reduces to eq.~(\ref{edla}).
The leading singularity governing
eq.~(\ref{edla}) can be written as
\begin{equation}
  G(z,y)\propto (z-z_{0}(y))^{-2}   \label{zz0}
\end{equation}
since differentiating $\ln G$ twice produces 
a second order pole on the left hand side which must be 
matched by a singularity of the same order on the right hand side. 
Since KNO scaling is valid in this approximation, 
the factorial moments do not depend on $y$ asymptotically, 
and the generating function depends on the 
product of $z$ and $\langle n\rangle $. 
The singularity is located at
\begin{equation}
  z_{0}^{DLA}\approx \frac {C}{\langle n\rangle }, \;\;\;\;\;\;  
  C\approx 2.552 , \label{zsin}
\end{equation}
i.e.~close to the singularity of the negative binomial 
distribution with $k$=2,
as it should be.
This singularity tends to $z$=0 at high multiplicities
as for the~NBD.

\begin{figure}[thb]
\vspace*{-2cm}
\begin{center}
  \begin{turn}{1}
  \epsfxsize=15cm
  \hspace*{-1cm}\epsffile{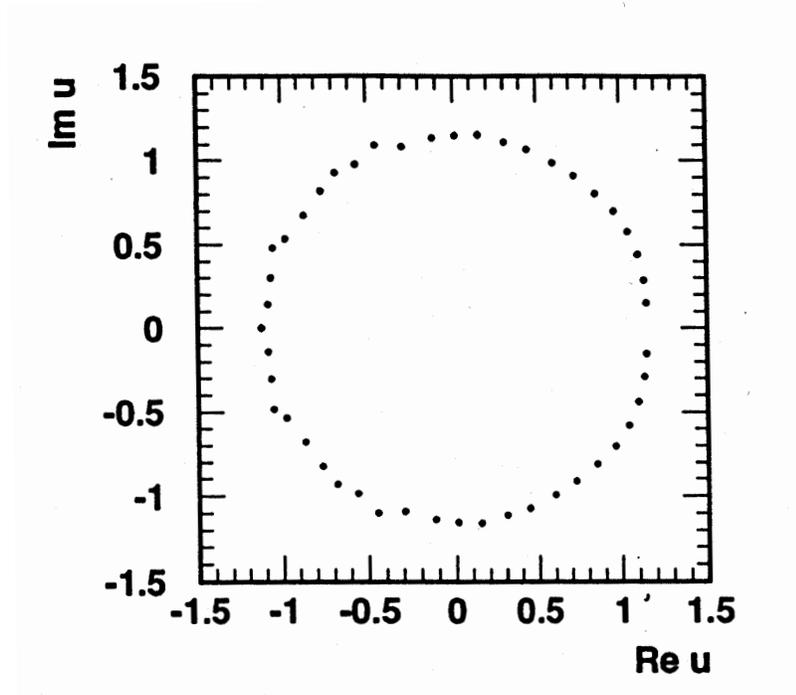}
  \end{turn}
\end{center}
\vspace*{-9cm}
\caption{Complex zeros of the charged particle multiplicity
distribution,
obtained~\cite{bib-eddi} from {\epem} Monte Carlo events
at {\ecm}=1~TeV,
where u$\,\equiv\,$$z$+1.
} 
\label{fig-dewolf}
\end{figure}

The overall structure of the singularities is more complicated
in DLA QCD than for the NBD, however,
because there are poles of second and first orders, 
logarithmic branch points,~etc. 
A more complete DLA result for 
$G(z,y)$ near the singularity is
\begin{equation}
  G(z,y)=\frac {2z_{0}^{2}}{(z-z_{0})^2}+\frac {2z_{0}}{z-z_{0}}-\frac {2}{3}
  \ln \frac {z_{0}-z}{z_{0}} +{\cal{O}}(1).   \label{gsin}
\end{equation}
In a certain sense this result is not very helpful since the DLA 
is a rather crude approximation to reality at present energies.
This result shows, however, 
that $z$=0 is still quite far from the singularity 
since the pole contributions cancel at this point. 
There is no cancelation in the derivatives of $G$, however,
which increase like factorials because of the leading singularity.
It is not yet known how this structure is modified
by higher orders.
Attempts to go beyond DLA are connected with the 
Taylor series expansion of $G$ discussed above. 
Then eq.~(\ref{edla}) acquires additional terms on the right hand 
side proportional to derivatives of $G(z,y)$. 
Each derivative exhibits a stronger singularity than $G$ itself.
Therefore such an approach cannot be applied near the singular points. 
This also demonstrates why the DLA has difficulties in practice. 
However, attempts to employ eq.~(\ref{63}) 
rather than eq.~(\ref{edla}) have not yet been done explicitly. 

In experiment, only the so-called truncated generating function
has been determined
\begin{equation}
  G^{tr}(y,z)=\sum _{n=0}^{n_{max}}P_{n}(1+z)^{n},   \label{trun}
\end{equation}
where the sum extends up to the maximum multiplicity 
$n_{max}$ available at the energy scale~$y$. 
Therefore the truncated generating function is a polynomial 
in $z$ and has no singularity at finite $\vert z\vert$. 
It possesses $n_{max}$ complex conjugated zeros. 
When $n_{max}$ tends to infinity these zeros should converge to
the singularity location of $G(y,z)$. 
The properties of zeros (the rate of convergence, etc.) 
can be used~\cite{dzer} to obtain some knowledge
about the singularities of the generating function.

The problem becomes even more intriguing if an analogy 
with statistical mechanics is invoked 
(see, e.g.~\cite{14, blan}). 
In this case $z$ is interpreted as fugacity, 
and $G(y,z)$ and $G^{tr}(y,z)$ play the roles of the canonical 
and grand canonical partition functions, respectively. 
Therefore their properties in the complex $z$-plane can be 
considered to reflect the statistical properties of multiparticle production. 
In statistical mechanics the motion of zeros to the real $z$ axis, 
``pinching it'' asymptotically,
implies a phase transition~\cite{leya, yale}. 
Experiment indicates that such a pinch occurs as the energy increases,
for various reactions initiated by particles and nuclei. 
The zeros tend to lie close to a circle of unit radius 
in the complex $z$ plane,
as is illustrated in Fig.~\ref{fig-dewolf}
using {\epem} annihilation Monte Carlo events 
generated at a cms energy of 1~TeV~\cite{bib-eddi}.
Similar results have been obtained using simulations 
of nucleus-nucleus collisions generated according 
to the dual parton model~\cite{cdnt}.
Moreover, the quantity analogous to the free energy exhibits 
its maximum in QCD~\cite{38}.
This fact is closely related to the 
minimum of the $H_q$ ratios
at $q$$\,\approx\,$5 discussed above. 
The interpretation in terms of a phase transition is the same here.

Nonetheless, we cannot rely too much on the analogy 
with statistical mechanics to obtain physics conclusions.
The properties of the zeros discussed above 
have been shown~\cite{bkt} in purely mathematical terms 
to be a consequence of the falloff of the multiplicity
distribution at large $n$ 
and its flattening with increasing energy.
This behavior forces all zeros to the unit circle in the complex $z$-plane, 
centered at $z$=$-1$, 
with the point $z$=0 removed. 
Experiment favors these tendencies. 
The physics implications of this behavior are 
therefore not entirely clear.

\section{Measurements of mean multiplicity, slopes,
and higher moments}
\label{sec-experiment}

We now explicitly turn our attention to experimental results.
In this, we concentrate on data which test
the QCD predictions discussed in
Sections~\ref{sec-gluodynamics}-\ref{sec-exactsolutions}.
We emphasize data from {\epem} annihilations.
The relative simplicity of hadronic events
from {\epem} collisions allows
a level of precision and conclusiveness difficult
to achieve in other types of reactions.
For example,
ep and pp collisions are characterized by initial-state
strong interactions and 
final-state remnants from the colliding hadrons.
These features 
--~absent in {\epem} events and the calculations~--
make a comparison to theory uncertain.
The comparison between experiment and theory is yet
more uncertain for nuclear collisions.
The available measurements of average
multiplicities in ep, pp and {\pp} collisions
are compiled in Ref.~\cite{pdg}.

To test the predictions in a meaningful manner,
the experimental definition of jets
should match the theoretical one.
The theoretical definition of jets is based on the production 
of a virtual quark-antiquark {\qq} pair (for quark jets) or 
gluon-gluon {\gluglu} pair
(for gluon jets) from a color singlet point source.
The inclusive multiplicity of these
{\qq} or {\gluglu} events defines the 
multiplicity of ``two-jet events.''
A single quark or gluon jet corresponds to a hemisphere
in these events.
Thus there is no selection of a specific event topology,
i.e.~jets are not selected using a mathematical algorithm 
such as the {\durham}~\cite{dur,cdotw} or
JADE~\cite{jade} jet finders,
in contrast to common experimental practice.
Jets defined according to the inclusive theoretical
prescription are called ``unbiased.''

It has proven difficult to measure gluon jet multiplicity 
in an unbiased manner
because {\gluglu} production from a color singlet point source
is a process which is practically unobserved in nature.
This difficulty impeded experimental progress in
the field for many years.
One channel where the experimental selection criteria
match the theoretical definition is the decays
of {\bbbar} bound states to $\gamma${\gluglu}:
events like this have been studied
through the selection of
$\Upsilon\rightarrow\gamma${\gluglu}$\rightarrow\,\gamma$+$hadrons$
events.
Another possibility is rare {\epem} hadronic annihilation events
in which the quark jets q and {\qbar} from the electroweak 
decay of the intermediate Z$^0$ or virtual $\gamma$
are approximately collinear:
the gluon jet hemisphere against which the 
q and {\qbar} recoil in these events
corresponds to a nearly unbiased gluon jet~\cite{dtro,bib-gary94}.
The two photon process 
{\epem}$\,\rightarrow\,${\epem}{\gluglu}$\rightarrow\,${\epem}+$hadrons$
has also been suggested as a source for high energy 
unbiased gluon jets~\cite{bib-valery93}:
this possibility has yet to be explored experimentally,
however.

In contrast to gluon jets,
it is easy to measure quark jet multiplicity in an unbiased manner,
using the charged particle multiplicity in hemispheres of
{\epem}$\rightarrow\,$$hadrons$ events.
{\epem} hadronic annihilations therefore provide a natural 
source for quark jets,
studied by many experiments.

For the results discussed here,
the data have been corrected for detector response
and initial-state photon radiation,
treating all charged and neutral particles with lifetimes greater
than \mbox{$3\times 10^{-10}$~s} as stable.
Hence charged particles from the decays of K$_{\mathrm S}^0$
and weakly decaying hyperons are included in the
definition of multiplicity.

Before proceeding to the main results,
presented in Sections~\ref{sec-meanmult}-\ref{sec-soft},
it is of interest to recount the original experimental efforts
to measure gluon jet multiplicity,
and in particular the multiplicity ratio
$r$=$\langle n_G\rangle/\langle n_F\rangle$
(see also Ref.~\cite{bib-gary94}).
The JADE Collaboration selected three-jet {\qq}g 
events in {\epem} annihilations~\cite{bib-jadegluon}.
The gluon jet was identified by assuming it had the
lowest energy of the three jets in an event.
Quark jets were the higher energy jets in the same events.
The UA2 Collaboration studied gluon jets produced in
two-jet gg final states from {\pp} collisions~\cite{bib-ua2gluon}.
These gluon jets were compared to lower energy quark jet
data from an {\epem} experiment~\cite{bib-tassoquark}.
For both studies,
qualitative indications were reported that the
gluon jet multiplicity was slightly larger.
The interpretation of these results was unclear, however,
since the gluon and quark jets were not identified using
the same criteria in either study.

The HRS Collaboration chose a different strategy,
selecting three-fold symmetric 
{\epem}$\,\rightarrow\,${\qq}g events
in which the quark and gluon jets were produced
with about the same energies and inter-jet 
angles~\cite{bib-hrsgluon}.
Thus in this case the gluon and quark jet identification
criteria were equivalent.
The probability that the gluon jet had a larger
multiplicity was tested by assuming a Poissonian
multiplicity distribution and independent production
of each of the three jets.
A value $r$=$1.29^{+0.29}_{-0.46}$ was derived,
the first quoted result for~$r$.
Although the largeness of the uncertainty precluded
any definite interpretation,
this result was considered at the time to indicate
that $r$ was much smaller than
the na\"{\i}ve expectation $r$$\,\approx\,$2
(see Section~\ref{sec-soft}).
The DELPHI Collaboration later applied a similar
analysis technique to a much larger data sample 
and obtained $r$$\,\approx\,$1~\cite{bib-delphi}, however,
thus demonstrating that the method was not sensitive
to gluon and quark jet differences.

The OPAL Collaboration selected quark and gluon jets
in one-fold symmetric {\qq}g {\epem} events in which 
the angle between the highest energy jet and each of 
the two lower energy jets was the same,
namely~150$^\circ$~\cite{bib-opalyevents}.
These events later became known as ``Y events''
since they are shaped like the letter~Y.
This analysis utilized b-quark tagging to identify
a high purity sample of gluon jets which were then
compared to a sample of about 50\% quark and
50\% gluon jets selected using the same criteria.
The result $r$=$1.27\pm 0.07$ was derived from these data,
the first positive observation of a larger
multiplicity in gluon jets compared to quark jets.
This result was later confirmed by other 
studies~\cite{del,bib-alephgluon}.
Although the results of this analysis represented 
a considerable success,
establishing a value of $r$ significantly larger
than unity for the first time,
the results were found to depend strongly 
on the jet finding algorithm used to select the events, 
see Refs.~\cite{del,bib-opalgluoncone}.
Since the jets were defined using a jet finder rather than
employing a hemisphere definition,
they could not be used for a 
quantitative test of the analytic results presented in
Sections~\ref{sec-gluodynamics}-\ref{sec-exactsolutions},
unlike the unbiased gluon jet measurements discussed below.

\subsection{Mean multiplicity}
\label{sec-meanmult}

\begin{figure}[tphb]
\vspace*{-.7cm}
\begin{center}
  \epsfxsize=11cm
  \hspace*{.2cm}\epsffile{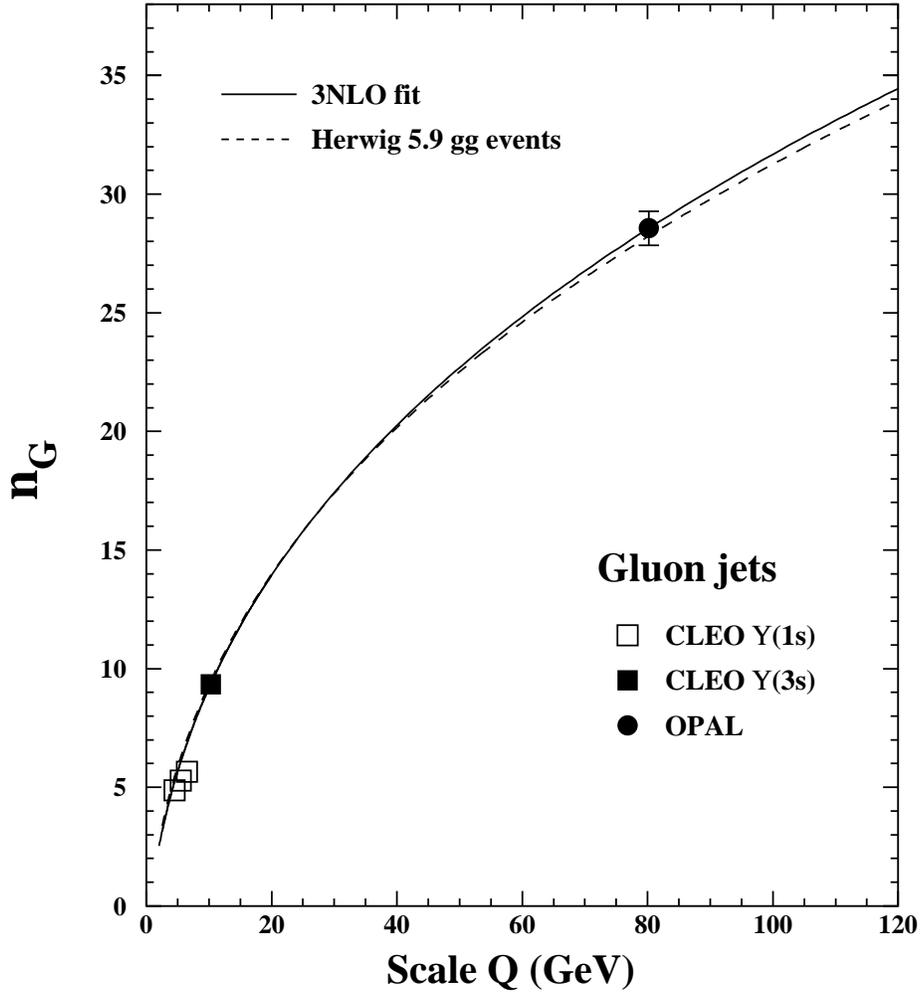}
\end{center}
\vspace*{-.6cm}
\caption{The mean charged particle multiplicity of 
{\gluglu} events from a color singlet point source
versus energy scale~Q.
The solid curve shows a fit of the 3NLO expression,
eq.~(\ref{eq-ng}), 
to the data, using $\nff$=3,
where $\nff$ is the number of active quark flavors.
The dashed curve shows the prediction of the HERWIG
Monte Carlo for the mean
charged particle multiplicity in {\gluglu} events.
} 
\label{fig-gluons}
\end{figure}

Experimental measurements of the inclusive charged
particle multiplicity of unbiased gluon jets
as a function of energy scale Q
are shown in Fig.~\ref{fig-gluons}.
The three data points at scale Q$\,\approx\,$5~GeV are
derived from the hadronic component of
$\Upsilon$(1S)$\rightarrow\gamma${\gluglu}
events~\cite{bib-cleo97}.
The virtuality Q is given by the invariant mass of
the hadronic system.
Similarly,
$\Upsilon$(3S)$\rightarrow\gamma\chi_{b2}
$(10.27)$\rightarrow\gamma${\gluglu} 
events provide the result at 
Q$\,\approx\,$10~GeV~\cite{bib-cleo92},
with the scale given by the $\chi_{b2}$ mass.
The measurement at 
Q$\,\approx\,$80~GeV~\cite{bib-opal96,bib-opal99}
is based on hadronic Z$^0$ decays:
Z$^0\rightarrow\,$q$\overline{\mathrm{q}}${\gincl},
in which {\gincl} refers to a gluon jet hemisphere
recoiling against two almost collinear quark jets q
and $\overline{\mathrm{q}}$
in the opposite hemisphere.
For the results shown here,
the {\gincl} hemisphere results in Ref.~\cite{bib-opal99}
have been multiplied by a factor of
two both for the multiplicity and energy scales so that
they correspond to {\gluglu} ``two-jet events''
analogous to the $\Upsilon$ data.
Although many studies of gluon jet multiplicity 
have been performed,
only the data in Fig.~\ref{fig-gluons} are based
on an unbiased definition of the jets.
These data are summarized in Table~\ref{tab-gnchtable}.

\begin{table}[t]
\centering
\begin{tabular}{|l|cc|}
 \hline
  &    &     \\[-2.4mm]
Experiment  & $Q$ (GeV) & {$\mnch$}, {\gluglu} events \\[2mm]
 \hline
 \hline
CLEO~\cite{bib-cleo97}  & 4.5  & $4.88\pm0.10$  \\
CLEO~\cite{bib-cleo97}  & 5.5  & $5.28\pm0.10$  \\
CLEO~\cite{bib-cleo97}  & 6.5  & $5.65\pm0.12$  \\
CLEO~\cite{bib-cleo92}  & 10.3 & $9.339\pm0.090\pm0.045$ \\
OPAL~\cite{bib-opal99}  & 80.2 & $28.56\pm0.36\pm0.62$ \\
 \hline
\end{tabular}
\caption{The mean charged particle multiplicity, {$\mnch$},
of unbiased ``two-jet'' {\gluglu} events.
For the OPAL data and the CLEO measurement at 10.3~GeV,
the first uncertainty is statistical and the second
is systematic.
For the CLEO measurements below 10~GeV,
only the statistical uncertainty is given.
}
\label{tab-gnchtable}
\end{table}

The solid curve in Fig.~\ref{fig-gluons}
shows the result of a $\chi^2$ fit of the 3NLO
expression for gluon jet multiplicity,
eq.~(\ref{eq-ng}), to the data,
with $y$=$\ln\,({\mathrm{Q}}/\Lambda)$,
where $\Lambda$ is a fitted parameter 
related to the perturbative cutoff~$Q_0$
by $\Lambda$=$Q_0/2$ (see Section~\ref{sec-equations}).
The other fitted parameter is 
the normalization constant~$K$ in eq.~(\ref{eq-ng}).
For this fit, {$\nff$}=3.
The $\Upsilon$(1S) measurements~\cite{bib-cleo97} 
are not included in the fit because a systematic uncertainty
was not provided for them:
the smallness of the statistical uncertainty for these data
would unduly bias the fit results.
Nonetheless,
these data are seen to lie near the fitted curve.
The results for the fitted parameters,
along with those found using {$\nff$}=4 and~5,
are given in the top portion of Table~\ref{tab-nchfit}.
The results of the fits with {$\nff$}=4 and 5 are
virtually indistinguishable from that shown by
the solid curve in Fig.~\ref{fig-gluons}.
The uncertainties given in Table~\ref{tab-nchfit} 
for the gluon jet parameters are defined by the
maximum deviations observed when the
gluon jet measurements are varied by their
one standard deviation uncertainties.

\begin{table}[tphb]
\centering
\begin{tabular}{|c|ccc|}
 \hline
  & & &  \\[-2.4mm]
  & $\Lambda$ (GeV) & $K$ & $\chi^2$/bins of data  \\[2mm]
 \hline
 \hline
 \multicolumn{4}{|l|}{}  \\[-2.4mm]
 \multicolumn{4}{|l|}{ (a) Gluon jets,
               eq.~(\ref{eq-ng})} \\[2mm]
 \hline
  & & &  \\[-2.4mm]
$\nff$=3 & $1.03\pm0.24$  &  $0.288\pm0.037$ & 0.01/2\\[2mm]
$\nff$=4 & $0.84\pm0.21$  &  $0.244\pm0.034$ & 0.01/2\\[2mm]
$\nff$=5 & $0.64\pm0.17$  &  $0.205\pm0.031$ & 0.01/2\\[2mm]
 \hline
 \multicolumn{4}{|l|}{}  \\[-2.4mm]
 \multicolumn{4}{|l|}{ (b) Quark jets,
               eq.~(\ref{eq-nq}), $K$=fixed} \\[2mm]
 \hline
  & & &  \\[-2.4mm]
$\nff$=3 & $0.670\pm0.036$  &  0.288 & 26.3/18\\[2mm]
$\nff$=4 & $0.579\pm0.034$  &  0.244 & 31.3/18\\[2mm]
$\nff$=5 & $0.469\pm0.031$  &  0.205 & 38.6/18\\[2mm]
 \hline
 \multicolumn{4}{|l|}{}  \\[-2.4mm]
 \multicolumn{4}{|l|}{ (c) Quark jets,
               eq.~(\ref{eq-nq})} \\[2mm]
 \hline
  & & &  \\[-2.4mm]
$\nff$=3 & $0.262\pm0.072$  &  $0.198\pm0.022$ & 6.6/18\\[2mm]
$\nff$=4 & $0.188\pm0.063$  &  $0.156\pm0.020$ & 6.8/18\\[2mm]
$\nff$=5 & $0.119\pm0.048$  &  $0.118\pm0.018$ & 7.1/18\\[2mm]
 \hline
\end{tabular}
\caption{
Results of a fit of QCD expressions 
for the scale evolution of event multiplicity 
to the measured mean charged particle multiplicities of
unbiased (a)~{\gluglu} (top),
and (b) and (c)~{\qq} (center and bottom) events.
For the gluon jets,
the fitted expression is the 3NLO result, eq.~(\ref{eq-ng}).
For the quark jets,
the expression is the 3NLO result, eq.~(\ref{eq-nq}).
In (b) the normalization $K$ is set equal to the
value found from the gluon jet fit;
a one parameter fit
of the scale parameter $\Lambda$ is performed.
In (c) a two parameter fit of $\Lambda$ and $K$
is performed.
For the quark jets,
the uncertainties are evaluated
by varying the fit range as described in the text.
}
\label{tab-nchfit}
\end{table}

From Fig.~\ref{fig-gluons} and the results in Table~\ref{tab-nchfit}
it is seen that the 3NLO expression provides a good description
of the growth of gluon jet multiplicity with energy
using a physically sensible value of the coupling strength,
corresponding to a value of $\Lambda$
in the range from about 100~MeV to 1~GeV.
For example the fitted result $\Lambda\,$=$\,0.64\pm0.17$~GeV
obtained for $n_f\,$=$\,5$
yields $\alpmz\,$=$\,0.142\pm0.008$,
compared to the world average value based on $\lms$ of
$\alpmz\,$$\approx$$\,0.120\pm0.002$~\cite{pdg}.
We remind the reader that $\Lambda$ is not in general
the same as~$\lms$.

\begin{figure}[tphb]
\vspace*{-2cm}
\begin{center}
  \epsfxsize=11cm
  \hspace*{.2cm}\epsffile{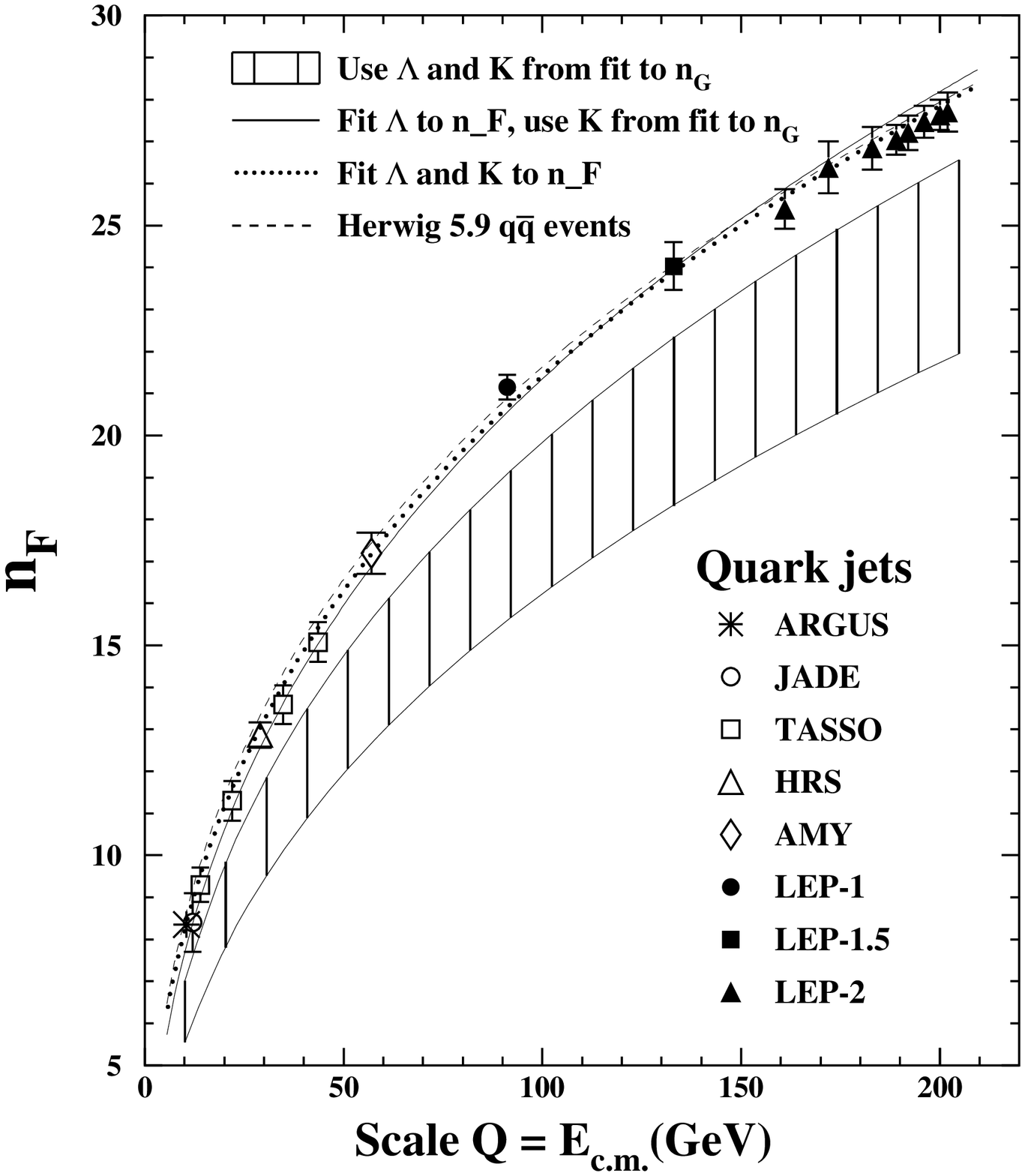}
\end{center}
\vspace*{-1cm}
\caption{
The mean charged particle multiplicity of {\epem}
hadronic annihilation events versus energy scale~Q={\ecm}.
The solid curve shows a fit of the 3NLO expression,
eq.~(\ref{eq-nq}), to the data,
using a one parameter fit of~$\Lambda$ (see text).
The dotted curve shows the corresponding result
from a two parameter fit of $\Lambda$ and~$K$.
The hatched band shows the 3NLO result
if the parameter values used in the fit of gluon jet
multiplicity are used.
The width of the band corresponds to the uncertainty
in the value of~$\Lambda$.
For the analytic curves, $n_f$=3.
The dashed curve shows the prediction of the HERWIG
Monte Carlo for the mean charged particle multiplicity in
{\epem}$\,\rightarrow\,${\qq} events.
} 
\label{fig-quarks}
\end{figure}

Experimental measurements of unbiased quark jet 
multiplicity are shown in Fig.~\ref{fig-quarks}.
These data are the inclusive charged particle
multiplicity values of
{\epem} hadronic annihilation events,
corresponding to ``two-jet events''
as for the data of Fig.~\ref{fig-gluons}.
The scale is Q={\ecm}.
The results shown for the LEP experiments
are combined values of ALEPH, DELPHI, L3 and OPAL.
The combined values are obtained using the
unconstrained averaging procedure described in~\cite{pdg},
for which a common systematic uncertainty is defined
by the unweighted mean of the systematic uncertainties
quoted by the experiments.
LEP-1 refers to data collected at the Z$^0$ peak,
\mbox{LEP-1.5} to data collected at {\ecm}$\approx$133~GeV,
and LEP-2 to data collected at or above the
threshold for W$^+$W$^-$ production.
The quark jet data are summarized in Table~\ref{tab-qnchtable}.

\begin{table}[t]
\centering
\begin{tabular}{|l|cc|}
 \hline
  &    &     \\[-2.4mm]
Experiment  & $Q$={\ecm} (GeV) & {$\mnch$} \\[2mm]
 \hline
 \hline
ARGUS~\cite{bib-argus92}    & 10.5 &  $8.35\pm0.02\pm0.20$   \\
JADE~\cite{bib-jade83}     & 12.0 &  $8.4\pm0.3\pm0.6$   \\
TASSO~\cite{bib-tasso89} & 14.0 & $9.30\pm0.06\pm0.41$   \\
TASSO~\cite{bib-tasso89} & 22.0  & $11.3\pm0.08\pm0.46$   \\
HRS~\cite{bib-hrs86}         & 29.0  & $12.87\pm0.03\pm0.30$   \\
TASSO~\cite{bib-tasso89} & 34.8  & $13.59\pm0.02\pm0.46$   \\
TASSO~\cite{bib-tasso89} & 43.6  & $15.08\pm0.06\pm0.47$   \\
AMY~\cite{bib-amy90}      & 57.0  & $17.19\pm0.07\pm0.48$   \\
LEP-1~\cite{bib-lep1nch}   
   & 91.2   & $21.15\pm0.01\pm0.29$   \\
LEP-1.5~\cite{bib-aleph98b,bib-lep1p5nch}
   & 133  & $24.03\pm0.20\pm0.53$   \\
LEP-2~\cite{bib-aleph98b,bib-dl161nch,bib-opal97}  & 161  
   & $25.39\pm0.23\pm0.41$   \\
LEP-2~\cite{bib-aleph98b,bib-dl161nch,bib-opal99b}
  & 172  & $26.38\pm0.26\pm0.56$  \\
LEP-2~\cite{bib-aleph98b,bib-l398,bib-opal99b}
  & 183  & $26.84\pm0.16\pm0.48$   \\
LEP-2~\cite{bib-aleph99,bib-l32000,bib-opal99b}
  & 189  & $27.04\pm0.11\pm0.34$   \\
LEP-2~\cite{bib-aleph2000,bib-l32000,bib-opal99c}
  & 192  & $27.21\pm0.25\pm0.32$   \\
LEP-2~\cite{bib-aleph2000,bib-l32000,bib-opal99c}
  & 196  & $27.47\pm0.18\pm0.33$   \\
LEP-2~\cite{bib-aleph2000,bib-l32000,bib-opal99c}
  & 200  & $27.63\pm0.17\pm0.32$   \\
LEP-2~\cite{bib-aleph2000,bib-l32000,bib-opal99c}
  & 202  & $27.70\pm0.25\pm0.38$   \\
 \hline
\end{tabular}
\caption{The mean charged particle multiplicity, {$\mnch$},
measured in {\epem} annihilations at various c.m. energies.
The first uncertainty is statistical and the second
is systematic.
A more complete compilation
for energies below 91~GeV
is given in~\cite{bib-nchcompilation}.
The results shown for LEP are combined values from 
ALEPH, DELPHI, L3 and OPAL (see text).
}
\label{tab-qnchtable}
\end{table}

The vertically striped band in Fig.~\ref{fig-quarks} shows the
3NLO prediction for quark jet multiplicity,
eq.~(\ref{eq-nq}), for $\nff$=3,
using the values of $\Lambda$ and $K$ found from
the fit to the gluon jet measurements 
(top portion of Table~\ref{tab-nchfit}).
The width of the band corresponds to the uncertainty in the value of
$\Lambda$ presented in 
the top portion of Table~\ref{tab-nchfit}.
Almost identical results to those shown by the band
are obtained using $\nff$=4 or~5.
The band lies 15-20\% below the data,
representing an inadequacy of the 3NLO calculations to simultaneously
describe the gluon and quark jet measurements using
precisely the same values of $\Lambda$ and~$K$.
This problem can mostly be attributed to
the analytic prediction for the multiplicity ratio $\ratio$,
discussed below in Section~\ref{sec-rexp}.

It is also of interest to {\it fit}
expression~(\ref{eq-nq}) to the quark jet data.
For this, we use the normalization $K$ found 
in the fit of the gluon jet multiplicity 
(compare eqs.~(\ref{eq-ng}) and~(\ref{eq-nq}))
and perform a one parameter fit of
$\Lambda$ to the quark jet data using $n_f$=3.
The result is shown by the solid curve in Fig.~\ref{fig-quarks}.
The results found using $\nff$=4 or~5 
are almost identical to that shown by this curve.
The fit is seen to provide a reasonable overall
description of the measurements.
The fitted values of $\Lambda$ for $n_f$=3, 4 and 5
are given in the central portion of Table~\ref{tab-nchfit}.
The uncertainties of the quark jet parameters in
Table~\ref{tab-nchfit} are defined by
the maximum difference between the results of the
standard fit and those found by
fitting only data between 29 and 202~GeV,
between 10.5 and 161~GeV,
or by excluding the \mbox{LEP-1} data point.

Comparing the parameter values obtained from 
the fits to the unbiased gluon and quark jet measurements
(top and central portions of Table~\ref{tab-nchfit}),
it is seen that the value of $\Lambda$ from 
gluon jets is about 40\% larger than
that from quark jets.
For $n_f$=5,
the quark jet result $\Lambda$=$0.469\pm0.031$~GeV
yields $\alpmz\,$=$\,0.135\pm0.002$,
to be compared to $\alpha_S$$\,\approx\,$$0.14\pm0.01$
found from the gluon jet data as
mentioned above.
Thus although the gluon and quark jet data are not well described
using exactly the same values of $\Lambda$ and~$K$,
they both yield physically acceptable results for~$\Lambda$,
leading to similar values of the coupling strength.
In this sense,
the analytic description of multiplicity 
in single gluon and quark jets is quite consistent.

For completeness,
we include in Fig.~\ref{fig-quarks} 
the result of a two parameter fit of $\Lambda$ and $K$
to the quark jet data, using $\nff$=3.
The result of this fit
is shown by the dotted curve in Fig.~\ref{fig-quarks}.
The corresponding parameter values are given in the
bottom portion of Table~\ref{tab-nchfit},
with systematic uncertainties defined as for the
quark jet parameters in the central portion of that table.
The results for $\Lambda$ 
are seen to be about 3 times smaller than if the 
normalization factor $K$ is required to be the same for
quark and gluon jets as described above.
For $n_f$=5,
the quark jet result $\Lambda$=$0.119\pm0.048$~GeV
yields $\alpmz\,$=$\,0.109\pm0.005$.

The dashed curves in Figs.~\ref{fig-gluons} and~\ref{fig-quarks}
show the prediction of the HERWIG Monte Carlo~\cite{bib-herwig}
for the mean inclusive charged particle multiplicity
of {\gluglu} and {\epem} annihilation events,
respectively.
The prediction of HERWIG is
generally similar to the fitted analytic results.

\subsection{Multiplicity ratio}
\label{sec-rexp}

A test of the QCD prediction for the ratio $r$ of 
mean multiplicities between gluon and quark jets is 
possible using the data in Table~\ref{tab-gnchtable}.
Such a test is in principle limited to these data 
since they are the only unbiased measurements
of gluon jet multiplicity currently available.
See also Refs.~\cite{cdnt8,lan}, however,
where results for $r$ based on gluon jets
from {\epem} three-jet events are also discussed.

To obtain a result for $r$ from their measurements of
gluon jet multiplicity in $\Upsilon$(1S) decays,
CLEO~\cite{bib-cleo97} divided the gluon jet multiplicity 
given in the top three rows of Table~\ref{tab-gnchtable}
by the charged particle multiplicity in radiative {\qq}$\gamma$ 
events with similar hadronic recoil mass values.
The three $\Upsilon$(1S)-based data points are shown
with their statistical uncertainties at
$y_c$$\,\approx\,$0.01 in Fig.~\ref{fig-eight}.
CLEO combined the three data points into a single value
at an effective scale of about 5.5~GeV~\cite{bib-cleo97}.
The resulting value for $r$ is given in the top row of Table~\ref{tab-ratio}.

\begin{figure}[tphb]
\vspace*{-1.5cm}
\begin{center}
  \epsfxsize=12cm
  \epsffile{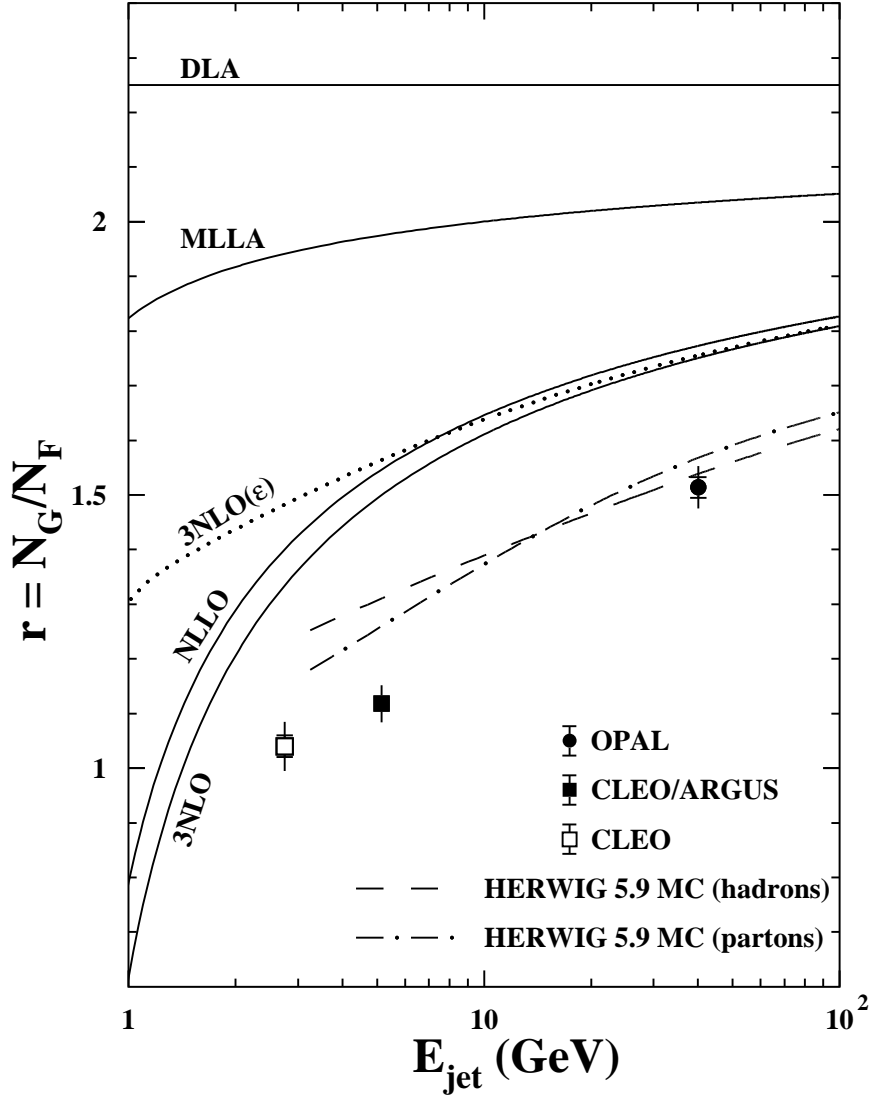}
\end{center}
\vspace*{-2cm}
\caption{Experimental results for the multiplicity ratio
of unbiased gluon and quark jets,
in comparison to QCD analytic predictions.
The curve marked 3NLO$(\epsilon)$
shows the analytic prediction at order 3NLO
with an approximate accounting for truncation of the
limits of integration at $e^{-y}$ and $1-e^{-y}$,
see Ref.~\cite{cdnt8}.
The dashed and dash-dotted curves show the prediction 
of the HERWIG Monte Carlo at the hadron and parton levels.
The experimental and HERWIG results at the hadron level
are based on charged particles only.
} 
\label{fig-ratio}
\end{figure}

\begin{table}
\begin{tabular}{|l|ccc|}
 \hline
  & & & \\[-2.4mm]
Experiment  & $Q$ (GeV)& {\ejet}(GeV) & $r$ \\[2mm]
 \hline
 \hline 
CLEO~\cite{bib-cleo97}  & 5.5 & 2.75  & $1.04\pm 0.02\pm 0.04$   \\
CLEO~\cite{bib-cleo92}, ARGUS~\cite{bib-argus92} 
     & 10.3 & 5.15 & $1.118\pm0.011\pm0.032$  \\
OPAL~\cite{bib-opal99}  & 80.2 & 40.1 & $1.514\pm 0.019\pm 0.034$  \\
 \hline
\end{tabular}
\caption{Measurements of the multiplicity ratio $r$ between unbiased
gluon and quark jets.
The first uncertainty is statistical and the second is systematic.
}
\label{tab-ratio}
\end{table}

OPAL~\cite{bib-opal99} divided their result for the multiplicity 
in gluon jet hemispheres at 40~GeV by the 
multiplicity in hemispheres of light quark (uds flavored) events
with about the same energy.
The reason for selecting light quark jets is that it provides
closer correspondence with the massless quark assumption
of the calculations.
The OPAL result for $r$ is shown at 
$y_c$$\,\approx\,$$3\times 10^{-5}$ in Fig.~\ref{fig-eight}
and is listed in the bottom row of Table~\ref{tab-ratio}.

CLEO did not report a result for $r$ based on their 
measurement~\cite{bib-cleo92} of gluon jet multiplicity 
in $\Upsilon$(3S) decays.
The effective scale of this result, 10.3~GeV,
is not too different from the scale of the lowest 
energy quark jet result in Table~\ref{tab-qnchtable}, however,
viz.~10.5~GeV.
Dividing the 10.3~GeV gluon jet result from CLEO
by the 10.5~GeV quark jet result from ARGUS
yields the result for $r$ given in the central row of Table~\ref{tab-ratio}.

The experimental results for $r$ are shown in Fig.~\ref{fig-ratio}.
The results are presented as a function of the jet energy {\ejet}
rather than the ``two-jet event'' scale $Q$ to correspond to
the analytic results.
Fig.~\ref{fig-ratio} includes the QCD analytic predictions for $r$
at various orders of perturbation theory.
The data are seen to lie substantially below the predictions
(Section~\ref{sec-rtheory}).
The theoretical predictions approach the experimental
results more closely, however,
as higher order terms are included.
At the scale of the~Z$^0$, the 3NLO expression, eq.~(\ref{Y}),
predicts $r$$\,\approx\,$1.7,
about 13\% larger than the 
OPAL result $r_{exp.}$$\,\approx\,$1.51.
At the scale of the $\Upsilon$,
the difference between theory and data is larger.
It is likely that uncalculated non-perturbative terms play
an important role at this low scale, however.
By itself, it is perhaps surprising to obtain any level of
agreement between theory and data at all
given the large value of the expansion parameter 
$\gamma_0\approx 0.5$.

The dashed and dash-dotted curves in Fig.~\ref{fig-ratio}
show the predictions of the HERWIG Monte Carlo for the ratio $r$
at the hadron and parton levels.
The HERWIG prediction is obtained by dividing the results for
hemispheres of {\gluglu} events by those for light quark {\qq} events.
The HERWIG results at the hadron level are based
on charged particles only.
HERWIG is seen to represent the OPAL measurement quite well
at both the hadron and parton levels,
implying a modest correction for hadronization
at the Z$^0$ energy.

The computer solution of the QCD equations,
discussed in Section~\ref{sec-computer},
leads to near perfect agreement with the OPAL result for~$r$,
as shown in Fig.~\ref{fig-eight}.
The computer solution differs from the analytic estimates 
by use of the explicit multiplicities in the integrals,
with no Taylor series expansion,
leading to a more accurate accounting of energy conservation,
as mentioned in Section~\ref{sec-computer}.
This fact becomes especially important for higher order terms
because of the large value of~$\gamma _0$.
The computer solution also implements the exact preasymptotic
limits of integration,
in contrast to the analytic solutions,
as was also mentioned in Section~\ref{sec-computer}.
This agreement between the experimental and theoretical
(computer)
results for $r$ at the scale of the Z$^0$ is an impressive
success for the QCD approach to multiplicity in jets.

In contrast to the result at the scale of the Z$^0$,
the computer prediction for $r$ at the scale of the $\Upsilon$
exceeds the data by about 25\% (see Fig.~\ref{fig-eight}).
For such low scales,
neglected non-perturbative effects are likely to be important,
as noted above.

Recently, results on the charged particle multiplicity
of jets in dijet events produced in {\pp} collisions 
have been reported by CDF Collaboration~\cite{bib-safonov}.
The dijet masses range from 80 to 630~GeV.
This substantially increases the range of accessible jet energies
compared with {\epem} collisions.
The multiplicity ratio $r$ has been estimated 
to be $1.7\pm 0.3$ from these data.
It should be noted that this result is obtained using a 
fit of an MLLA expression to the data.
Since in MLLA quark and gluon jets differ
only by the constant factor $r$=9/4,
it is unclear how the experimental result $r\,$$\approx\,$1.7
should be interpreted.
Also, the large uncertainty
precludes a precise differentiation between the different
theoretical predictions in this region.

\subsection{Slopes}

According to eq.~(\ref{Y}),
the ratio $r$ becomes smaller as the scale decreases
since $\gamma _0$ increases. 
The same trend is observed experimentally
(see Fig.~\ref{fig-eight} or~\ref{fig-ratio}). 
Thus the data and theory are in qualitative agreement.
From Figs.~\ref{fig-eight} and~\ref{fig-ratio}
it is seen that the theoretical predictions decrease with
scale more slowly than the data, however.
For example,
the computer solution for $r$~\cite{lo1}
agrees perfectly with the data at the scale of the Z$^0$,
but it exceeds the experimental result by about 25\% 
at the scale of the $\Upsilon$ (Fig.~\ref{fig-eight}),
as discussed in Section~\ref{sec-rexp}.
Thus there is a significant difference between the
theoretical and experimental results for the slope~$r^{'}$
(see eq.~(\ref{rpri})),
at least for energies below the~Z$^0$.
From Fig.~2 in~\cite{lo1}, 
the slope predicted by the computer solution can be estimated 
to be about~0.096 at the Z$^0$.
Using the data in Fig.~\ref{fig-eight},
the corresponding result is about~0.174.
Both these values differ from the analytic prediction
of~0.06 given in Section~\ref{sec-rtheory} (cf.~eq.~(\ref{rpri})).
This implies a strong influence of higher order 
perturbative corrections,
even at the Z$^0$,
and of non-perturbative terms at lower energies.

The ratio of slopes $r^{(1)}$, eq.~(\ref{r12}),
is less sensitive to higher order corrections than $r^{'}$ 
and is better approximated by the MLLA expression than $r$,
as discussed in Section~\ref{sec-rtheory}.
Experimental results for $r^{(1)}$ are
available from the DELPHI~\cite{lan}
and OPAL~\cite{bib-opaleta} Collaborations.
These data are based on three-jet events selected 
using a jet finder and thus
--~unlike the other {\epem} data presented
in Section~\ref{sec-experiment}~--
do not employ the unbiased (hemisphere) definition of a jet
utilized by the calculations.
This makes a quantitative test of the QCD prediction uncertain.
Nonetheless
we proceed to a comparison of experiment with theory 
with an eye towards qualitative agreement.

The energy scale of the experimental results~\cite{lan,bib-opaleta}
for $r^{(1)}$
is presented in terms of the so-called jet hardness
$\kappa$=$2E\sin (\Theta /2)$ (see e.g.~\cite{5}), 
where $\Theta$ is the opening angle between the
two lowest energy jets.
For small~$\Theta$, $y$$\,\approx\,$$\ln (\kappa /Q_0)$. 
The jet hardness has been shown~\cite{lan} to be a
more appropriate scale than the jet energy when comparing
jets embedded in a three-jet environment.

\begin{figure}[tphb]
\begin{center}
  \epsfxsize=12cm
  \epsffile{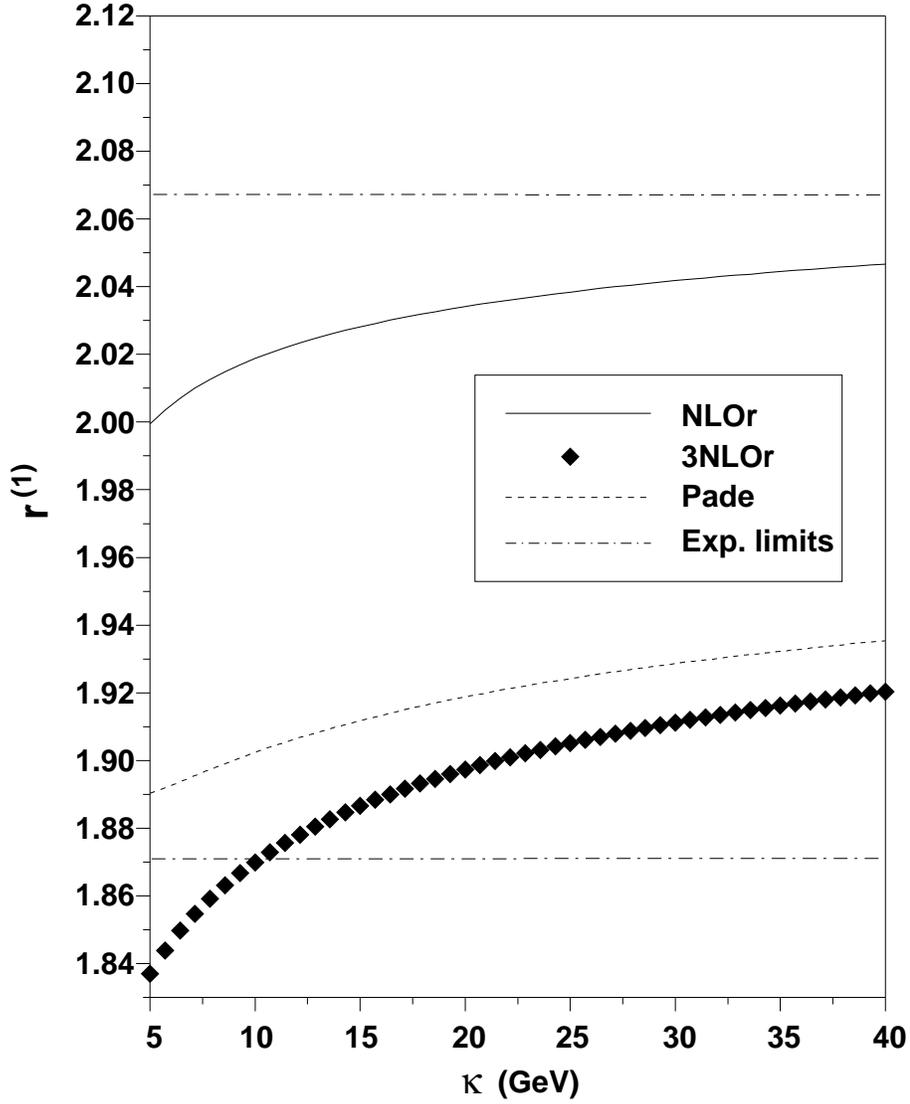}
\end{center}
\vspace*{-.5cm}
\caption{The ratio of slopes of the average multiplicities 
in gluon and quark jets, $r^{(1)}$,
in the MLLA and 3NLO (with its Pad\'{e} expression) 
approximations.
The theoretical results are obtained using $n_f$=4.
Experimental limits (statistical uncertainties only)
from Ref.~\cite{lan} 
are shown by the dash-dotted lines.
} 
\label{fig-slopes}
\end{figure}

Both DELPHI and OPAL observe the experimental value 
of $r^{(1)}_{exp}$ to depend very little on scale.
The OPAL measurement
$r^{(1)}_{exp}$=$2.27\pm0.09\,$(stat.) $\pm0.27$(syst.)
is about one standard deviation higher than the 
DELPHI result 1.97$\pm$0.10~(stat.).
Note that the systematic uncertainty quoted by OPAL
is mostly due to the jet selection bias.
The experimental results agree with the MLLA
prediction $r^{(1)}_{MLLA}$$\,\approx\,$2.01-2.03. 
The expressions (\ref{Y}), (\ref{rh1}) and (\ref{cor3}) yield
$r^{(1)}_{3NLO}$$\,\approx\,$1.86-1.92 in the energy range
from the $\Upsilon$ to the~Z$^0$,
in agreement with DELPHI but about 1.3
standard deviations of the total uncertainty below OPAL.
Fig.~\ref{fig-slopes} displays the values of $r^{(1)}$ 
in the MLLA and 3NLO approximations for $n_f$=4
in comparison to the experimental limits from DELPHI. 

A recent study~\cite{bib-delphiopen}
attempts to correct for the jet bias
introduced by the three-jet event selection
by subtracting the biased quark jet component from the
multiplicity of the {\qq}g events.
The result,
$r^{(1)}$=$1.77\pm 0.03\,$(stat.) is
substantially smaller than the DELPHI and OPAL results
presented above.
There are a number theoretical and phenomenological assumptions
which enter this analysis, however.

From the DELPHI result~\cite{lan} for $r^{(1)}_{exp}$,
the measured values of $\rho_{1,exp} \approx r_{exp}/2$ can be 
determined to range from 0.53 at the $\Upsilon$ to 0.75 at the Z$^0$. 
Its perturbative values according to
eq.~(\ref{cor3}) vary from 0.85 to~0.9. 
Either the next order perturbative or else non-perturbative terms 
are needed~\cite{d98} to reproduce the large difference
$d_1$=$r^{(1)}_{exp}-r_{exp}$$\,\approx\,$$2(1-\rho_{1,exp})$ 
observed in experiment.
The simplest Pad\'{e} expression~(\ref{rpad}) 
improves the situation only slightly.

\subsection{Higher moments}
\label{sec-highermoments}

\begin{figure}[tphb]
\begin{center}
\begin{tabular}{cc}
  \epsfxsize=7.2cm\epsffile{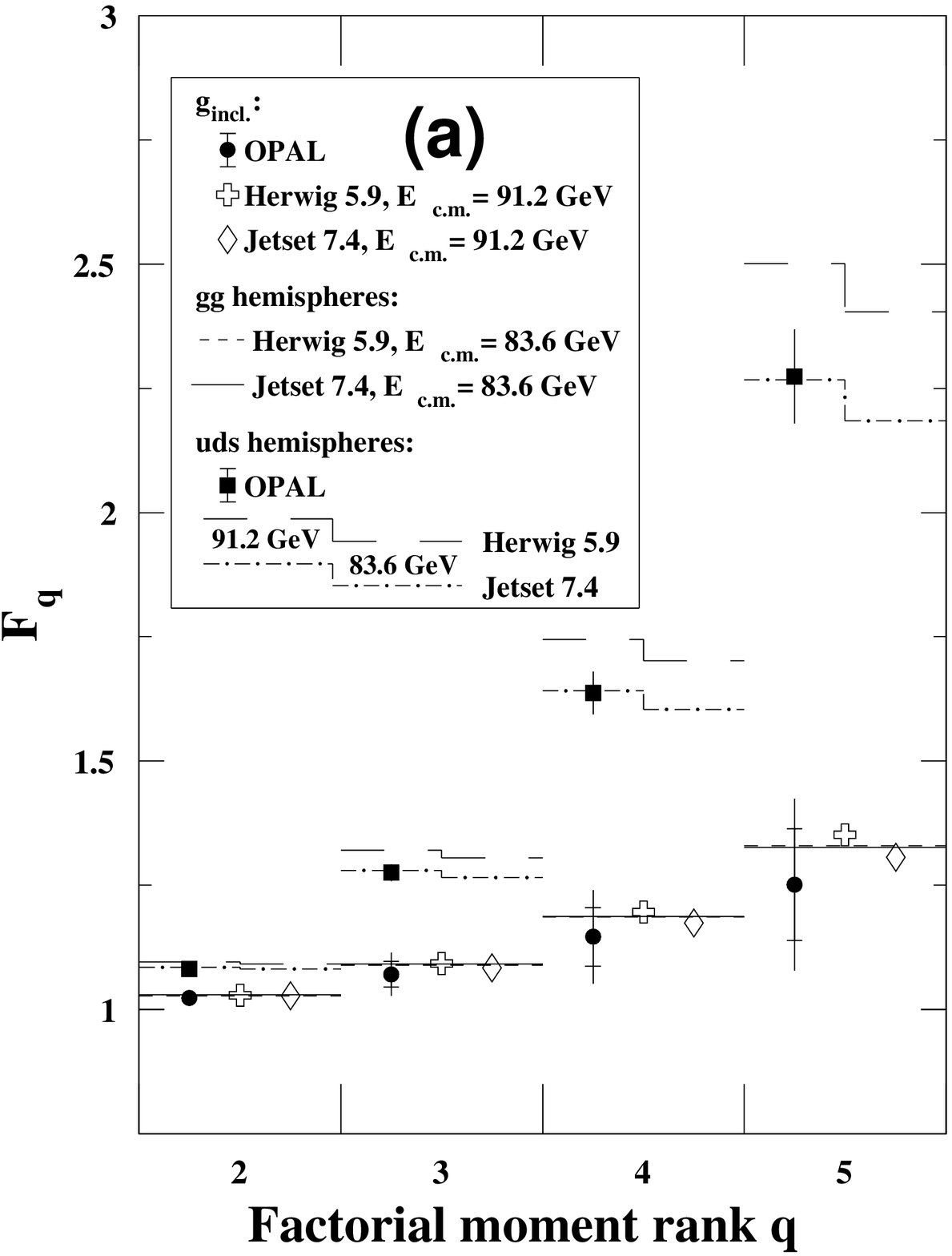} &
  \hspace*{-.5cm}
  \epsfxsize=7.2cm\epsffile{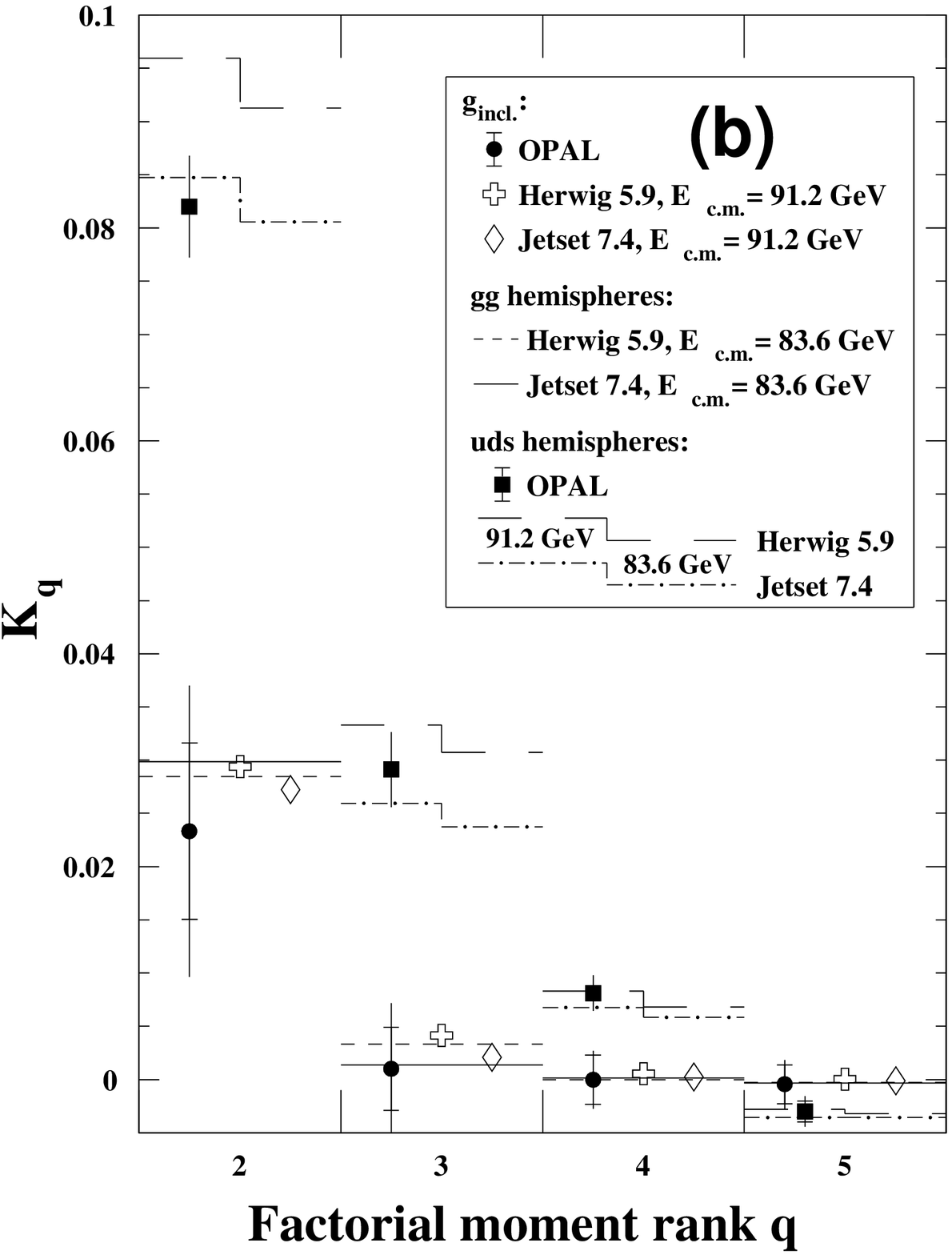} \\
\end{tabular}
\end{center}
\caption{The (a) factorial ($F_q$) 
and (b) cumulant ($K_q$) moments of the charged particle
multiplicity distributions of unbiased single gluon and quark jets,
for 41.8~GeV {\gincl} gluon jets
and 45.6~GeV uds quark jets~\cite{bib-opal98},
as a function of the rank~$q$.
The predictions of the HERWIG and JETSET parton shower 
Monte Carlo event generators are also shown. 
} 
\label{fig-expmoments}
\end{figure}

Measurements of higher moments of the multiplicity
distribution in unbiased gluon and quark jets are
presented in Ref.~\cite{bib-opal98}.
The data are collected at the~Z$^0$.
The experimental results are given for
factorial and cumulant moments with ranks 2-5.
These data are shown in Figs.~\ref{fig-expmoments}(a) 
and~\ref{fig-expmoments}(b)
in comparison to the predictions of QCD Monte Carlo programs,
and are listed in Table~\ref{tab-expmoments}.
The quark jet moments are observed to be
larger than those of gluon jets
in accordance with theoretical expectation
(Section~\ref{sec-psolutions}),
i.e.~event-to-event fluctuations are larger
for quark jets than for gluon jets.
Note that factorial moments of different rank are
highly correlated with each other statistically.
In contrast,
cumulant moments of different rank
are not (see e.g. Ref.~\cite{bib-opal98}).
Thus there is a high degree of bin-to-bin correlation
in Fig.~\ref{fig-expmoments}(a) 
but not in Fig.~\ref{fig-expmoments}(b).
It is worth noting that cumulants of ranks 2, 3 and 4
are directly correlated 
to dispersion, skew and kurtosis, respectively,
i.e. to the ordinary moments,
cf.~Ref.~\cite{bib-opal98}.

\begin{table}[tphb]
\centering
\begin{tabular}{|c|cc|}
 \hline
  & & \\[-2.4mm]
 $q$ & $F_q$, gluon jets & $F_q$, quark jets \\[2mm]
 \hline
2 & $1.023\pm0.008\pm0.011$ & $1.0820\pm0.0006\pm0.0046$
        \\
3 & $1.071\pm0.026\pm0.034$ & $1.275\pm0.002\pm0.017$
        \\
4 & $1.146\pm0.059\pm0.074$ & $1.637\pm0.005\pm0.042$
        \\
5 & $1.25\pm0.11\pm0.13$   & $2.274\pm0.014\pm0.093$
        \\
 \hline
 \hline
  & & \\[-2.4mm]
 $q$ & $K_q$, gluon jets & $K_q$, quark jets \\[2mm]
 \hline
2 & $0.0233\pm0.0083\pm0.0109$ & $0.0820\pm0.0006\pm0.0048$
        \\
3 & $0.0010\pm0.0039\pm0.0048$ & $0.0291\pm0.0006\pm0.0035$
        \\
4 & $0.0000\pm0.0023\pm0.0015$ & $0.0081\pm0.0007\pm0.0015$
        \\
5 & $-0.0005\pm0.0018\pm0.0014$ & $-0.00300\pm0.00096\pm0.00095$
        \\
 \hline
\end{tabular}
\caption{
The factorial ($F_q$) and cumulant ($K_q$) moments 
of the charged particle multiplicity distribution of
unbiased 41.8~GeV \protect{\gincl} gluon jets and
45.6~GeV uds quark jets~\cite{bib-opal98}
as a function of the rank~$q$.
The first uncertainty is statistical and the 
second is systematic.
}
\label{tab-expmoments}
\end{table}

The experimental results for 41.8~GeV gluon jets, 
$F_2^G\,$=1.023,
and for 45.6~GeV uds quark jets, $F_2^F\,$=1.082, 
are much smaller than the DLA predictions,
viz.~1.33 and~1.75, respectively (see eq.~(\ref{fas})).
The analytic results in MLLA,
$F_2^G(MLLA)$$\,\approx\,$1.11 and $F_2^F(MLLA)$$\,\approx\,$1.22
(for $n_f$=4), 
agree much better with the data.
This is analogous to the situation described 
in Section~\ref{sec-rexp} for the multiplicity ratio~$r$,
i.e.~the inclusion of higher orders results in a marked improvement 
in the description of the data compared to leading order.
If one accepts the effective value of $\alpha_S$
averaged over all the energies of the partons
during the jet evolution to be $\alpha_S\,$$\approx\,$0.2, 
one obtains
$F_2^G(MLLA)$$\,\approx\,$1.039 and $F_2^F(MLLA)$$\,\approx\,$1.068,
which are quite close to the experimental results. 
In this sense the MLLA prediction can be said to describe the widths 
of the gluon and quark jet multiplicity distributions
at the Z$^0$ energy to within 10\% accuracy.

Unfortunately the NNLO and 3NLO terms
worsen the agreement with data compared to MLLA
(but not compared to DLA),
as was already mentioned in Section~\ref{sec-widths}.
The analogy with~$r$,
for which each subsequent correction brings the theory
closer to data, does not hold here.
The NNLO calculation yields slightly larger values for
$F_2^G$ and $F_2^F$ than in MLLA,
resulting in a slight worsening of the description of experiment.
The 3NLO corrections are large and negative,
yielding $F_2^G(3NLO)$$\,\approx\,$1.01 
and $F_2^F(3NLO)$$\,\approx\,$0.94 for $n_f$=4.
These latter results are in qualitative conflict with the data
in that they imply $F_2^F$$\,<\,$$F_2^G$.
The problem in the analytic description of higher moments
can be traced to an inappropriate treatment of soft particles,
as discussed in Section~\ref{sec-widths}.

\begin{figure}[tphb]
\vspace*{-1.2cm}
\begin{center}
  \epsfxsize=13cm
  \epsffile{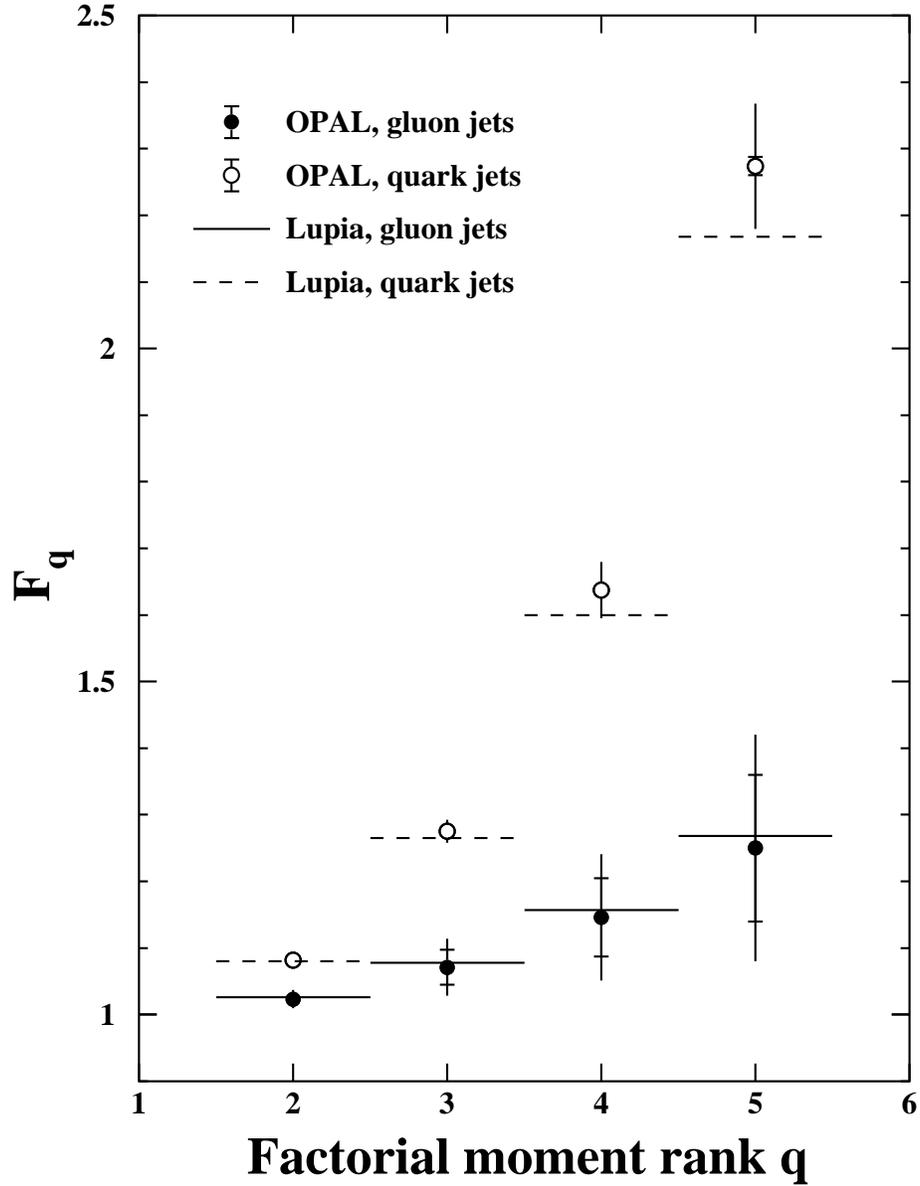}\hspace*{.3cm}
\end{center}
\vspace*{-2cm}
\caption{The measured, normalized factorial moments of the charged
particle multiplicity distributions of unbiased, separated
gluon and quark jets~\cite{bib-opal98},
in comparison to the corresponding QCD predictions 
obtained from a computer solution of the equations for
the generating functions~\cite{lo2}.
} 
\label{fig-lupia}
\end{figure}

Given this problem with the analytic approach,
it is remarkable that the computer solution of the QCD equations
provides a near perfect description of the higher moments.
This is shown in Fig.~\ref{fig-lupia},
which displays the computer solutions for the factorial moments
of single gluon and quark jets~\cite{lo2} in comparison to
the corresponding data from Table~\ref{tab-expmoments}.
Overall, excellent agreement is observed between experiment
and theory for ranks up to $q$$\,\approx\,$5.
This suggests that the failure of the analytic approach
to describe the widths of single quark and gluon jets
is mainly a technical issue.
The success of the computer solution again
emphasizes the importance of an exact treatment 
of energy conservation and the limits of integration.

Further insight into this issue can be gained by examining
the predictions of the ARIADNE Monte 
Carlo~\cite{bib-ariadne} at the parton level,
as was done in~\cite{bib-ochs99} using Monte Carlo parameter 
values consistent with~LPHD.\footnote{Essentially
this means using a small value of the cutoff parameter,
$Q_0$$\,\approx\,$0.2~GeV~\cite{bib-ochs99}.}
At the parton level,
ARIADNE is roughly equivalent to a DLA analytic calculation
except that it includes exact conservation of
both energy and momentum.
The parton level predictions of ARIADNE for the factorial moments 
of uds quark jet hemispheres~\cite{bib-ochs99}
are listed in the rightmost column of Table~\ref{tab-aridanemoments}.
These values are remarkably similar
to the corresponding measurements in Table~\ref{tab-expmoments}.
This suggests that an exact treatment 
of energy-momentum conservation is essential
for an accurate prediction of the higher moments.
The corresponding ARIADNE results for hemispheres
of {\gluglu} events are also listed in Table~\ref{tab-aridanemoments}.
The agreement with the corresponding data in Table~\ref{tab-expmoments}
is not as good as for quark jets.
Nonetheless, the ARIADNE results are much more similar to the data 
than either the LO or 3NLO analytic results discussed above,
for both quark and gluon jets.

\begin{table}[tphb]
\centering
\begin{tabular}{|c|cc|}
 \hline
  & & \\[-2.4mm]
 $q$ & $F_q$, gluon jets & $F_q$, quark jets \\[2mm]
 \hline
2 & 1.044 & 1.092 \\
3 & 1.132 & 1.310 \\
4 & 1.271 & 1.720 \\
5 & 1.474 & 2.444 \\
 \hline
\end{tabular}
\caption{
The factorial ($F_q$) moments of the parton level multiplicity
distributions of hemispheres of {\gluglu} and uds {\qq}
events generated using the ARIADNE multihadronic
Monte Carlo event generator,
as a function of the rank~$q$.
}
\label{tab-aridanemoments}
\end{table}


Qualitatively, 
the results shown in Fig.~\ref{fig-lam} for the energy dependence
of the second factorial moments are similar to 
experimental trends reported at LEP~\cite{del}. 
The NNLO terms tend to slightly violate the level of
agreement with experiment while the agreement is even worse 
if the 3NLO terms are included,
analogous to the situation discussed above in connection with the
data of Table~\ref{tab-expmoments}.

\subsection{Oscillating $H_q$ moments}
\label{sec-oscillating}

\begin{figure}[tphb]
\begin{center}
  \epsfxsize=12cm
  \epsffile{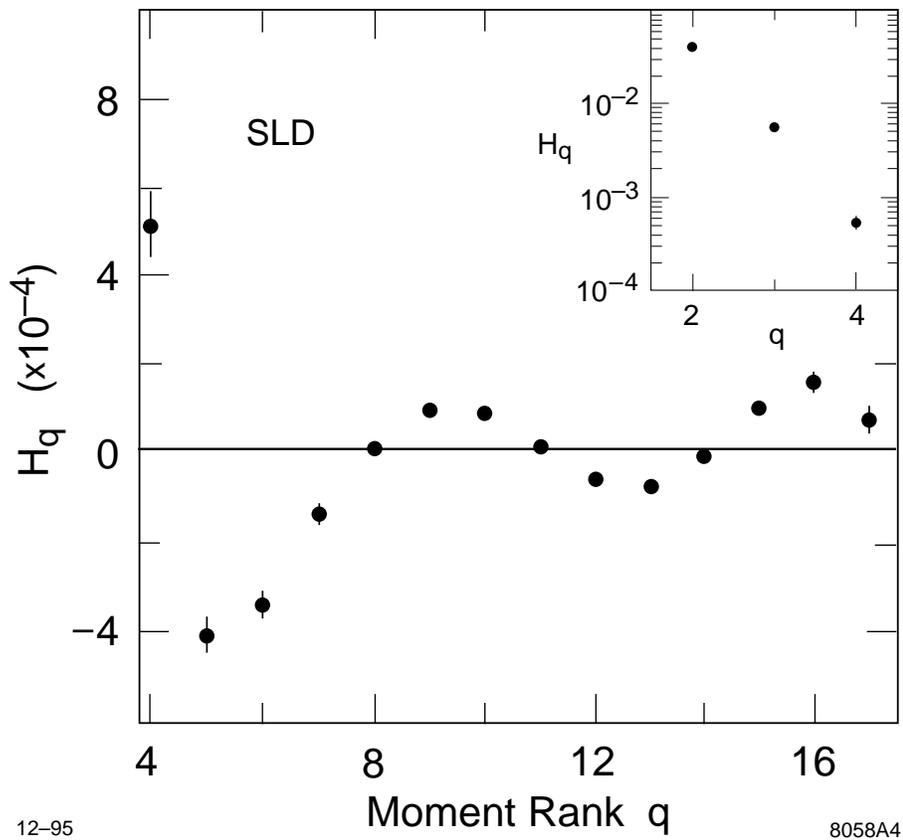}
\end{center}
\caption{Measured ratio of cumulant to factorial moments,
$H_q$, as a function of the rank~$q$,
for the charged particle multiplicity distribution
in {\epem} hadronic Z$^0$ decays~\cite{bib-sldhq}.
The uncertainties are statistical only.
} 
\label{fig-hqdata}
\end{figure}

The $H_q$ moments of quark jets,
eq.~(\ref{13}),
have been experimentally studied up to rank $q$$\,\approx\,$17 
in {\epem} annihilations~\cite{bib-sldhq,bib-l3hq}.
Results from the SLD Collaboration
based on the charged particle multiplicity distribution
from hadronic Z$^0$ decays 
are shown in Fig.~\ref{fig-hqdata}~\cite{bib-sldhq}.
Note that these results employ full
inclusive {\epem} hadronic annihilation events
and thus correspond to unbiased ``two-jet'' {\qq} configurations,
unlike the data in Table~\ref{tab-expmoments}
and Fig.~\ref{fig-expmoments} which are based on 
event hemispheres, i.e.~single jets.

\begin{figure}[thb]
\vspace*{-.5cm}
\begin{center}
  \begin{turn}{-90}
    \epsfxsize=7.7cm
    \epsffile{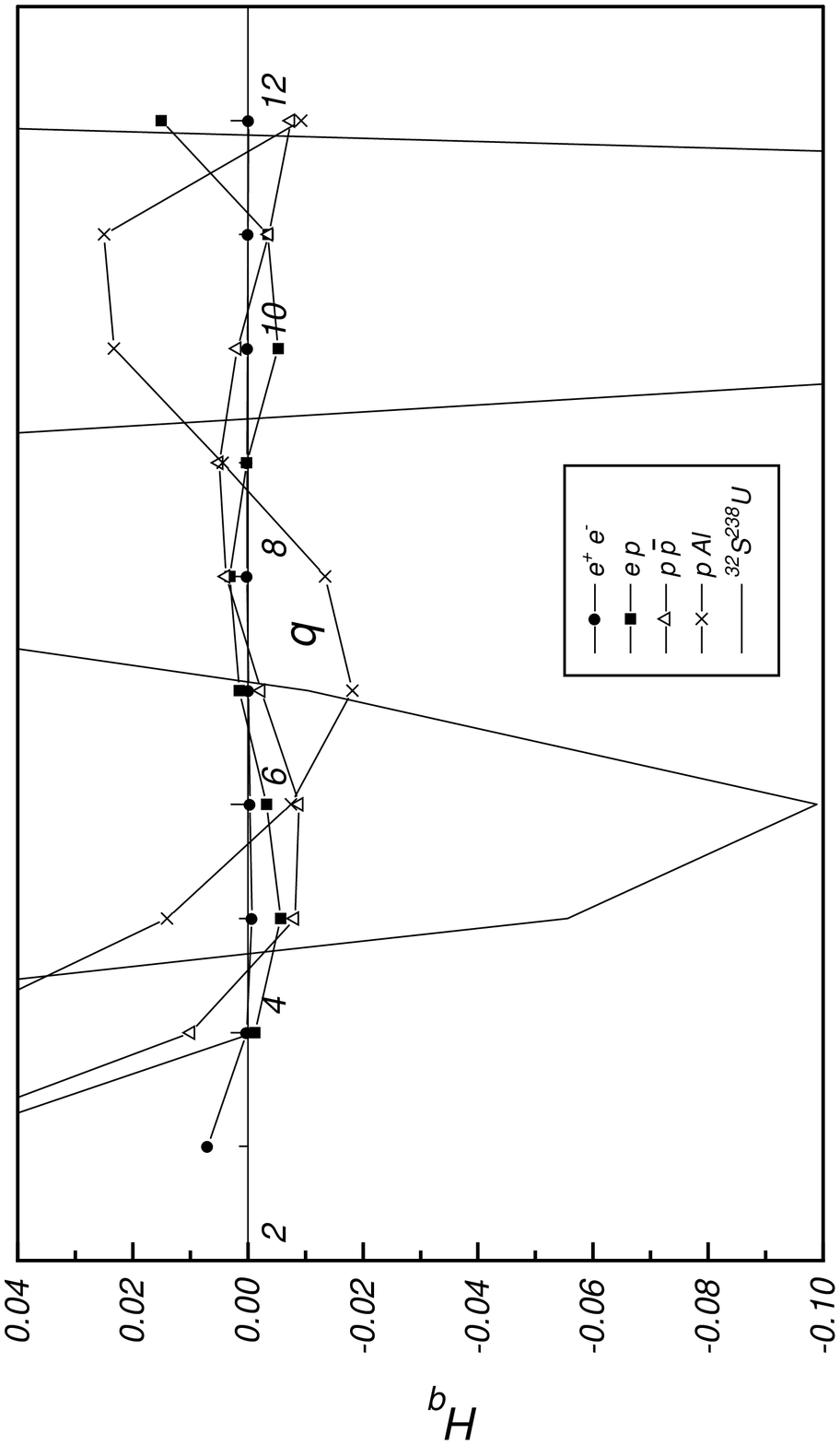}
  \end{turn}
\end{center}
\vspace*{-.1cm}
\caption{
The ratio $H_q$ for {\epem}, ep, {\pp}, pA and AA collisions~\cite{dnbs},
where A denotes a heavy ion.
} 
\label{fig-hionshq}
\end{figure}

The results in Fig.~\ref{fig-hqdata} are strikingly similar
to the QCD analytic predictions discussed
in Sections~\ref{sec-hoapprox} and~\ref{sec-widthshigher},
viz.~the $H_q$ moments oscillate with increasing rank,
with the occurrence of the first minimum at rank \mbox{$q$$\,\approx\,$5.}
We again note the essential role that energy conservation plays
for the theoretical prediction.
Similar oscillations also have been observed in
ep, {\pp}, and heavy ion collisions~\cite{dabg,dnbs},
see~Fig.~\ref{fig-hionshq}.
From Fig.~\ref{fig-hionshq} it is seen that
the amplitude of the oscillations increases as the structure
of the colliding particle becomes more complicated.
The large positive values for
ranks less than about 3 are due to
strong correlations between particles from resonance decays. 
The positive values for q$\,\approx\,$7 or 8 imply the production of
clusters (or mini-jets) at this multiplicity scale. 
The negative cumulants at q$\,\approx\,$5 suggest anti-correlations
between resonances (and particles) inside these mini-jets.
The behavior at larger ranks arises from a
complicated mixture of attractive and repulsive forces 
inside higher multiplicity groups (jets). 
It is interesting that these features are rather
universal in nature.

In a study by the L3 Collaboration~\cite{bib-l3hq},
it is noted that Monte Carlo simulations provide a good
representation of the oscillations in the data.
This is found to be true even for simulations
based on the MLLA approximation,
despite the fact that analytic calculations 
predict oscillations only at order NNLO and above
(Section~\ref{sec-hoapprox}).
This situation can easily be understood
to be a consequence of energy conservation.
Unlike the MLLA analytic results,
Monte Carlo results based on MLLA formulas 
incorporate exact energy-momentum conservation.
However, one of the principal differences between analytic
results at NNLO compared to lower orders is that energy
conservation is included.
Thus Monte Carlo results based on the MLLA
effectively incorporate terms at NNLO and above.
This provides an explanation of why MLLA based
Monte Carlo event generators exhibit oscillations of $H_q$
in agreement with the data.

\subsection{Soft particles}
\label{sec-soft}

\begin{figure}[thbp]
\begin{center}
    \epsfxsize=8cm
    \epsffile{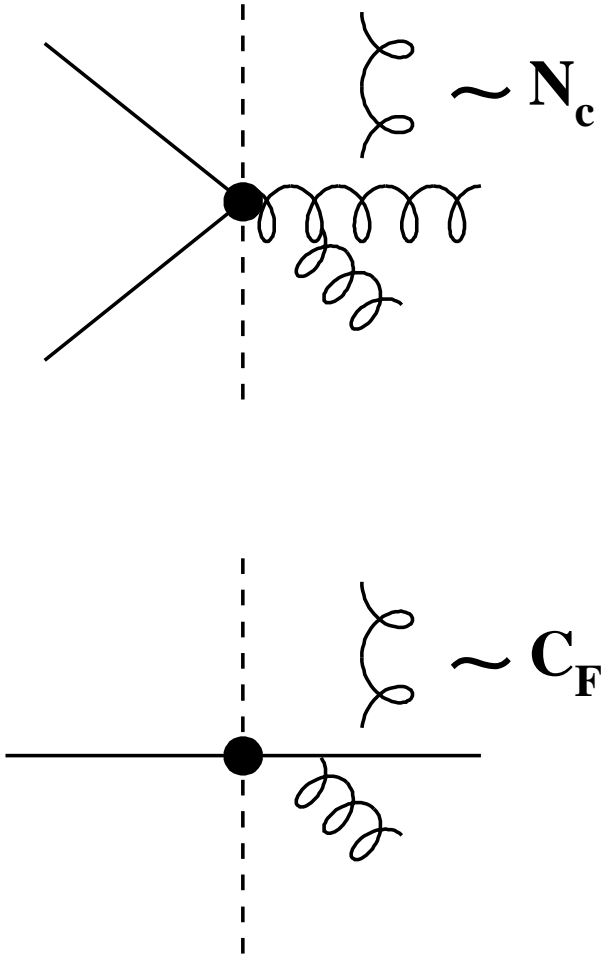}
\end{center}
\caption{
Soft gluons radiated from a gluon (top) or quark (bottom) jet
cannot resolve the color substructure of the jet because of
their long wavelength.
They therefore couple to gluon and quark jets with relative
strengths given by the respective color factors
$N_{c}$=3 and $C_{F}$=4/3,
see Refs.~\cite{bib-klo98,bib-opal99,bib-klo96}.
} 
\label{fig-softcacf}
\end{figure}

So far,
our discussion has focused exclusively on multiplicity
in full phase space, i.e.~within an entire jet.
It is of considerable interest to also consider multiplicity
within a limited region of phase space.
In this section,
we consider the limited phase space volume
defined by soft particles,
i.e. particles with an energy or momentum below
some cutoff.
See also Section~\ref{sec-intermittency} where multiplicity
in limited phase space is discussed in the context
of intermittency.

Particle multiplicity in limited phase space is not described
by the analytic approach based on generating functions
presented above in 
Sections~\ref{sec-equations}-\ref{sec-exactsolutions}.
Nonetheless soft particle multiplicity possesses a simple
theoretical interpretation when considered
in terms of the ratio between unbiased gluon and quark jets
(see below).
Recent theoretical and experimental progress in this area
merits discussion in our review.

In QCD,
the ratio of mean multiplicities between gluon and
quark jets, $r$,
equals $r_0$=$N_{c}/C_{F}$=2.25
in the asymptotic limit {\ejet}$\,\rightarrow\infty$,
as discussed in Section~\ref{sec-rtheory}.
The approach to this limit is very slow, however.
For finite energies such as {\ejet}$\,\approx\,$M$_{\mathrm{Z}}$,
preasymptotic corrections from higher order terms
and conservation laws limit $r$ to values of approximately
1.5--1.7 (Sections~\ref{sec-rtheory} and~\ref{sec-computer}).
Energy and momentum
conservation only apply to full phase space, however.
In limited phase space,
conservation laws are not a constraint.
Furthermore,
the asymptotic condition {\ejet}$\,\rightarrow\infty$
may effectively be replaced by the condition that the energy
of the particles being considered, $E$, 
satisfy $Q_0$$\,<\,$$E$$\,<<\,${\ejet}.
By selecting {\it soft} particles,
one therefore fulfils the asymptotic condition in
at least an approximate manner,
opening the possibility of observing
$r$$\,\rightarrow r_0$=2.25
even for finite energies.

\begin{figure}[tbp]
\vspace*{-.5cm}
\begin{center}
    \epsfxsize=11cm
    \epsffile{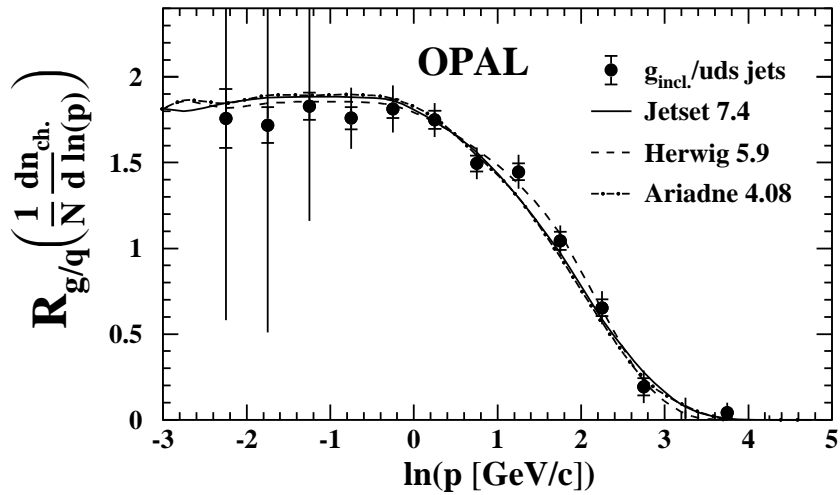}
\end{center}
\vspace*{-.2cm}
\caption{
The ratio of mean charged particle multiplicities between
unbiased gluon and quark jets as a function of $\ln p$,
with $p$ particle momentum~\cite{bib-opal99}.
The results are shown in comparison to the predictions
of QCD Monte Carlo programs.
} 
\label{fig-opallnp}
\end{figure}

These ideas are developed in 
Refs.~\cite{bib-klo98,bib-klo96}.
Because of color coherence in the radiation of soft gluons,
the low energy end of the particle spectrum is found 
to be nearly independent of the energy scale~{\ejet}.
Qualitatively,
this arises because soft gluons have long wavelengths
and cannot resolve individual partons within a jet.
Instead,
they couple to the jet with a strength proportional to the 
color charge of the parton which initiated it,
given by $N_{c}$ for gluon jets and $C_{F}$ for
quark jets,
see Fig.~\ref{fig-softcacf}.
Thus for {\it soft} particles,
the ratio of mean multiplicities between gluon and
quark jets, {\rsoft},
is predicted to approach
{\rsoft}$\,\rightarrow N_{c}/C_{F}$
as $E$$\,\rightarrow Q_0$
irrespective of the energy scale.

Using their samples of unbiased gluon 
(``\gincl'') and quark jets
(see Section~\ref{sec-meanmult}),
the OPAL Collaboration tested this prediction by measuring $r$ 
as a function of particle momenta~$p$~\cite{bib-opal99}.
The results are shown in Fig.~\ref{fig-opallnp}.
As $p$ becomes small,
$r$ is seen to saturate at a value of
approximately~1.8.
This result is larger than the result $r$$\,\approx\,$1.5
obtained in full phase space (Table~\ref{tab-ratio}).
However it is
considerably smaller than the theoretical prediction
{\rsoft}$\,\approx\,$2.25 discussed in the previous paragraph.


\begin{figure}[tbp]
\vspace*{-.5cm}
\begin{center}
    \epsfxsize=11cm
    \epsffile{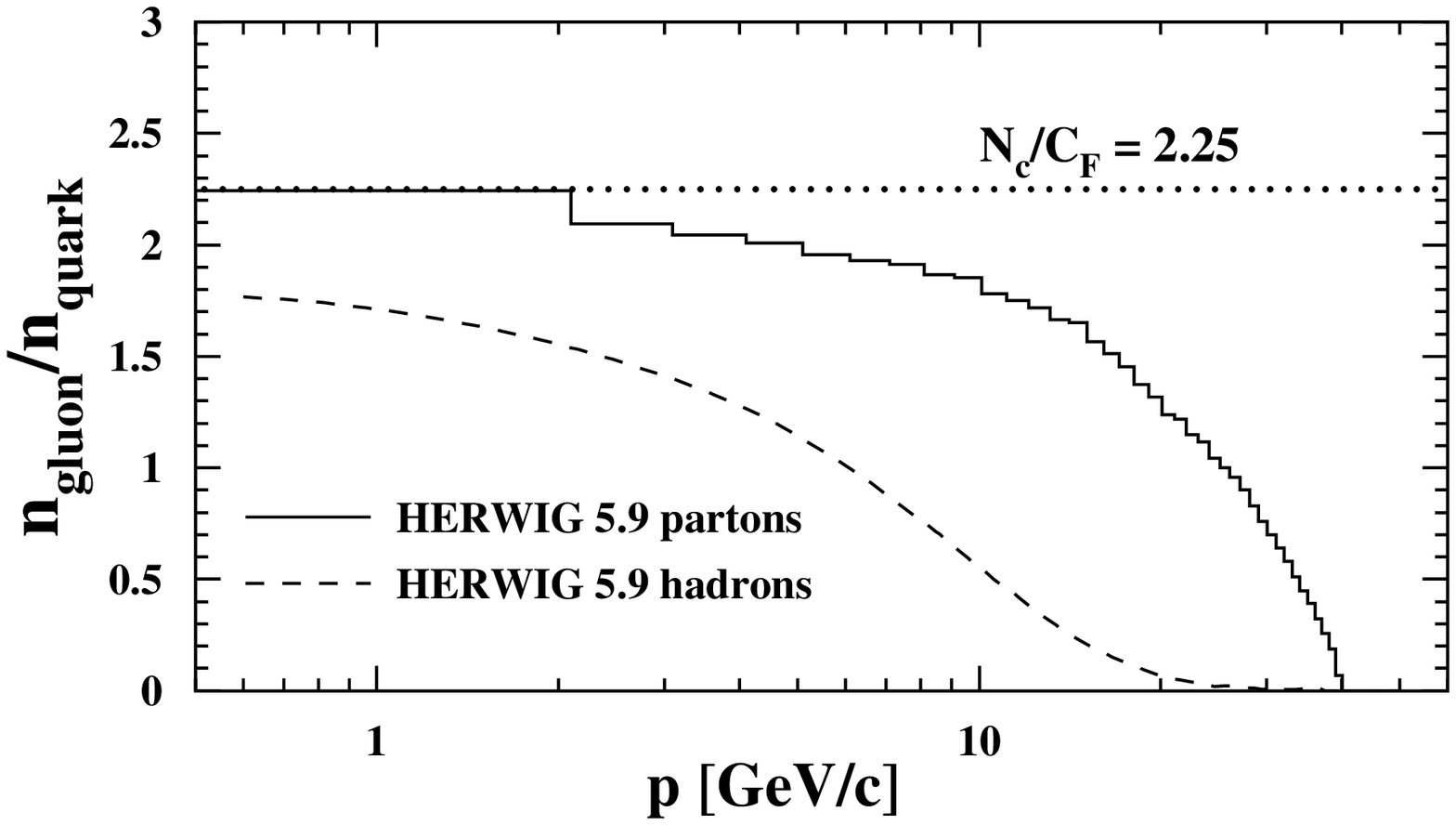}
\end{center}
\vspace*{-.2cm}
\caption{
Predictions of the HERWIG Monte Carlo~\cite{bib-herwig}
for the ratio of mean particle multiplicities between
unbiased gluon and quark jets as a function of
particle momentum $p$,
for partons and charged hadrons.
} 
\label{fig-herwigpsoft}
\end{figure}

Insight into this difference between the measured
and theoretical values of {\rsoft} can be obtained by examining
the predictions of the HERWIG Monte Carlo.
The predictions of HERWIG for $r$ versus $p$ at the
hadron and parton levels are shown
in Fig.~\ref{fig-herwigpsoft}.
For small momenta, the hadron level result (dashed curve) 
converges to {\rsoft}$\,\approx\,$1.8,
in agreement with OPAL
(see also the Monte Carlo curves in Fig.~\ref{fig-opallnp}).
The parton level prediction (solid histogram) 
converges to {\rsoft}$\,\approx\,$2.25,
in agreement with the theoretical expectation~\cite{bib-klo98}.
Thus the difference between the measured and 
theoretical values of {\rsoft} can be
attributed to the effects of hadronization.

\begin{figure}[tbp]
\vspace*{-.5cm}
\begin{center}
    \epsfxsize=11cm
    \epsffile{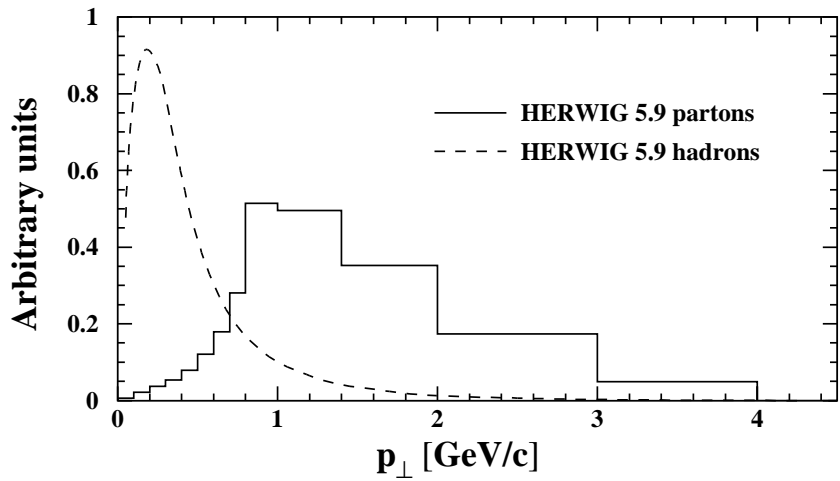}
\end{center}
\vspace*{-.2cm}
\caption{
Predictions of the HERWIG Monte Carlo~\cite{bib-herwig}
for the transverse momentum distributions of 
partons and charged hadrons in unbiased uds quark jets.
The transverse momentum {\pperp} is defined 
with respect to the jet axis.
The ordinate units are arbitrary.
} 
\label{fig-herwigptq}
\end{figure}

For soft particles,
hadronization is most important within the core of a jet,
corresponding to small transverse momenta 
{\pperp} with respect to the jet axis.
For example,
Fig.~\ref{fig-herwigptq} shows the predictions of HERWIG
for the {\pperp} distributions of particles in quark jets,
for partons (solid histogram)
and charged hadrons (dashed curve).
Partons are overwhelmingly produced with {\pperp} values
larger than 0.8~GeV/$c$
(because of the cutoff $Q_0$ in the Monte Carlo)
in contrast to hadrons which predominantly appear at
smaller~{\pperp}.
On the basis of this information,
OPAL repeated their analysis selecting particles 
with {\pperp}$\,>\,$0.8~GeV/$c$ only,
i.e.~they restricted their analysis to the
perturbative region of phase space 
as predicted by HERWIG.

\begin{figure}[tbp]
\vspace*{-.5cm}
\begin{center}
    \epsfxsize=13cm
    \epsffile{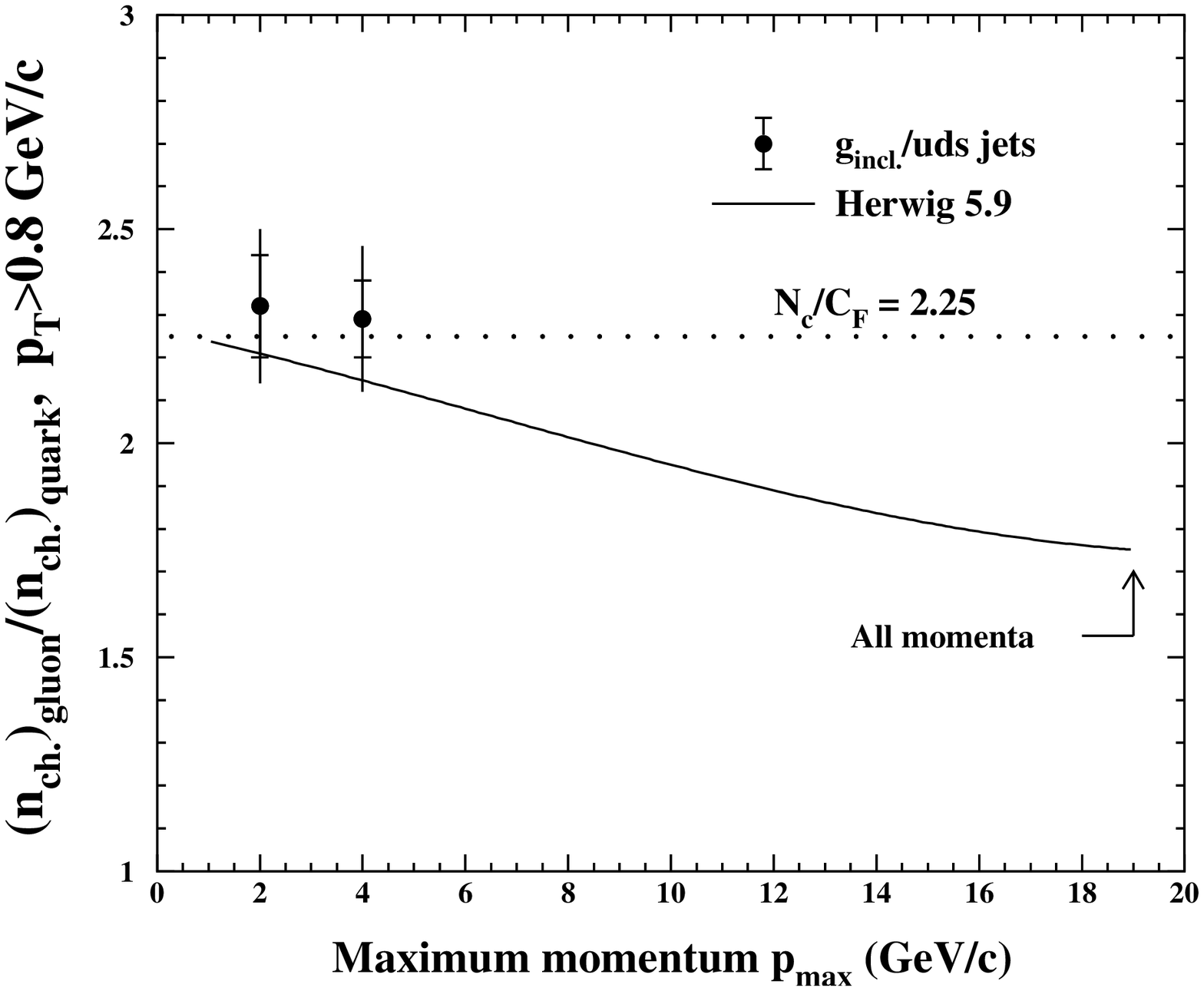}
\end{center}
\vspace*{-.2cm}
\caption{
Ratio of charged particle multiplicities between
unbiased gluon and quark jets for particles with large transverse
momenta to the jet axis defined by $p_\perp$$\,>\,$0.8~GeV/$c$.
The results are shown as a function of the softness of the
the particles,
defined by the maximum particle momentum $p_{\mathrm{max.}}$
used to determine~$r$.
} 
\label{fig-ptgt0p8}
\end{figure}

The results are shown in Fig.~\ref{fig-ptgt0p8}.
The charged particle multiplicity ratio $r$ is shown as
as a function of the softness of the particles.
The softness of particles is
defined by the maximum particle momentum $p_{\mathrm{max.}}$
considered when determining~$r$.
Unlike the results in Fig.~\ref{fig-opallnp},
particles are required to have $p_\perp$$\,>\,$0.8~GeV/$c$
as explained in the previous paragraph.
The solid curve shows the prediction of HERWIG.

With no explicit cut on $p_{\mathrm{max.}}$
(``All momenta''),
the multiplicity ratio is predicted to be about~1.8.
As softer and softer particles are selected
($p_{\mathrm{max.}}$ is decreased),
the HERWIG prediction approaches the QCD result
{\rsoft}$\,\approx\,$2.25 in conformity with 
theory~\cite{bib-klo98}.
Thus a measurement of {\rsoft} for particles with
$p_\perp$$\,>\,$0.8~GeV/$c$ effectively yields a measurement
of the color factor ratio~$N_{c}/C_{F}$.
OPAL results are shown for $p_{\mathrm{max.}}$=2~GeV/$c$
and 4~GeV/$c$.
The result using $p_{\mathrm{max.}}$=4~GeV/$c$
is $r$=$2.29\pm0.017 \;\mathrm{(stat.+syst.)}$~\cite{bib-opal99}
which provides one of the most accurate
experimental determinations of $N_{c}/C_{F}$ 
currently available.
Note that unlike all other measurements of $N_{c}/C_{F}$,
this result is not based on a fit of a QCD motivated 
expression
--~in which $N_{c}/C_{F}$ is extracted as a fitted
parameter~--
but is the ratio of directly measured quantities.
This OPAL result confirms one of the oldest,
previously unverified
predictions for multiplicities in gluon and quark jets,
that the particle multiplicity ratio between gluon
and quark jets should equal the color factor difference
(in an appropriate region of phase space),
dating back to Refs.~\cite{brgu,kuv78}.

\section{Jet and subjet multiplicities}
\label{sec-subjets}

To this point, 
we have considered only the final result of
well developed cascades, 
namely hadron multiplicities. 
It is also of interest to study the intermediate stages 
of the cascade evolution, 
i.e.~the jet substructure.
Recalling the phrase about 
``whorls inside whorls inside whorls'' from the field of turbulence, 
one can ask what is predicted by pQCD for the
structure of ``jets inside jets inside jets''
in {\epem} events. 
The ordering of emissions by transverse momentum
$(k_t)$ is at the heart of this issue. 
Thus one can ask about the number of jets or subjets 
(the ``jets inside jets'') 
as the angular $(k_t)$ resolution scale is varied.
With very low resolution one obtains two jets,
corresponding to the condition eq.~(\ref{72}) imposed 
on the equations for the generating functions,
eqs.~(\ref{50}) and~(\ref{51}),
which requires that a single quark-antiquark pair be
initially created in an {\epem} annihilation. 
A three-jet structure appears when a gluon with large
transverse momentum is emitted by the quark or antiquark. 
Three-jet event production is suppressed by an additional 
factor of $\alpha _S$,
which is small for large transferred momenta. 
Therefore, this process can be calculated perturbatively.
The well known exact three-jet matrix element~\cite{ert} 
can then be directly compared with the results of the
generating function approach described above.
At relatively low transferred momenta,
the jet evolves to angular (or $k_t$) ordered subjets. 
By increasing the resolution, 
more and more subjets are observed.
In the ideal case,
each subjet is an individual final-state 
hadron in the limit of very small resolution scales.
The resolution criteria must be chosen in a manner 
to provide infrared safe results. 
Then the perturbative expansion gives rise to finite answers 
which can be confronted to experimental data,
so long as the hadronization stage does not 
significantly alter the situation
as is assumed by local parton-hadron duality.

Details of the resolution criteria are not too 
important if ratios are considered. 
In particular, one can form the ratio of subjet 
multiplicities in three- and two-jet events.
Na\"{\i}vely,
this ratio should equal the corresponding ratio of Casimir factors:
\begin{equation}
  \frac {\langle n_{3}^{sj}\rangle }{\langle n_{2}^{sj}\rangle }=
  \frac {2C_{F}+C_{A}}{2C_{F}}=\frac {17}{8}.  \label{sj}
\end{equation}
However soft gluon coherence suppresses this ratio
to a value below~1.5 (see Ref.~\cite{36}),
demonstrating the importance of correlations
when dealing with interfering jets.

The first proposal to eliminate collinear and mass singularities 
when resolving subjets was formulated in Ref.~\cite{sw}. 
According to this proposal, 
one should consider the relative angles between any two partons 
$\theta _{kl}$ and their fractional energies 
$x_k$ and $x_l$ which exceed some finite values: 
$\theta _{kl}$$\,>\,$$\delta$ and $x_{k,l}$$\,>\,$$\epsilon$. 
The algorithm can be simplified by replacing these two
conditions by a single one,
as is accomplished by the JADE and Durham proposals. 
In the JADE algorithm~\cite{jade}, 
the relative invariant mass of two partons $M_{kl}/s^{1/2}$ 
is required to exceed some value ($M^{2}_{kl}/s$$\,>\,$$\xi _c$). 
In the Durham (or $k_t$) algorithm~\cite{dur,cdotw},
a similar inequality is imposed on the relative transverse momentum,
or, more precisely,
\begin{equation}
  \xi _{kl}=2(1-\cos \theta _{kl})
    \min \frac{(E_{k}^{2}, E_{l}^{2})}{s} >\xi _c, \label{dur}
\end{equation}
where $E_{k,l}$
are the energies of partons $k$ and~$l$.
For small relative angles $\theta _{kl}$,
the parameter $\xi_c$ approximates the rescaled transverse momentum 
$k_{t, kl}^{2}/Q^2$ of the jet of lower momentum with respect 
to that of higher momentum. 

\begin{figure}[tphb]
\begin{center}
  \epsfxsize=13cm
  \epsffile{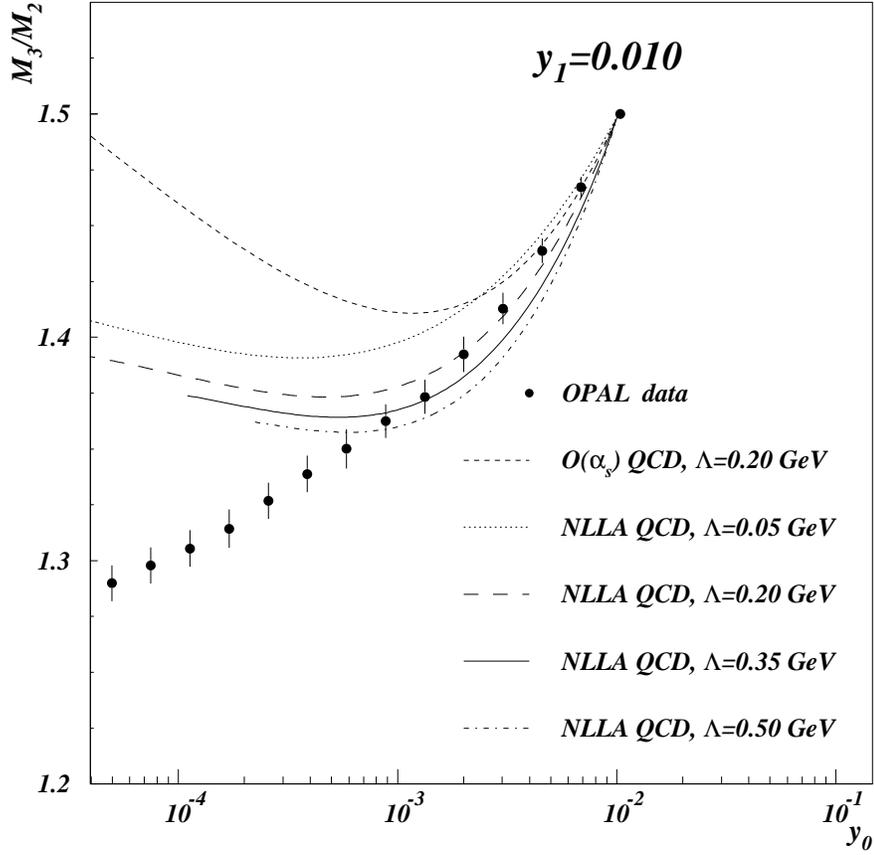}
\end{center}
\caption{The measured ratio of subjet multiplicities between 
three- and two-jet events as a function of
the subjet resolution scale~$y_0$~\cite{bib-opalsj},
in comparison to analytic calculations~\cite{36}.
The results are obtained from {\epem} Z$^0$
data using the Durham jet finder
with an initial clustering scale $y_1$=0.010.
} 
\label{fig-opalsj}
\end{figure}

The Durham algorithm is well suited for analytic calculations and 
gives rise to small hadronization corrections
according to Monte Carlo schemes. 
Therefore it is widely used for comparison of theory with experiment. 
The two limiting cases of $\xi_c$$\,\rightarrow\,$1 and
$\xi_c$$\,\rightarrow\,$0 correspond to low resolution,
for which the minimum number of jets are 
resolved\footnote{$\langle n_{e^{+}e^{-}}^{(jet)}(\xi _{c}$=1$)\rangle$ 
corresponds to two initial quark jets in {\epem} collisions.},
and to the final parton (or particle) limit of the highest available resolution,
discussed in detail in the previous sections. 
In terms of virtualities, $\xi _c$=$Q_{c}^{2}/Q^2$. 
Thus for $Q_c$=$Q$ one deals with the initial stage of the process, 
while ever smaller values of $Q_c$ correspond to its later subjet stages, 
and final hadrons appear at values of $Q_c$ approaching the 
non-perturbative scale $Q_0$ of a few hundred~MeV. 
The number of jets increases as $\xi _c$ decreases,
as calculated in Ref.~\cite{32}.

Jet multiplicities in {\epem} processes are 
usually fitted by the formula
\begin{equation}
  \langle n^{jet}_{e^{+}e^{-}}(Q_{c},Q)\rangle =2\langle n_{F}\left(
\frac {Q}{Q_c},\frac {Q_c}{\Lambda }\right) \rangle, \label{njet}
\end{equation}
whereas the multiplicity of charged hadrons has been 
approximated by a formula with two additional constants:
\begin{equation}
  \langle n^{ch}_{e^{+}e^{-}}(Q_{0},Q)\rangle =2K_{ch}\langle n_{F}\left( 
  \frac {Q}{Q_0}, \frac {Q_0}{\Lambda }\right) \rangle + constant 
     \label{nch}
\end{equation}
It is interesting to determine if eqs.~(\ref{50}) and~(\ref{51})
can describe jet, subjet and hadron multiplicities 
in {\epem} annihilations in a unified manner. 
The answer is yes~\cite{lo2,lo1},
so long as the equations are solved numerically 
using the proper limits of integration. 
By proper limits of integration,
we mean a lower limit $x_c$ and an upper limit $1-x_c$ 
instead of 0 and~1,
with $x_c$=$Q_{c}\sqrt 2 /Q$ as has been
suggested in Refs.~\cite{lo2, lo1}. 
The numerical solutions presented in Refs.~\cite{lo2, lo1} 
describe the average multiplicities of jets, 
subjets and hadrons using a common normalization 
corresponding to ``one parton=one hadron''
without additional constants,
over a wide energy interval from the
threshold region to the Z$^0$ energy. 
This result is especially remarkable given that only two
hard partons are present at the initial stage,
whereas up to 60 charged particles appear
in the final stage at the Z$^0$ energy,
some of which originate from the decays
of intermediate resonances. 
The most common approach consists of inserting a
so-called $K$-factor, $K_{ch}$ (see eq.~(\ref{nch})),
i.e.~some constant relating the parton and hadron distributions 
which usually differs from unity (typically, $K_{ch}$$\,\approx\,$1.2). 
This difference from unity is ascribed to the confinement process.
In Monte Carlo models, this stage is fitted by some
phenomenological formulas which are not yet explained in the
framework of the more theoretical approaches discussed above. 

The ratio of hadron multiplicities in gluon and quark jets
is found~\cite{lo1} to be even smaller than in the 
analytic solutions of the evolution equations,
bringing it into closer agreement with experiment,
see Fig.~\ref{fig-eight} and
the discussion in Section~\ref{sec-rexp}.
The running property of the coupling constant is a
crucial consideration for this result. 
Even more astonishing is that high moments 
of the multiplicity distributions and their oscillations fit 
the experimental data using the same normalization
as used for~$r$~\cite{lo2},
see Figs.~\ref{fig-nine} and~\ref{fig-lupia}.
This implies that power corrections due to energy conservation,
obtained by using the correct integration limits,
are essential,
raising questions about the potential
importance of other possible power corrections,
due to instantons,~etc., as discussed above.

We have briefly described the latest theoretical
developments in the problem,
leaving aside the corresponding formulas which can be found in
Refs.~\cite{khoc,36,cdfw,lo2,lo1,cdotw,dtro,36,bkss}.

Subjet multiplicities have been studied experimentally
by a number of groups~\cite{bib-l3sj}-\cite{bib-alephsj}.
Measured results for the ratio~(\ref{sj}) as a function of the
subjet resolution scale~$y_0$$\equiv$$\xi _c$ are shown
in Fig.~\ref{fig-opalsj}.
These results are obtained using the Durham jet finder
with an initial clustering scale $\xi _c$$\equiv$$y_1$=0.010
to select the two- and three-jet events.
The ratio has a value of 1.5 for $y_0$=$y_1$,
as required from its construction,
and decreases rapidly at smaller resolution scales~$y_0$.
The results are far below the na\"{\i}ve prediction of 17/8
for all values of~$y_0$.
Fig.~\ref{fig-opalsj} also shows the perturbative predictions
of Ref.~\cite{36},
both for fixed order ${\cal{O}}(\alpha_S)$ calculations and
including resummed terms in the MLLA approximation.
The MLLA results are shown for different choices of
the scale parameter~$\Lambda$.
Both the fixed order and MLLA results reproduce
the qualitative behavior of the data for $y_0\approx y_1$,
i.e.~they exhibit a falloff from the value of 1.5
with about the same slope as the data
as $y_0$ decreases.

\begin{figure}[tphb]
\begin{center}
\vspace*{-.5cm}
\begin{tabular}{cc}
  \hspace*{-.5cm}\epsfxsize=7cm\epsffile{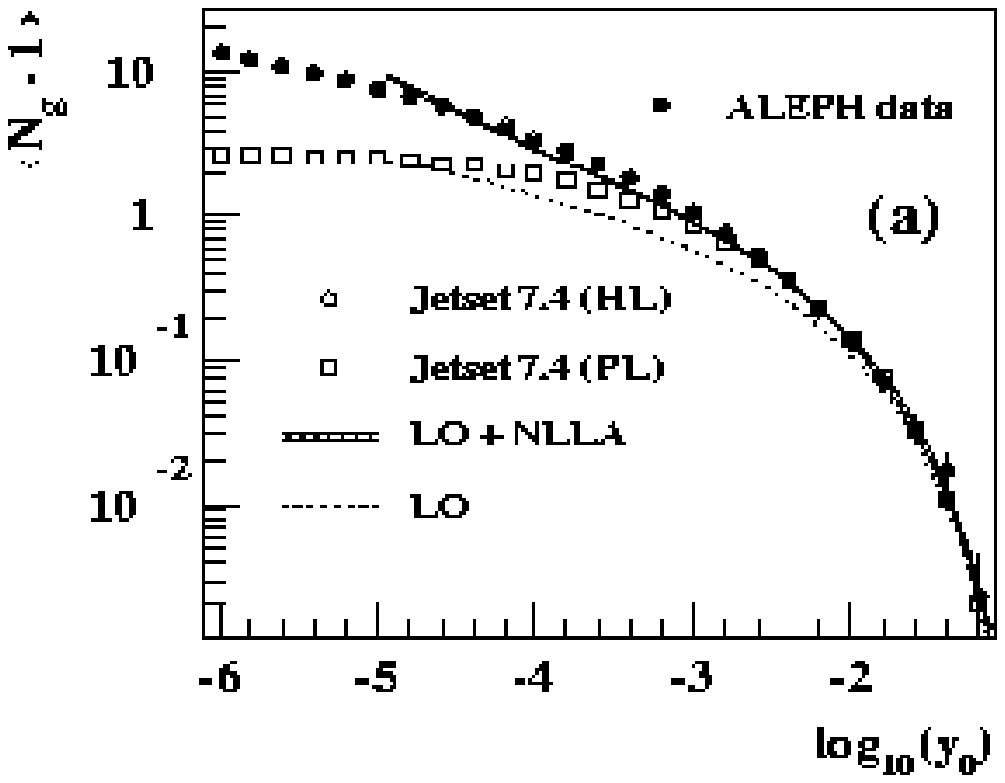} &
  \hspace*{-.5cm}
  \epsfxsize=7cm\epsffile{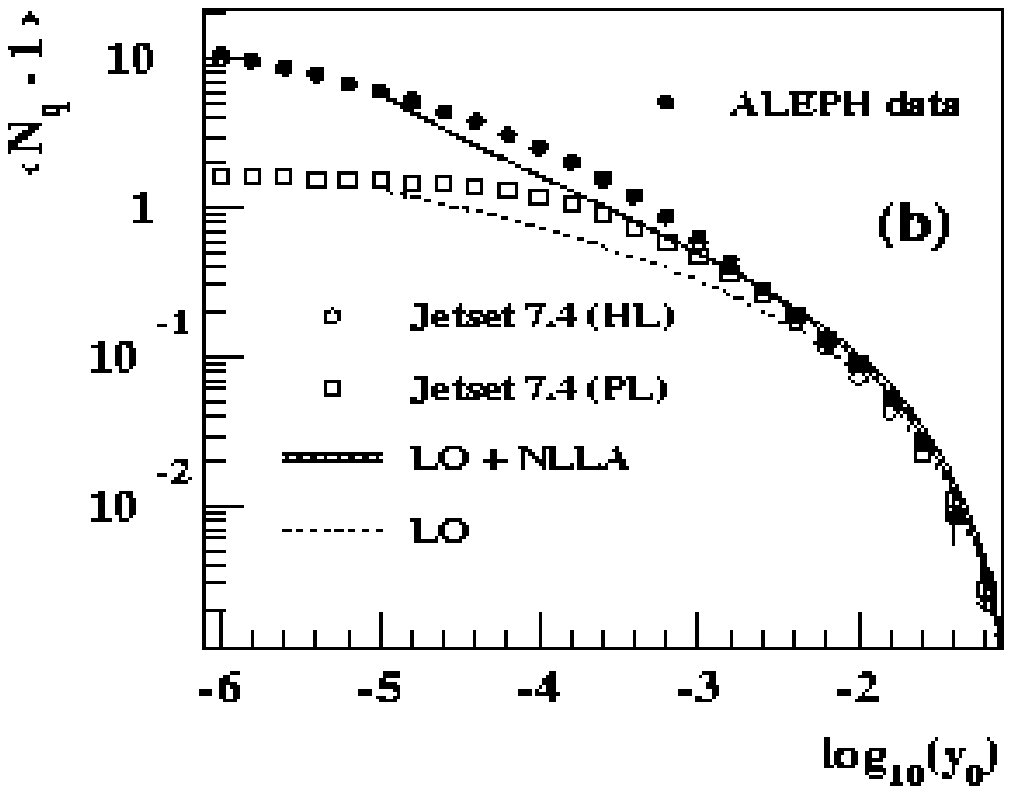} \\
\end{tabular}
\end{center}
\vspace*{-.5cm}
\caption{The subjet multiplicities of separated (a)~gluon and
(b)~quark jets~\cite{bib-alephsj}
in comparison to analytic results from Ref.~\cite{bib-seymoursj}
and to Monte Carlo predictions.
} 
\label{fig-aleph5a-5b}
\end{figure}

Subjet multiplicities have also been studied in
separated gluon and quark jets,
both experimentally~\cite{bib-alephsj}
and theoretically~\cite{bib-seymoursj}.
An example is shown in Fig.~\ref{fig-aleph5a-5b}.
The jets in this example are defined using the
Durham algorithm.
The analytic results~\cite{bib-seymoursj} are seen to
represent the data fairly well for large values of the
subjet resolution scale~$y_0$.

\section{Multiplicity of three-jet events}
\label{sec-threejets}

The multiplicity of two-jet events was defined from the 
product of the generating functions for single jets,
eq.~(\ref{72}),
implying their independence. 
The assumption of independence is valid because 
the angular separation of the jets is large, i.e. about~$180^\circ$.
For three-jet events,
various angular combinations are possible.
Mutual interference between the jets cannot be neglected.
Thus the particle multiplicity in three-jet events depends
on the angular topology of the events.

\begin{figure}[tphb]
\vspace*{-.6cm}
\begin{center}
  \epsfxsize=10cm
  \hspace*{.1cm}\epsffile{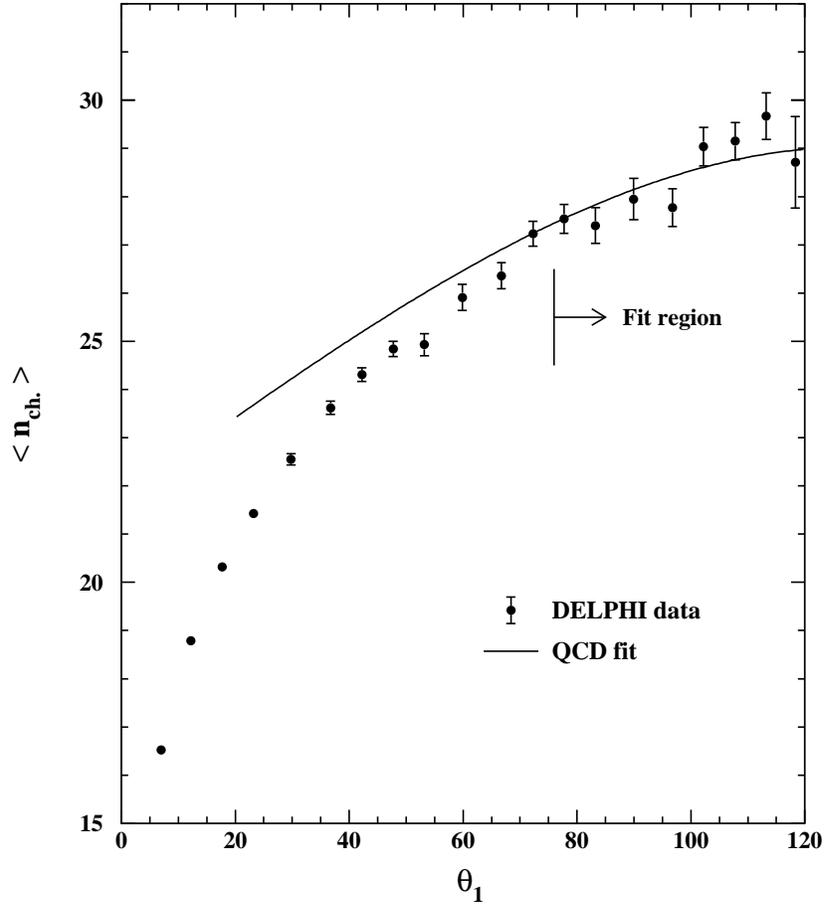}
\end{center}
\vspace*{-1.3cm}
\caption{Measurements~\cite{bib-delphi} of the
mean charged particle
multiplicity of three-jet Y events~\cite{bib-opal91}
as a function of the opening angle $\theta_1$ 
between the two lower energy jets,
for {\ecm}=91~GeV.
The events are selected using the {\durham} jet finder.
The solid curve shows the result of a one parameter
fit~\cite{bib-jwg} of eq.~(\ref{nfne}) 
to the data within the fit region shown.
} 
\label{fig-jwg3jet}
\end{figure}

To describe this effect it is necessary to go beyond the 
equations presented in this work,
to consider angular correlations between jets 
in the framework of the equations for the generating functionals 
or directly in terms of Feynman graphs. 
Therefore we refer to Refs.~\cite{khoc,dtro} 
where this problem is treated in detail. 
Up to further correction terms of order ${\cal{O}}(\alpha_S)$, 
the multiplicity of three-jet events can be approximately
written as the sum of the multiplicities of independent jets 
at properly defined energy scales:
\begin{equation}
  \langle n_{F\overline{F}G}\rangle \approx 2\langle n_{F}(y_{F})\rangle +
  \langle n_{G}(y_{G})\rangle ,  \label{nffg}
\end{equation}
where $y_{F}$=$\ln (E^{*}/\Lambda )$ and $y_{G}$=$\ln (p_{t}/2\Lambda )$,
with $E^{*}$ the quark energy in the quark-antiquark cms
and $p_t$ the gluon transverse momentum in this system. 
This expression has the correct limit when $p_t$$\,\rightarrow\,$0 
and the two jet configuration is restored.

Eq.~(\ref{nffg}) can be re-expressed using directly observable 
quantities such as the multiplicity of {\epem} events and the 
ratio $r$ of multiplicities between gluon and quark jets:
\begin{equation}
  \langle n_{F\overline{F}G}\rangle \approx 
    \langle n_{e^{+}e^{-}}(2E^{*})\rangle
  +\frac {1}{2}r(p_{t})\langle n_{e^{+}e^{-}}(p_{t})\rangle. 
  \label{nfne}
\end{equation}
These formulas also can be written in forms 
suitable for studies of gluon jets or to relate 
multiplicities at different scales (see Ref.~\cite{khoc}).

Expression~(\ref{nfne}) has been tested in
several experimental studies~\cite{bib-delphi,bib-jwg}.
As an example,
the data in Fig.~\ref{fig-jwg3jet} show the
charged particle multiplicity of so-called 
``Y events''\footnote{Y events
were first studied in~\cite{bib-opal91}.},
namely three-jet events for which
the angle between the highest energy jet and each of
the two other jets is about the same,
versus the opening angle $\theta_1$
between the two lower energy jets.
In Ref.~\cite{bib-jwg} it is found that eq.~(\ref{nfne}) describes 
this data accurately only if the angular separation between 
the jets is about 80$^\circ$ or larger,
as shown by the solid curve in Fig.~\ref{fig-jwg3jet}.
In Ref.~\cite{bib-edenkhoze},
this effect is explained to arise as a consequence of the 
bias in quark jet multiplicity introduced by the three-jet event
selection procedure.
A modified version of eq.~(\ref{nfne}) is suggested~\cite{bib-edenkhoze}
based on the dipole model of QCD (see Ref.\cite{bib-gustafsoneden}
and references therein),
which implicitly includes non-perturbative string effects.
The modified prediction is found to
yield a better description of the experimental measurements,
see Ref.~\cite{bib-delphiopen}.

\section{Evolution of distributions with decreasing phase space:
intermittency and fractality}
\label{sec-intermittency}

Multiplicity distributions can be studied not only in total phase space 
(as discussed in the previous sections
for very large phase space volumes) 
but within any subset of it. 
For a homogeneous distribution of particles, 
the average multiplicity decreases in proportion to the
considered volume,
whereas the fluctuations increase. 
The most interesting problem here is the law governing 
the growth of the fluctuations
and its possible deviation from a purely statistical law
related to the decrease of the average multiplicity. 
Such a deviation necessarily would be a consequence 
of the dynamics of the underlying interaction. 
In particular, it has been proposed~\cite{66} to search for
a power law behavior of the factorial moments 
in small rapidity intervals~$\delta y$:
\begin{equation}
  F_{q}\sim (\delta y)^{-\phi (q)}  \;\;\;\;\;\;\; (\delta y \rightarrow 0) ,
  \label{100}
\end{equation}
where $\phi (q)$$\,>\,$0. 
This proposal is motivated by an analogy to
turbulence in hydrodynamics, 
where a similar behavior is known as intermittency
and the $\phi (q)$ are called intermittency exponents. 

The power-like dependence in eq.~(\ref{100})
can be related to fractal properties 
of particle distributions in the available phase space, 
as was first discussed in Ref.~\cite{drje}. 
Earlier, the fractal properties of branching processes and, 
in particular, of QCD jets in {\epem} annihilation
were considered in Refs.~\cite{fey}-\cite{ven}. 
From the point of view of multiplicity,
intermittency implies a rather strong increase 
of the distribution's width,
and a longer tail at high multiplicities.

Experimental data from various processes over a 
wide energy range reveal a power law dependence
in agreement with eq.~(\ref{100})
and thus suggest the relevance of intermittency
to high energy multiparticle processes.
Simple branching models have been 
proposed to explain this phenomenon~\cite{abrp}-\cite{chhw}. 
Calculations based on perturbative QCD provide 
more specific predictions than these models, however.
The current state of affairs is reviewed in Ref.~\cite{14}. 
In the following we show how Quantum Chromodynamics
produces intermittency~\cite{38},~\cite{67}-\cite{71} 
and then compare the theoretical predictions
to recent experimental results.

We again stress that QCD deals 
with partons (quarks and gluons),
in contrast to experiment which
provides results based on hadrons. 
Local parton-hadron duality implies the proportionality of
inclusive distributions.
The validity of LPHD is not obvious for correlations,
however,
and is sometimes not fulfilled in Monte Carlo schemes. 
For phenomena such as intermittency,
one can therefore anticipate a qualitative agreement between
theory and data, at best.

In contrast to the previous sections, 
here we rely on the diagrammatic approach
rather than on the equations for the generating functions. 
This is because we must now deal with a small part of the 
parton content of a well developed jet, 
namely with those partons which are present in 
a small phase space volume.
The pre-history of the jet as a whole 
is significant for the subjet under consideration.
This is most readily addressed using the diagrammatic technique, viz.:
\begin{enumerate}
  \item the primary quark emits a hard gluon with energy $E$ in
    the direction of the angular interval $\theta$, 
    but not necessarily hitting the window;
  \item the emitted gluon produces a jet of partons with parton splitting
    angles larger than the window size;
  \item among those partons there exists a parton with energy $k$
    which initiates a subjet which hits the window;
  \item all decay products of the subjet exactly cover the angle $\theta$.
\end{enumerate}
This picture dictates the rules of calculation 
of the $q$th correlator of the whole jet. 
The $q$th correlator of the subjet 
$\Delta N^{(q)}(k\theta )$ should be averaged
over all possible ways it can be produced, 
i.e.~convoluted with the inclusive spectra of partons 
$D^{\theta }$ in the whole jet and with the probability 
of creation of the jet ($\alpha _{S}K_{F}^{G}$).
Analytically, this is represented by
\begin{equation}
  \Delta N^{(q)}\left ( Q\theta _{0}, 
     \frac {\theta _{0}}{\theta }\right ) \propto
  \int ^{Q} \frac {dE}{E} 
    \frac {\alpha _{S}}{2\pi }K_{F}^{G}\left (\frac {E}{Q}
  \right )\int ^{E}\frac {dk}{k}D^{\theta }\left (\frac {E}{k};E\theta _{0},
  k\theta \right )\Delta N^{(q)}(k\theta ) ,  \label{101}
\end{equation}
where $\Delta N^{(q)}$$\,\equiv\,$$F_{q}\langle n \rangle ^{q}$ 
is the unnormalized factorial moment 
(on the left hand side for the whole jet, and on the right hand
side for the parton subjet with momentum $k$ within the angle $\theta $).
Since the unnormalized moments increase with energy,
whereas the parton energy spectrum decreases,
the product $D^{\theta }\Delta N^{(q)}(k\theta )$ 
has a maximum at some energy,
and the integral over momenta can be calculated 
using the method of steepest descent. 
Leaving aside the details (see Ref.~\cite{38}), 
we describe the general structure of the correlator 
in the case of a fixed coupling constant, $\gamma_{0}$=constant:
\begin{equation}
  \Delta N^{(q)} \propto \Delta \Omega \left 
    ( \frac {\theta _{0}}
   {\theta }\right )^{\frac {\gamma _{0}}{q}}\left 
    ( \frac {E\theta }{\mu }\right )^{q\gamma _{0}} ,
  \;\;\;\;\;\;  (\mu = constant ) ,   \label{102}
\end{equation}
where the three factors represent the phase space, 
the energy spectrum factor, 
and the $q$th power of the average multiplicity. 
To obtain the normalized moment, 
expression~(\ref{102}) should be divided by the $q$th power 
of that part of the mean multiplicity of the whole jet which appears 
inside the window $\theta$, 
i.e.~by the proportion of the total average multiplicity 
corresponding to the phase space volume $\Delta \Omega $:
\begin{equation}
  \Delta N(\theta ) \sim \Delta \Omega \Delta N(\theta _{0}) .  \label{103}
\end{equation}
If the analysis is performed in $D$-dimensional space, 
the phase space volume obeys
\begin{equation}
  \Delta \Omega \sim \theta ^{D} ,  \label{104}
\end{equation}
where $\theta$ corresponds to the minimum linear size 
of the $D$-dimensional window
which stems from the singular behavior of parton propagators in
Quantum Chromodynamics (see Ref.~\cite{38}). 
This allows the factorial moments to be represented 
as the product of a purely kinematic factor depending on the
dimension of the analyzed space, 
and of a dynamic factor not related to that dimension
which depends on the coupling constant,~i.e.
\begin{equation}
  F_{q} \sim \theta ^{-D(q-1)}\theta ^{\frac {q^{2}-1}{q}\gamma _{0}}. 
  \label{105}
\end{equation}
For small angular windows, $\theta \sim \delta y$. 
The intermittency indices defined by eq.~(\ref{100}) are then
\begin{equation}
  \phi (q) = D(q-1) - \frac {q^{2}-1}{q}\gamma _{0} .  \label{106}
\end{equation}

Expression~(\ref{106}) is valid for moderately small windows, 
for which the condition $\alpha _{S}\ln \theta _{0}/\theta <1$ is fulfilled. 
For extremely small windows it is necessary to account 
for the running property of the coupling constant.
In this case the constant $\gamma _{0}$ should be replaced 
by an effective value $\langle \gamma \rangle$, 
which depends logarithmically on the width of the
window $\theta$ and may be approximated by~\cite{38}
\begin{equation}
  \langle \gamma \rangle \approx
    \gamma _{0}(1+\frac {q^{2}+1}{4q^2}\epsilon),
  \label{107}
\end{equation}
where
\begin{equation}
  \epsilon = \frac {\ln \theta _{0}/\theta }{\ln (E\theta_{0}/\mu )} \leq 1.
\label{108}
\end{equation}
Thus intermittency indices for very small windows are markedly smaller 
than for intermediate sized windows (fixed coupling regime),
especially for low rank moments. 
Moreover, the simple power law, eq.~(\ref{100}),
becomes modified by logarithmic corrections, 
and the intermittency indices depend 
on the size of the chosen interval. 
The resulting curve of $\ln F_{q}$ versus $-\ln \theta $ 
is characterized by two distinct regions with different slopes. 
A rather steep linear increase is predicted
for moderately small windows $\theta$,
with a slope given by eq.~(\ref{106}).
For smaller window sizes,
this behavior is replaced by a much slower quasi-linear 
increase given by eqs.~(\ref{107}) and~(\ref{108}). 
It is easy to determine the location of the transition point 
between the two regimes and to show that this point shifts 
to smaller window sizes at larger values of $q$.
In any case, the factorial moments of any rank increase 
as the size of the interval decreases. 
This demonstrates that the fluctuations of the
multiplicity distribution become stronger for smaller intervals and, 
more importantly, 
that they significantly exceed Poisson fluctuations.

Above, we have described the results of the 
double-logarithmic approximation,
with corrections from the running 
coupling constant, eq.~(\ref{107}).
Corrections from 
the modified-leading logarithm approximation
(see eq.~(\ref{geff}) below) are comparatively small,
i.e.~about 10\%. 
The MLLA terms move the transition point between the power law 
and quasi-power law regimes to slightly smaller windows for all
moments except the second one, 
which moves to somewhat larger windows.
This tendency can be attributed to the mutual influence 
of the energy spectrum and the average multiplicity. 

Of greater interest are qualitative effects introduced in MLLA
which yield a functional dependence on the rank~$q$ 
from terms proportional to~$q\gamma$. 
Consider an analogy to statistical mechanics,
presented in Ref.~\cite{38}, where the quantity
\begin{equation}
  \Phi = 1-\frac {\phi (q)+1}{q}   \label{phif}
\end{equation}
is interpreted as the free energy
and the rank $q$ as the inverse temperature $\beta =1/kT$.
In the lowest order approximation,
$\Phi$ increases monotonically with $q$.
Higher order corrections produce a maximum in $\Phi$
at values of $q$ where $H_{q}$ exhibits its first minimum 
(see~eq.~(\ref{71})), i.e.~at
\begin{equation}
  q_{cr} \approx \frac {1}{h_{1}\gamma _{0}}+\frac {1}{2} \approx 5 .  
  \label{109}
\end{equation}
In statistical mechanics, this corresponds to zero entropy, 
i.e.~to a phase transition. 
In Quantum Chromodynamics
it just indicates the role of the new parameter~$q\gamma$. 
This feature is related to the singularities of the 
generating function and to the pinching behavior of the zeros 
of the truncated generating function discussed
in Section~\ref{sec-singularities}.

The above results can be restated in terms of fractals. 
The power-like behavior of factorial moments suggests 
fractal properties of particle (parton) distributions in phase space. 
According to the general theory of fractals 
(see Ref.~\cite{14} and references therein), 
intermittency indices are related to fractal (R\'{e}nyi) 
dimensions $D_{q}$ by the formula
\begin{equation}
  \phi (q) = (q-1)(D-D_{q}) .  \label{110}
\end{equation}
In the double-logarithmic approximation 
taking into account eqs.~(\ref{106}) and~(\ref{107}),
one obtains
\begin{equation}
  D_{q} = \frac {q+1}{q}\langle \gamma \rangle = (\gamma _{0} + 
  \frac {\gamma _{0}}{q})(1+\frac {q^{2}+1}{4q^2}\epsilon ) .\label{111}
\end{equation}
The first term in the first bracket corresponds to mono-fractal 
behavior in the case of fixed coupling and is due to the 
increase in average multiplicity. 
The second term in the first bracket corresponds to
multi-fractal properties and is related 
to the falloff of the energy spectrum. 
One can easily obtain the multi-fractal spectral 
function in this case (see Ref.~\cite{38}). 
It is found that the fractality in Quantum Chromodynamics 
has a purely dynamical origin ($D_{q}\sim \gamma _{0}$) 
related to the cascade nature of the process, 
and that the kinematic factor in relation~(\ref{105}) 
has an integer dimension.
The term with $\epsilon$ in eq.~(\ref{111}) accounts for
the running property of the coupling constant.
This term slightly violates pure multi-fractal behavior, 
giving rise to a slower increase of factorial moments at small angles.
The fixed coupling regime corresponds to $\epsilon$=0, yielding
\begin{equation}
  D_q = \gamma_0 \frac{q+1}{q} .
  \label{eq-dq}
\end{equation}
Further corrections to eq.~(\ref{111}) appear in MLLA~\cite{38}. 
The general form of eq.~(\ref{111}) remains valid,
but $\gamma _{0}(Q)$ is replaced by an
effective $\gamma _{0}^{eff}(Q)$ which depends on~$q$: 
\begin{equation}
  \gamma _{0}^{eff}(Q)=\gamma _{0}(Q)+
  \gamma _{0}^{2}(Q)\frac {\beta_0}{4N_c}\nonumber \\
  \left[-B_0 \frac {q-1}{2(q+1)}+\frac {q-1}{2(q+1)(q^{2}+1)}
     +\frac {1}{4}\right],
  \label{geff}
\end{equation}
where
\begin{equation}
  B_0 =\frac {1}{\beta _{0}}\left[ \frac {11N_c}{3}
   +\frac {2n_f}{3N_{c}^{2}}\right].
  \label{B}
\end{equation}
For large ranks $q$ one obtains a negative shift:
\begin{equation}
  \gamma _{0}^{eff}=\gamma _{0}
   -\gamma _{0}^{2}\frac {\beta _{0}(2B_{0}-1)}{16N_c}.
\label{gas}
\end{equation}
This corresponds to the general tendency of 
negative shifts from MLLA corrections.  
However, this represents only a partial accounting for
higher order effects. 
Other corrections stemming from energy conservation 
are more difficult to include
and are not incorporated into the above equations.

An alternative approximation for the $\epsilon$ dependence of 
$D_q$ in DLA has been suggested in Ref.~\cite{70}:
\begin{equation}
  D_{q}=2\gamma _{0}(Q)\frac {q+1}{q}\cdot \frac {1-\sqrt {1-\epsilon }}
  {\epsilon }. \label{bmpd}
\end{equation}
A somewhat different expression for $D_q$ was derived 
in Refs.~\cite{68, 69} starting from the expression for cumulant moments,
supposing that they converge to factorial moments at high energies
up to a factor of $q^{-2}$ as discussed above:
\begin{equation}
  D_{q}=2\gamma _{0}(Q)\frac {q}{q-1}\cdot \frac {1-\sqrt {1-\epsilon }\left( 
  1-\frac {\ln (1-\epsilon )}{2q^2}\right)}{\epsilon }. \label{wojw}
\end{equation}

The variety of proposed forms for $D_q$ suggests that
there are neglected effects which are important for a
proper analytic treatment. 
In particular, energy conservation constraints are just as severe 
in small phase space windows as in total phase space (see~\cite{jlmp}). 
This problem has not yet been solved.
Thus the above formulas should be compared to experiment
on a qualitative level only.
In addition, variables such as rapidity or pseudo-rapidity should be used.
Other variables, such as the invariant mass $Q^2$,
are more suitable for studies of Bose-Einstein correlations 
than of intermittency (for more details, see~\cite{abds}).

In general, the qualitative trends predicted by the QCD
equations are observed in experiment. 
Intermittency is observed to be stronger in {\epem} annihilations
than in reactions with initial hadrons,
and weaker yet in nucleus-initiated processes~\cite{14}.
This is related to the increasing number of competing 
subprocesses (sources) in reactions with complicated structure,
which smoothes out the intermittent behavior of individual sources. 
Due to the stronger effect and higher statistics,
intermittency studies in {\epem} collisions are more conclusive. 
Recent studies from LEP~\cite{opi,L3i,dei} represent
considerable improvements upon earlier results
(summarized in Ref.~\cite{14}). 
In the new studies,
the experimental data are compared with
QCD predictions for local multiplicity fluctuations 
in 1-, 2- and 3-dimensional intervals at 
\mbox{LEP-1} and \mbox{LEP-2} energies. 
The general predictions of QCD are confirmed by the data.
The factorial moments rise in an approximately linear manner
for large bins and level off for smaller ones.
The dependence on the bin dimension, 
the order of the factorial moments,
and on energy are qualitatively reproduced, 
i.e.~the moments increase as the bin dimension increases,
become larger for higher orders,
and increase with energy. 
The analytic predictions depend sensitively 
on the QCD parameter $Q_0$,
as is also true for the predictions of 
the QCD Monte Carlo models.

\begin{figure}[tphb]
\vspace*{-3cm}
\begin{center}
  \epsfxsize=14.5cm
  \hspace*{-.7cm}\epsffile{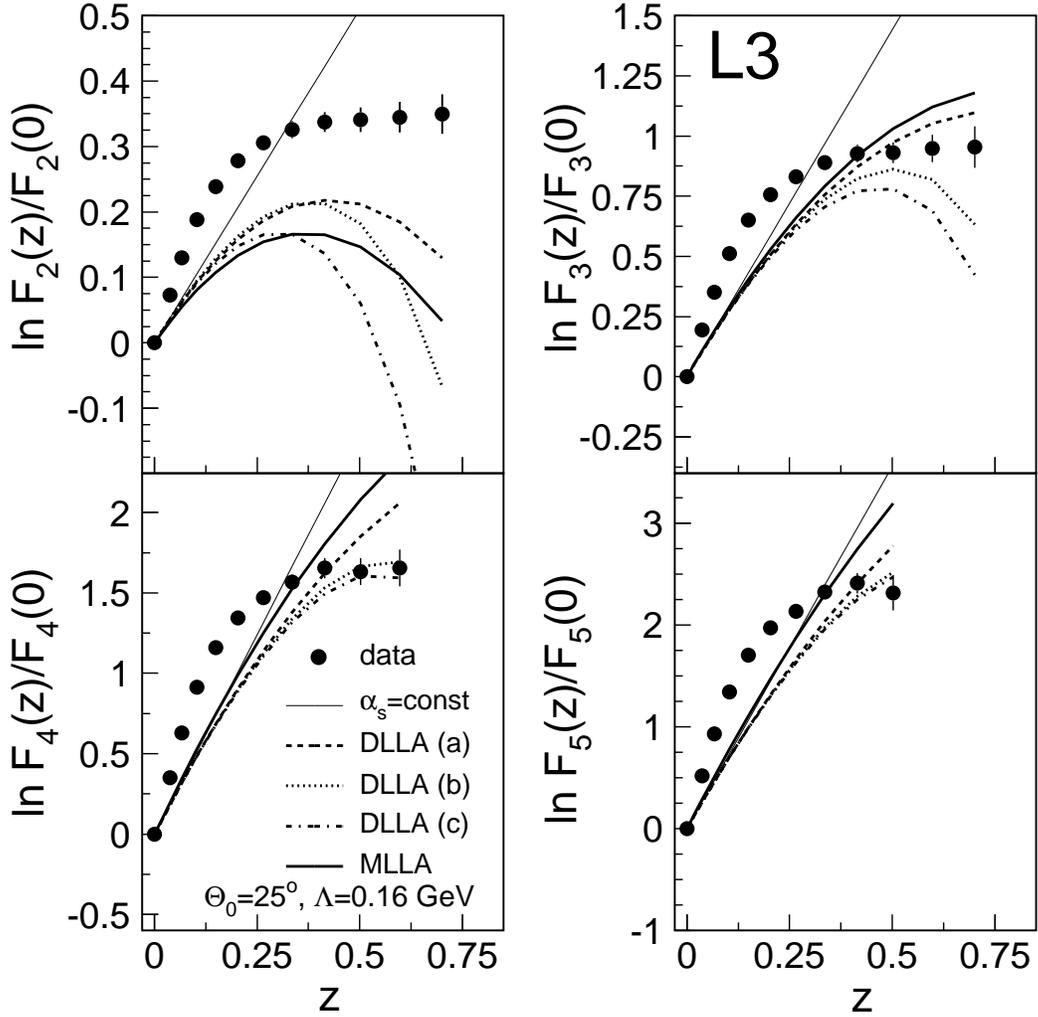}
\end{center}
\vspace*{-.5cm}
\caption{Measurements of the ratio $F_q(z)/F_q(0)$ 
from {\epem} Z$^0$ data
for \mbox{$q$=2--5,}
as a function of
the scaling variable~$z$$\,\equiv\,$$\epsilon$~\cite{L3i}.
The data are compared to QCD analytic calculations 
with fixed coupling strength~$\alpha_S$ (eq.~(\ref{eq-dq})):
DLA~(a) (eq.~(\ref{111})), DLA~(b) (eq.~(\ref{bmpd})), 
DLA~(c) (eq.~(\ref{wojw})), 
and MLLA (eq.~(\ref{geff})).
Note that DLA is referred to as ``DLLA'' in
the figure legend.
} 
\label{fig-l3int}
\end{figure}

On a quantitative level,
significant differences are observed 
between theory and data, however.
The factorial moments increase faster in experiment than in
the calculations as the phase space interval decreases,
and generally level off at larger bin sizes than the theory as
the intervals continue to decrease.
An example is shown in Fig.~\ref{fig-l3int}~\cite{L3i},
for which $z$$\,\propto\,$$\ln(\Theta_0/\Theta)$ measures the size
of the phase space window~$\Theta$$\,\leq\,$$\Theta_0$.
Thus $z$=0 corresponds to the maximum phase space region
$\Theta$=$\Theta_0$ (where $\Theta_0$=25$^\circ$ in this example),
while $z$ increases as the window size decreases.
Variation of the QCD parameters, e.g.~$Q_0$,
does not improve the level of agreement between
theory and data.

Given the quantitative discrepancies between theory and data,
an evaluation of QCD parameters using intermittency
is not possible at present.
The discrepancies are not surprising given the crucial role played by
energy conservation in total phase space,
see Sections~\ref{sec-gluodynamics}-\ref{sec-exactsolutions},
and the absence of energy conservation from the
analytic description here.
Energy conservation is included in Monte Carlo studies, 
and they provide a reasonably good description
of the data as is shown in Refs.~\cite{opi,L3i,dei}.

The increase of correlations for smaller bin sizes, 
as described by the behavior of the cumulants, 
reveals the important role of genuine four- 
and five-particle correlations in {\epem} annihilations~\cite{opi}.
This is in contrast to nucleus collisions
where it has not yet proved possible to observe 
correlations for groups of more than two particles.
The fractal nature of the particle distributions in {\epem} events
has been also demonstrated~\cite{L3b}
using the so-called bunching parameters~\cite{ck, ckk}.

The fractality of particle distributions in 
limited phase space volumes suggests fractal properties 
for colliding objects in ordinary space.
Surely, owing to its dynamical origin, 
such a structure would itself be dynamical, 
i.e.~rapidly evolving in space and time. 
There are two reasons to favor this interpretation. 
First, the evolution of the parton shower 
in ordinary space must give rise to a tree-like
structure of a fractal type which should evolve 
due to the cascade evolution.
Second, lattice computations in SU(2) gluodynamics~\cite{72} 
have shown that the system of gluons in the vicinity 
of the phase transition is fractal in the sense that it fills 
a volume $V$ bounded by a surface $S$
which are related by the formula $V\sim S^{1.12}$. 
This is typical of fractal objects:
the exponent would equal 1.5 for ordinary three-dimensional objects.

The geometric fractality of macroscopic bodies has been 
revealed by measurements of the power-like shape 
of their structure functions when point-like particles 
(photons, electrons, neutrons, etc.) are scattered from them. 
Using this example, one might try to measure~\cite{73}
structure functions in deep inelastic processes,
in order to determine the fractal dimensions of the scattered partners. 
Theoretically, this has been considered only at the level of models.
Experimental difficulties have prevented 
direct tests of these models. 
The question of the fractal geometry 
of particles in ordinary space-time therefore remains open.

\section{Heavy quark jets}
\label{sec-heavy}

The multiplicity of heavy quark jets is of special interest 
because it exhibits two peculiar effects with analogues in QED. 
It is well known from QED that photon emission from muons 
is strongly suppressed compared to emission from electrons 
(as the ratio of corresponding masses squared). 
This suppression is related to the different masses in the
propagators and to the vector nature of the emitted photons. 
The second effect,
known from cosmic ray studies and named after Chudakov, 
is that two opposite charges in an electron-positron pair 
screen each other if they are not well separated,
i.e.~until the distance between them becomes larger than
the inverse transverse momentum of the resolving quanta.

In QCD, the emission of gluons by heavy quarks 
is suppressed in a similar manner.
At relatively small angles ($\theta \ll 1$),
gluons are emitted according to the formula
\begin{equation}
   \frac {\theta ^{2}}{(\theta ^{2}+\theta _{M}^{2})^2},   \label{thet}
\end{equation}
where $\theta _{M}=M/E_Q$,
with $M$ and $E_Q$ the mass and energy of a heavy quark. 
The form of the numerator in eq.~(\ref{thet})
is due to the vector nature of the radiated quanta. 
The denominator equals the quark propagator squared.
The larger the value of $M$, 
the stronger the suppression of forward 
($\theta<\theta _M $) emissions.
This gives rise to ``ring-like'' events
as advocated in Refs.~\cite{80}-\cite{82} 
using somewhat different but similar arguments,
or to the ``dead cone'' phenomenon~\cite{79,83}. 
Note that the suppression of gluon emission at small angles 
does not necessarily imply a suppression of emissions with low
transverse momenta. 
The origin of the ring-like events is the 
large mass of heavy quarks,
which is not the same as $k_t$~ordering.

There is a difference between QED and QCD with respect 
to the properties of the emitted quanta, however. 
While photons do not have electric charge, 
gluons carry color. 
Radiated photons do not multiplicate,
in contrast to gluons which produce jets. 
Gluons emitted at rather large angles ($\theta\geq\theta _M)$ 
multiplicate in the usual manner. 
For gluons emitted at smaller angles, 
it is necessary to account for interference between the secondary
emissions and possible emission by the quark. 
This situation is analogous to the Chudakov effect. 
In QCD, the analogous effect
for partons moving close to one another
is called color transparency.

However, there is a substantial difference for heavy quarks 
(for a more detailed discussion, see Ref.~\cite{khoc}). 
Because its mass is large, the heavy quark moves
more slowly than the emitted gluon, 
and the two become separated more quickly
than is the case when the quark is light.
Therefore color charge screening is not
as effective for heavy quarks as it is for light quarks
and interference effects are less important.
It can be shown that the primary gluon can be treated as
an incoherent state with respect to the quark,
and that, as a consequence,
gluon emission by massive quarks is suppressed 
for angles $\theta <\theta _M $ compared to light quarks.

To obtain an expression for the multiplicity of partons 
accompanying a heavy quark pair in {\epem} annihilations, 
the distribution of gluons emitted by the quark,
eq.~(\ref{thet}),
should be convoluted 
with the multiplicity of a gluon subjet. 
According to the above discussion,
this expression can be integrated over all transverse momenta 
exceeding a fixed value $Q_0$ to obtain
\begin{equation}
  \langle n_{Q\overline{Q}}(W)\rangle =2\int _{M^2}^{W^2}\frac {d\kappa ^{2}}
  {\kappa ^{2}}\left( 1-\frac {M^2}{\kappa ^2}\right) \int _{Q_0}^{\kappa }
  \frac {dk_t}{k_t} K _{F}^{G}(x)\frac {\alpha _{S}(k_t )}{4\pi }
  \langle n_{G}(k_t )\rangle ,   \label{nqw}
\end{equation}
with $K _{F}^{G}(x)$ given by eq.~(\ref{55}), 
$k_{t}\,$$\,=x\kappa$,
and $\kappa^{2}\,$=$\,W^{2}[\sin^{2}(\theta /2)+\theta_{M}^{2}/4]$. 
Eq.~(\ref{nqw}) yields the 
relation between the multiplicities accompanying 
quark-antiquark pairs with heavy and light components 
produced in {\epem} processes~\cite{79}:
\begin{equation}
  \langle n_{Q\overline {Q}}(W)\rangle =\langle n_{q\overline {q}}(W)\rangle 
  -\langle n_{q\overline {q}}(\sqrt {e}M)\rangle ,
  \label{nqq}
\end{equation}
which is valid up to terms of order ${\cal{O}}(\alpha_S$). 
Therefore the difference 
between the total multiplicities 
of events with heavy and light quarks is
\begin{equation}
  \langle n_{Q\overline {Q}}^{tot}(W)\rangle 
     -\langle n_{q\overline {q}}^{tot}(W)
  \rangle =\langle n^{dc}_{Q\overline {Q}}(M)\rangle 
     -\langle n_{q\overline {q}}
  ^{tot}(\sqrt {e}M)\rangle , \label{QQqq}
\end{equation}
where $\langle n^{dc}\rangle$ denotes the decay multiplicity of a 
heavy quark with mass~$M$.  
The difference does not depend on the primary energy $W$ but only
on the heavy quark mass.

\begin{figure}[tphb]
\begin{center}
  \epsfxsize=13cm
 \hspace*{2cm} \epsffile{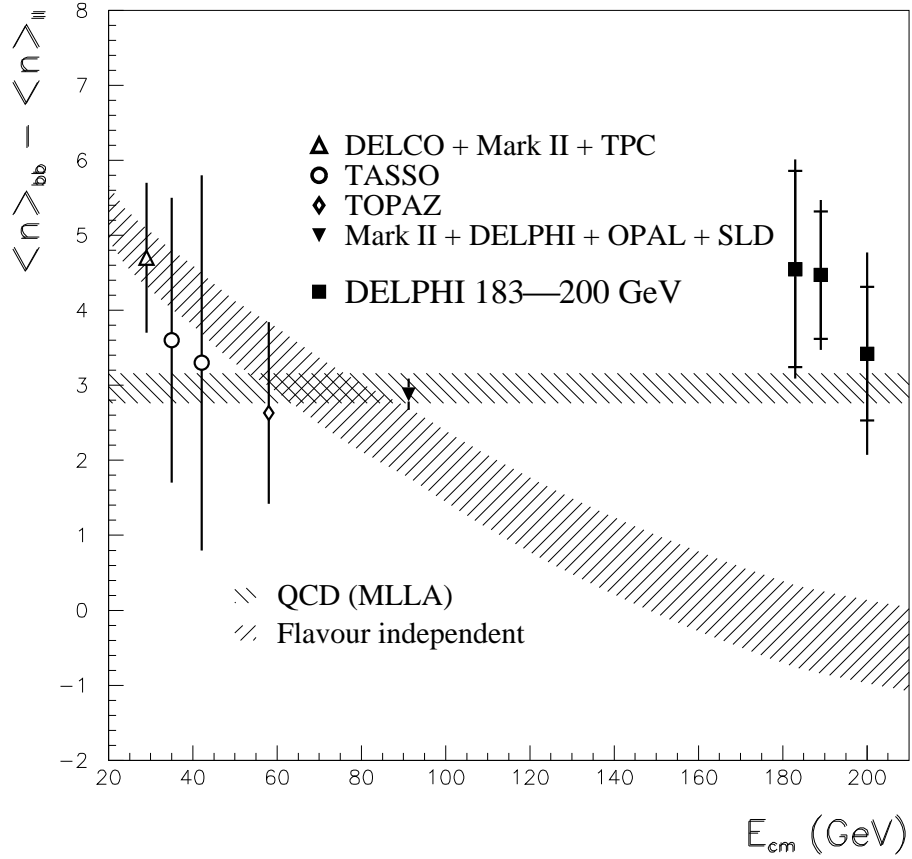}
\end{center}
\caption{Measurements of the difference $\delta_{Q\ell}$ between
the total event multiplicities of heavy and light quark events
in {\epem} annihilations,
versus the center-of-mass energy,
compared to a prediction motivated by MLLA QCD
and the prediction of the
so-called na\"{\i}ve model of energy rescaling
(the na\"{\i}ve model is referred to as ``Flavour independent''
in the figure legend).
The figure is taken from Ref.~\cite{bib-delphibvsl}.
} 
\label{fig-heavy}
\end{figure}

In contrast to eq.~(\ref{QQqq}),
the so-called na\"{\i}ve model~\cite{85}-\cite{87} considers the
difference between heavy and light quark event multiplicities
to be the consequence of energy rescaling only:
\begin{equation}
   \langle n_{Q\overline {Q}}(W)\rangle =\langle n_{q\overline {q}}
   [(1-\langle x_{Q}\rangle )W]\rangle , \label{wres}
\end{equation}
where $\langle x_{Q}\rangle $ denotes the mean energy fraction of
the heavy quark,
$\langle x_{Q}\rangle$$\,\equiv\,$$2\langle E_Q\rangle/W$.
In this case the difference in total multiplicities between
heavy and light quark events is energy dependent,
decreasing towards zero as $W$ becomes larger.

Measurements of the difference
$\delta_{Q\ell}\,$$\,\equiv\,$$\,\langle n_{Q\overline {Q}}^{tot}(W)\rangle 
-\langle n_{q\overline {q}}^{tot}(W)\rangle$
between heavy and light quark event multiplicities
in {\epem} annihilations
are shown as a function of the center-of-mass energy
in Fig.~\ref{fig-heavy}~\cite{bib-delphibvsl}.
The heavy and light quark jet samples consist of
b and uds flavored events, respectively.
The data support the QCD hypothesis that
$\delta_{Q\ell}$ is independent of energy, eq.~(\ref{QQqq}),
over the energy dependent hypothesis of the na\"{\i}ve model.
Note the essential contribution that the highest energy points
(from LEP-2) provide for this conclusion.
The initial QCD estimates~\cite{79} predicted a rather large
value $\delta_{Q\ell}$$\,\approx\,$5.5.
Later this prediction was reduced to
$\delta_{Q\ell}$$\,\approx\,$3.7~\cite{87}.
The horizontal band in Fig.~\ref{fig-heavy}
at $\delta_{Q\ell}$$\,\approx\,$3~\cite{bib-delphibvsl} is not,
strictly speaking,
a theoretical estimate,
but an admixture of the the MLLA statement that
$\delta_{Q\ell}$ is constant together with
the experimental measurement from the Z$^0$ pole.

We will also briefly mention an interesting feature of 
heavy quark fragmentation.
Because of their large masses, heavy quark hadrons tend to
retain a substantial portion of the primary quark energy,
which can be described
in the framework of perturbative QCD 
(for more details, see Ref.~\cite{khoc}). 
The comparatively long lifetime of heavy quark hadrons plays
an important role at extremely high energies in 
the spatial multiplicity distribution of cosmic ray showers~\cite{drya}. 
Because of relativistic effects and smaller interaction cross sections, 
high energy heavy hadrons created in cosmic ray showers
penetrate more deeply into detectors on average
than light quark hadrons,
giving rise to so-called long flying cascades. 
The multiplicity of the main cascade in these events
is somewhat reduced,
while particles appear at longer distances 
from the primary interaction point,
compared to cascades without heavy quarks.
Thus the mean length of the cascade is larger than normal,
from which term ``long flying'' is derived.

\section{Conclusions}

The distribution of particle multiplicity in
high energy inelastic processes provides essential information
on the dynamics of strong interactions. 
Its importance stems from the fact 
that it contains all the correlations
between final-state hadrons in an integrated form.
It is comparatively easy to measure,
in contrast to differential correlation 
functions which require much more detailed analysis.

Higher order perturbative solutions of the equations
for the generating functions
based on the running coupling constant,
or the exact solution based on a fixed coupling,
exhibit qualitative features
which are absent in lowest order.
Features predicted for partons,
such as the oscillations of the $H_q$ ratios or the larger 
factorial moments of quark jets compared to gluon jets,
are experimentally observed for hadrons.
The shape of the multiplicity distribution is mostly defined 
by soft particles created during the final stages of
development of the cascade. 
The success of the theoretical methods implies the applicability 
of higher order perturbative techniques to the soft stages of evolution,
and the suitability of local parton-hadron duality
to multiplicity based measurements.
The principal features of multiplicity predicted by QCD
are observed qualitatively both
in {\epem} and hadron-initiated processes. 
This prompts speculation about the similarity of the production 
mechanisms in the two cases. 
This similarity probably occurs because the parton 
wave functions of the colliding particles in hadron collisions
are prepared long before the scattering,
and are described by a space-like bremsstrahlung cascade
with angular ordering similar to {\epem} case.

The historical development of the perturbative approach 
to multiplicity is summarized in the introduction. 
The initial excitement created by the first applications
of QCD to multiplicity
gave way to disappointment over the extremely 
wide multiplicity distribution predicted by the theory 
at lowest order, 
although it was soon realized that the corrections
were rather large. 
Now the importance of higher orders,
up to NNLO and beyond, is well recognized, 
and it is possible to account for them in a consistent manner.
The results reveal approximate $F$ (or KNO) scaling,
with some dependence on the coupling constant 
in the preasymptotic regime. 

The long standing discrepancy between theory and experiment
for the value of the ratio of the average multiplicities 
in gluon and quark jets has now been effectively resolved.
Measurements of the higher moments of the multiplicity
distribution are found to be in excellent agreement
with the computer solutions of the QCD equations.
The rise in average multiplicity with energy is
described in a satisfactory manner,
using a sensible value of the strong coupling strength $\alpha_S$
for both quark and gluon jets.
The evolution of the distributions in smaller phase space 
regions has also been successfully described,
at least at a qualitative level. 
Energy conservation in the calculations
and a proper unbiased definition of jets in the experiments
are crucial for the quantitative tests of the QCD results,
and are the principal developments which allowed
recent progress in the field.
The astonishing success of the computer solutions 
also demonstrates
the importance of using the correct limits of integration
over the parton energy splitting variables,
i.e.~the importance of the treatment of the
boundary between the perturbative and non-perturbative regions.

In combination with predictions for inclusive spectra 
and various correlation functions, 
the above results suggest that 
Quantum Chromodynamics can be successfully applied 
to predict qualitative features of soft processes. 
Angular ordering inside each jet and the collective
nature of the inter-jet flows are essential considerations
for multiplicity and its energy evolution in individual events.
Analytic approaches provide a means to predict qualitative
new effects and to gain insight into the behavior of the solutions
(dependence on energy scale,
the number of active quark flavors,~etc.).
Recent progress in that direction, described at some length above, 
gives hope for further success.

\section{Acknowledgments}
\vspace{2mm}
We are very much indebted to A. Capella, Yu.L. Dokshitzer, 
G. Gianini, R.C. Hwa,
C.S. Lam, B.B. Levtchenko, V.A. Nechitailo and
J. Tran Thanh Van for collaboration and discussions.

This work was supported by the Russian Fund for Basic Research,
by INTAS,
and by the US Department of Energy under grant DE-FG03-94ER40837.

\end{document}